\documentclass{aa}  

\usepackage{graphicx}
\usepackage{natbib}
\usepackage{upgreek}
\usepackage[switch]{lineno}
\usepackage{subfig}
\usepackage{float}
\usepackage{placeins}

\bibpunct{(}{)}{;}{a}{}{,} 
\usepackage{txfonts}
\usepackage[colorlinks=true, backref=page]{hyperref}
\hypersetup{allcolors=blue}

\newcommand{\hi}{H$\,\scriptstyle{\mathrm{I}}$\xspace}
\newcommand{\hii}{H$\,\scriptstyle{\mathrm{II}}$\xspace}
\newcommand{\nhi}{$N(\mathrm{H\,\scriptstyle{I}})$\xspace}

\newcommand{\nhd}{$N(\mathrm{H_2})$\xspace}
\newcommand{\wco}{$W_\mathrm{CO}$\xspace}
\newcommand{\hd}{$\mathrm{H}_2$\xspace}
\newcommand{\xco}{$X_\mathrm{CO}$\xspace}

\newcommand{\cmmnt}[1]{}

\newcommand{\F}{\textit{Fermi}\xspace}

\newcommand{\halo}{FCES~G78.74$+$1.56\xspace}
\newcommand{\ig}{FCES~G80.00$+$0.50\xspace}
\newcommand{\sw}{FCES~G78.83$+$3.57\xspace}
\newcommand{\ob}{FCES~G85.00$-$1.78\xspace}
\newcommand{\NNhalo}{CoExt\xspace}
\newcommand{\NNig}{CoCent\xspace}
\newcommand{\NNsw}{CoWest\xspace}
\newcommand{\NNob}{OffExc\xspace}

\begin{document}

   \title{Multiple emission components in the Cygnus cocoon\\ detected from \textit{Fermi}-LAT observations
   \thanks{The template used to model source \ig, the map of excess counts in figure~\ref{fig:DataVsExcess} (right), the spectral points from section~\ref{sec:spectra}, the map of total neutral hydrogen column density in the local arm (Figure~\ref{fig:totcoldens}), and the intensity and emissivity profiles discussed in section~\ref{sec:discussion:model} are available in electronic form
at the CDS via anonymous ftp to cdsarc.cds.unistra.fr (130.79.128.5)
or via https://cdsarc.cds.unistra.fr/cgi-bin/qcat?J/A+A/.}
   }


   \author{X. Astiasarain\inst{1}\fnmsep\thanks{xan.astiasarain@irap.omp.eu}
          \and
          L. Tibaldo\inst{1}\fnmsep\thanks{luigi.tibaldo@irap.omp.eu}
          \and
          P. Martin\inst{1}\fnmsep\thanks{pierrick.martin@irap.omp.eu}
          \and
          J. Kn\"odlseder\inst{1}
          \and
          Q. Remy\inst{2}
          }

   \institute{IRAP, Universit\'e de Toulouse, CNRS, CNES, UPS, 9 avenue Colonel Roche, 31028 Toulouse, Cedex 4, France \and
                  Max Planck Institut f\"ur Kernphysik, Saupfercheckweg 1, 69117 Heidelberg, Germany}

   \date{Received 29 November 2022; Accepted 06 January 2023}

 
  \abstract
  {Star-forming regions may play an important role in the life cycle of Galactic cosmic rays (CRs), notably as home to specific acceleration mechanisms and transport conditions.
  Gamma-ray observations of Cygnus~X have revealed the presence of an excess of hard-spectrum gamma-ray emission, possibly related to a cocoon of freshly accelerated particles.}
  {We seek an improved description of the gamma-ray emission from the cocoon using $\sim$13~years of observations with the \textit{Fermi}-Large Area Telescope (LAT) and use it to further constrain the processes and objects responsible for the young CR population.}
  {We developed an emission model for a large region of interest, including a description of interstellar emission from the background population of CRs and recent models for other gamma-ray sources in the field. Thus, we performed an improved spectro-morphological characterisation of the residual emission including the cocoon.}
{The best-fit model for the cocoon includes two main emission components: an extended component \halo, described by a 2D Gaussian of extension $r_{68} = 4.4\degr \pm 0.1\degr\,^{+0.1\degr}_{-0.1\degr}$ and a smooth broken power law spectrum with spectral indices $1.67 \pm 0.05^{+0.02}_{-0.01}$ and $2.12 \pm 0.02^{+0.00}_{-0.01}$ below and above $3.0 \pm 0.6^{+0.0}_{-0.2}$~GeV, respectively; and a central component \ig, traced by the distribution of ionised gas within the borders of the photo-dissociation regions and with a power law spectrum of index $2.19 \pm 0.03^{+0.00}_{-0.01}$ that is significantly different from the spectrum of \halo. An additional extended emission component \sw, located on the edge of the central cavities in Cygnus X and with a spectrum compatible with that of \ig, is likely related to the cocoon.
For the two brightest components \ig and \halo, spectra and radial-azimuthal profiles of the emission can be accounted for in a diffusion-loss framework involving one single population of non-thermal particles with a flat injection spectrum. Particles span the full extent of \halo as a result of diffusion from a central source, and give rise to source \ig by interacting with ionised gas in the innermost region.}
{For this simple diffusion-loss model, viable setups can be very different in terms of energetics, transport conditions, and timescales involved, and both hadronic and leptonic scenarios are possible. The solutions range from long-lasting particle acceleration, possibly in prominent star clusters such as Cyg~OB2 and NGC~6910, to a more recent and short-lived release of particles within the last $10-100$\,kyr, likely from a supernova remnant. The observables extracted from our analysis can be used to perform detailed comparisons with advanced models of particle acceleration and transport in star-forming regions.}


   \maketitle
%

\section{Introduction}


There is firm evidence that cosmic rays (CRs) at energies below 1\,PeV originate from the Milky Way. Supernova remnants (SNRs) remain the leading candidate as sources of the majority of Galactic CRs, most likely through the process of diffusive shock acceleration, while alternative source classes including massive star-forming regions, the Galactic centre, pulsar wind nebulae (PWNe), and compact binary systems may bring complementary contributions over specific parts of the extended CR spectrum \citep[see, for instance][ and references therein]{gabici2019}.
   
Massive star-forming regions are of particular interest in this context \citep[for instance][]{bykov2020}. The clusters of OB stars at their centres are the progenitors of a variety of particle acceleration sites such as SNRs, pulsars, and PWNe, or compact binary systems. In addition, the collective action of powerful stellar winds and, after a few million to a few tens of million years, the explosion of massive stars into supernovae lead to the formation of super-bubbles (SBs), which are large cavities filled by a highly dynamical medium that, as a whole, may play a specific role in the life cycle of CRs. The isotopic abundances measured in CRs suggest that at least a fraction of the CR material is sourced from the winds of massive stars \citep{binns2008,tatischeff2021}. 

Accelerated particles in distant locations can be revealed via the gamma-ray emission produced when they interact with interstellar gas, through inelastic collisions for nuclei or Bremsstrahlung for leptons, and radiation fields, through the inverse-Compton (IC) scattering by leptons. Therefore, star-forming regions are expected to be bright gamma-ray sources from the interactions of particles with the large masses of interstellar gas and the intense radiation fields available in these environments. Gamma-ray emission in the GeV and TeV energy ranges is detected towards a growing number of massive star-forming regions \citep[for a review see for instance][]{tibaldo2021}, and taken as evidence in favour of in situ CR acceleration. However, the clustering of energetic objects and interstellar clouds combined with the limited resolution of gamma-ray telescopes makes it difficult to firmly identify the acceleration sites and mechanisms, and to understand how particles propagate and interact through the region and eventually escape to merge into the large-scale CR population in the Galaxy.

Observational progress is matched by a flourishing development of models of particle acceleration and transport by stellar winds   \citep{gupta2018,bykov2020,morlino2021} and SBs \citep{bykov2001,ferrand2010,tolksdorf2019,vieu2022}. The models show that these objects can be efficient particle accelerators and make a contribution to Galactic CRs. They predict a number of morphological and spectral signatures that can be looked for to test the physical processes at the origin of the observed gamma-ray signals. 

Cygnus~X is one of the best studied massive star-forming regions in the Milky Way. Cygnus X contains Cygnus OB2 that, with 78 confirmed O stars \citep{berlanas2020}, is among the largest associations of massive stars in the Milky Way. It is composed of multiple substructures with a main group at $\sim$1.76~kpc from the Earth and a foreground group at $\sim$1.35~kpc \citep{berlanas2019} and at least two star-forming bursts $\sim$3 and $\sim$5~Myr ago \citep{berlanas2020}. A second prominent massive stellar cluster in Cygnus X is NGC~6910 at a distance of $\sim$1.73~kpc \citep{cantat-gaudin2020}, an age in the range from 5 to 10~Myr \citep{delgado2000,cantat-gaudin2020}, and a flat mass function pointing to a large number of massive stars \citep{kaur2020}.

The Large Area Telescope (LAT) aboard the \textit{Fermi Gamma-ray Space Telescope} \citep{LATpaper} unveiled a hard gamma-ray excess towards Cygnus~X with an extension\footnote{Throughout the paper we refer to a source extension as its 68\% containment radius $r_{68}$. For a 2D Gaussian intensity distribution, $r_{68} = 1.51\sigma$.} $r_{68} = 3.0 \degr \pm 0.3\degr$ \citep{ackermann2011}. The excess was interpreted as the signature of a cocoon of freshly accelerated particles. Gamma-ray emission in the energy range from hundreds of GeV to hundreds of TeV from the Cygnus cocoon was subsequently detected using ARGO-YBJ, HAWC, and LHAASO \citep{cocoon_argo,cocoon_hawc,cao2021,cocoon_lhaaso}. The most common interpretation involves nuclei accelerated by Cygnus OB2, possibly up to PeV energies. The radial gamma-ray emission profile above 10 GeV was taken as indication of diffusion following continuous CR injection over a few million years \citep{aharonian2019}.

In this paper, we present a new study of the Cygnus cocoon based on more than 13 years of \F-LAT observations with the aim of improving the morphological and spectral characterisation of the emission in order to constrain particle acceleration and propagation scenarios in the region. The characterisation of the cocoon requires a careful modelling of the interstellar gas distribution in the region that is presented in Section ~\ref{sec:gas}, while we describe the analysis of gamma-ray data, including morphological, spectral, and spectro-morphological characterisation of the cocoon emission in  Section ~\ref{sec:gamma}. The observables we derived are then discussed and interpreted in Section ~\ref{sec:discussion} and our conclusions are presented in Section ~\ref{sec:conclusions}.
   
\section{Construction of interstellar gas maps}\label{sec:gas}

The distribution of interstellar gas towards the region of interest is a key ingredient of our analysis for two reasons: 1) it is necessary to model the strong foreground and background gamma-ray emission from the interactions of the large-scale Galactic CR population with interstellar gas in the direction of Cygnus, and thus be able to extract and characterise the emission of the cocoon; 2) it is used in the interpretation of the gamma-ray signal in terms of the underlying CR populations.

\subsection{Atomic and molecular gas}\label{sec:hico}

We trace atomic gas using the $21$~cm emission line from the hyperfine transition of atomic hydrogen \hi. We use data from the Canadian Galactic Plane Survey (CGPS, \citealt{CGPS}) with an angular resolution of $1'$ and a velocity resolution of 1.3 km~s$^{-1}$ in the region with Galactic longitude $75.5\degr < l <90\degr$ and Galactic latitude $-3\degr < b < 5\degr$. Outside this region we use data from the all-sky HI4PI survey from Effelsberg and Parkes observations \citep{HI4PI} with a lower angular resolution of 0.27\degr\ and velocity resolution of 1.49 km~s$^{-1}$. We checked the consistency of the two surveys by comparing the data in the region covered by the CGPS.

We derived column densities \nhi under the hypothesis of a uniform spin temperature. All results are shown for the reference spin temperature of 250~K suggested by emission-absorption spectrum pairs in the CGPS area \citep{dickey2009} and that was also found to best reproduce gamma-ray observations of the Cygnus region based on an earlier analysis \citep{Ackermann2012}. This is a highly uncertain parameter that is not expected to be uniform along lines of sight and across the region. Therefore, the analysis was also performed for alternative uniform spin temperatures of 100~K (lower bound set by the brightness temperatures observed in the region), 400~K, and the optically thin case, which are used to set systematic uncertainties on relevant quantities.

Molecular hydrogen \hd cannot be traced directly. We use the $^{12}$CO $J_{1\rightarrow0}$ rotational line at 2.6~mm as surrogate tracer, under the usual hypothesis that \nhd column densities are directly proportional to the CO intensity (velocity-integrated brightness temperature) \wco through a constant known as $X_\mathrm{CO} \equiv N(\mathrm{H_2})/W_\mathrm{CO}$. We use CO data from the composite survey by \citet{CfA_CO} with an angular resolution of 0.125\degr\ in the area considered in this paper and a velocity resolution of 1.3 km~s$^{-1}$. Data were noise-filtered using the moment-masking technique \citep{dame2011}.

\begin{figure}[!htbp]
    \centering
    \includegraphics[width=0.50\textwidth]{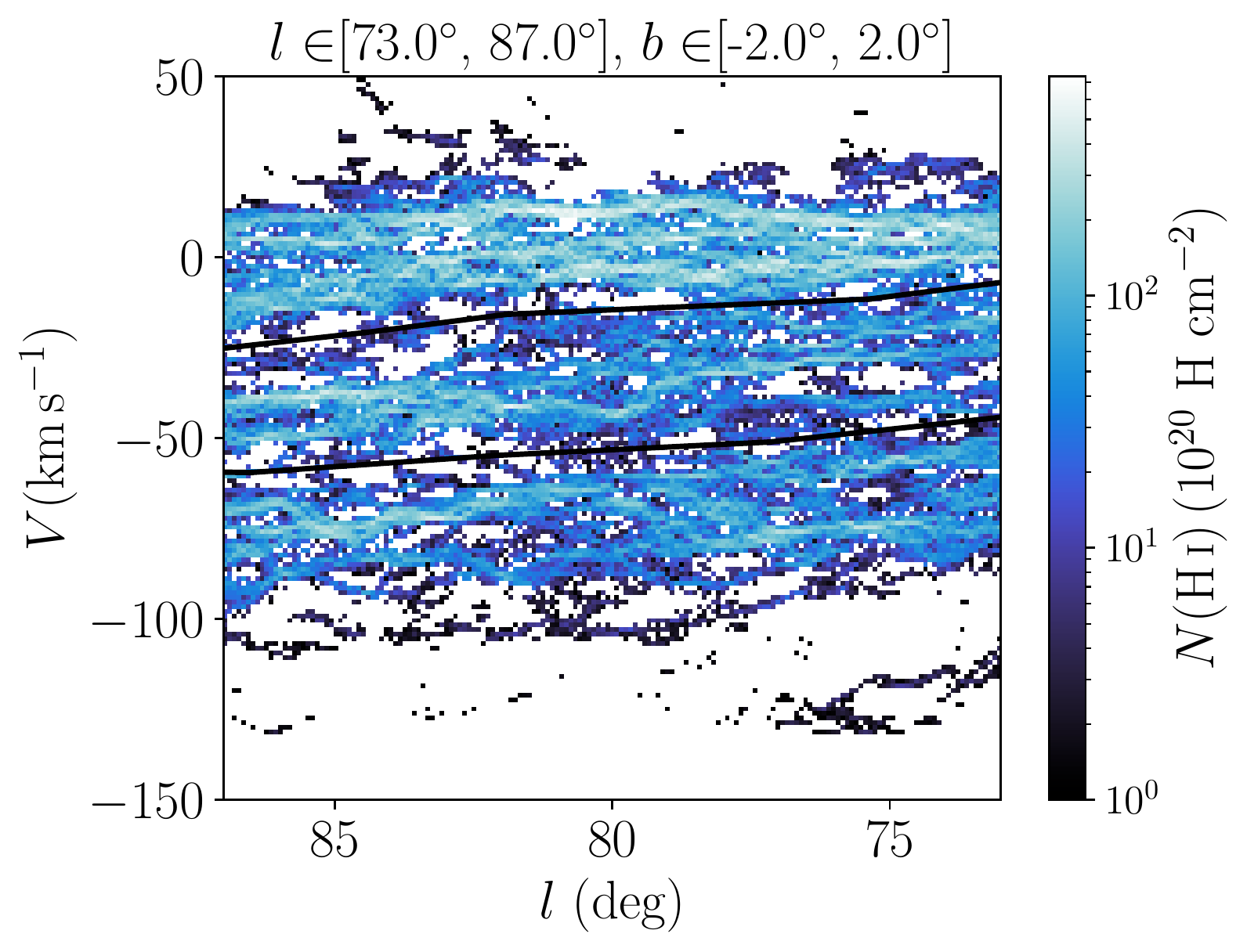}
    \caption{\hi column density as a function of Doppler-shift velocity and Galactic longitude summed over $-2\degr < b < 2\degr$. Total column densities from each pseudo-Voigt profile were assigned to the velocity of the peak. The black lines show the boundaries that we defined to separate the three structures along the line of sight: local arm, Perseus arm, and outer Arm and beyond (from top to bottom).}
    \label{fig:cuts}
\end{figure}

The Doppler shift of the lines can be used to infer the gas velocity along the line of sight due to Galactic rotation, and therefore separate multiple structures. However, intrinsic velocity dispersion can cause biases in the estimates of gas column densities across adjacent structures. To address this problem, we used the line profile fitting technique described in \citet{remy2017} to decompose emission from each line of sight into a combination of pseudo-Voigt functions. We built longitude-velocity and latitude-velocity diagrams based on the fit results for \hi and CO, and defined in the longitude-latitude-velocity space boundaries that separate the gas into three structures along the line of sight, namely the local arm (including the Cygnus complex), the Perseus arm, and the outer arm and beyond. Figure~\ref{fig:cuts} shows an example of longitude-velocity diagram in the longitude range $73\degr < l < 87\degr$ that is used in the following analysis. 



\begin{figure*}[!hbtp]
    \centering
    \includegraphics[width=0.8\textwidth]{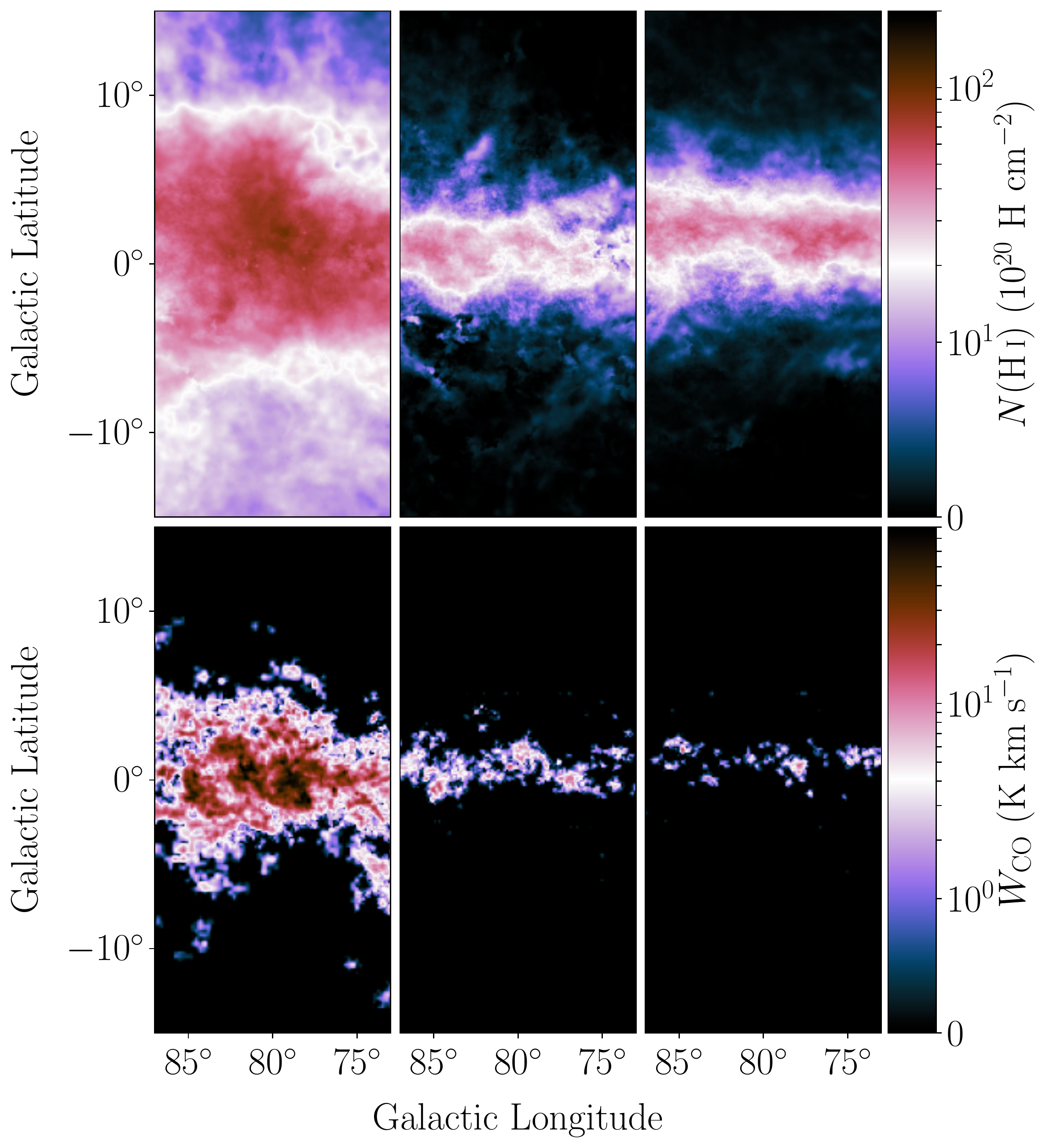}
    \caption{\hi column densities for a spin temperature of 250~K (top row) and \wco intensities (bottom row) for the local arm, Perseus arm, and outer arm and beyond (from left to right).}
    \label{fig:gasmaps}
\end{figure*}

The final \hi and CO maps for the three regions along the line of sight are shown in Figure~\ref{fig:gasmaps}. Since all surveys have different angular resolution, the maps were re-binned on a common grid of $1.875'$.

\subsection{Dark neutral medium (DNM)}\label{sec:dnm}

A significant fraction of neutral interstellar gas cannot be traced by the \hi $21$~cm line nor by the $^{12}$CO $J_{1\rightarrow0}$ rotational line \citep{grenier2005} and is therefore missing in the maps described above. It can be referred to as the dark neutral medium (DNM) and it is thought to be a combination of opaque \hi and diffuse \hd at the atomic-molecular interface of clouds, or dense \hd at the core of molecular clouds \citep{remy2017}.

If dust and gas in the interstellar medium (ISM) were well mixed and the dust grains physical and chemical properties were the same everywhere, dust thermal emission would be proportional to total gas column densities along the line of sight. Therefore, we can derive a DNM map by subtracting from the dust thermal emission the components correlated with \hi and CO. We use a map of the dust optical depth at $353$~GHz obtained from component separation of \textit{Planck} and \textit{IRAS} data \citep{GNILC} with an effective angular resolution of $5'$ in high signal-to-noise regions.  

\begin{figure}[!htbp]
    \centering
    \includegraphics[width=0.45\textwidth]{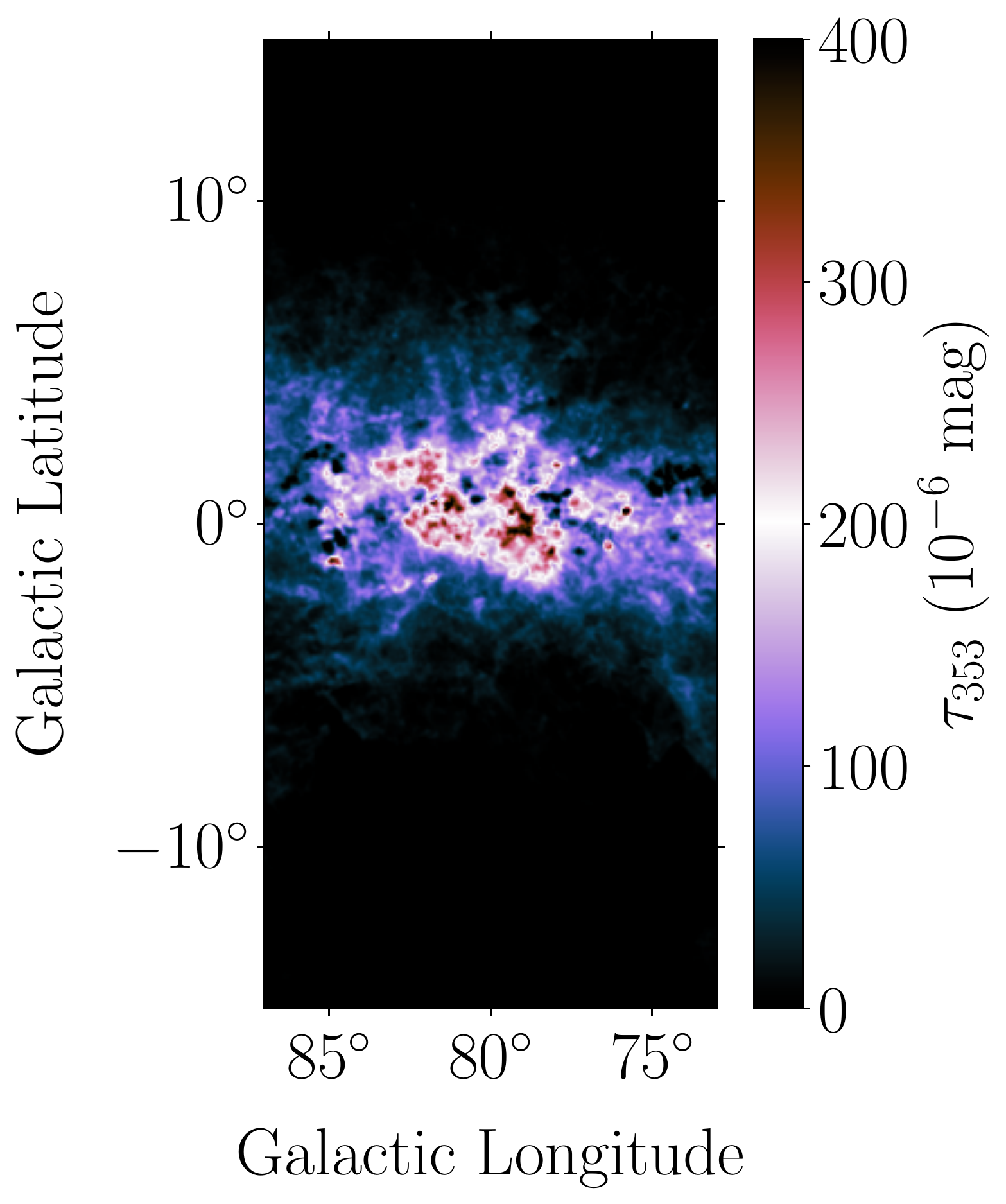}
    \caption{Excess dust optical depth associated to the DNM obtained using the procedure described in the text for the reference \hi spin temperature of 250~K.}
    \label{fig:DNM}
\end{figure}

To avoid biases from the missing DNM component in the determination of the components correlated with \hi and CO, we used the iterative fitting procedure described in \citet{Tibaldo2015}. Briefly, the procedure consists in an iterative fitting of the gas maps to the dust map where the positive part of the residuals is re-injected in the model at each iteration to compute unbiased values of the fit parameters and obtain an estimation of the missing  DNM component. A DNM map was calculated for each uniform spin temperature considered. Figure~\ref{fig:DNM} shows the DNM map obtained for the reference spin temperature value of $250$~K.

\subsection{Ionised gas}\label{sec:ig}

We derived an ionised gas column density map from the free-free emission measure $EM(l, b)$ extracted from component separation of \textit{Planck}, WMAP, and 408~MHz data by \citet{freefree}.
The free-free emission measure from Cygnus~X is dominated by two strong peaks that, as indicated by 8~$\upmu$m emission from dust, lie inside the cavities carved in the ISM by the intense star-forming activity in the region.

We calculated \hii column densities under the assumption that ionised gas fills a sphere of  radius $3.5\degr$ corresponding to $\sim$100~pc at a distance of $1.7$~kpc and with an uniform density along each line of sight. The sphere is meant to model the ionised cavities at the hearth of Cygnus X.
With $r$ the radius of the sphere and $d$ the distance to Cygnus X the electron volume density is:

\begin{equation}
    n_e(l,b) = \left[ \frac{EM(l,b)}{2 \,\sqrt{r^2 -  d^2\,\sin^2 (l - l_0) - d^2\,\sin^2 (b - b_0)}} \right]^{1/2}.
\end{equation}
Therefore, for the column density we obtain:
\begin{equation}
    N_{\mathrm{H\,\scriptstyle{II}}} (l, b) = n_e(l, b) \times 2 \,\sqrt{r^2 - d^2\,\sin^2 (l - l_0) - d^2\,\sin^2 (b - b_0)},
\end{equation}
where $l_0$ and $b_0$ are the position of the sphere's centre and $EM(l,b)$ the emission measure in a given direction.

\begin{figure}[!htbp]
    \centering
    \includegraphics[width=0.5\textwidth]{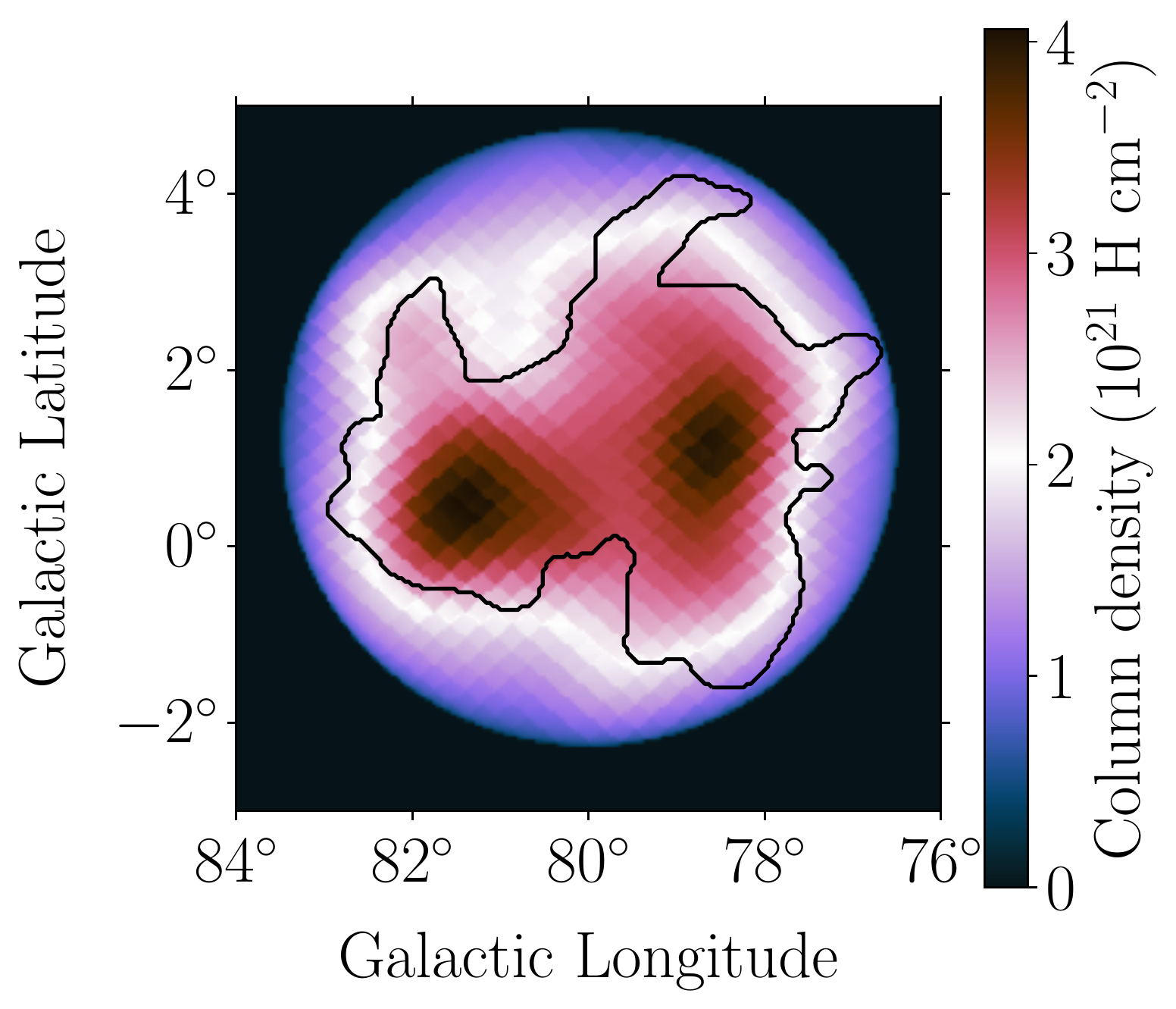}
    \caption{Ionised gas column densities based on the hypothesis of spherical geometry with a uniform density along each line of sight. The black contours delineate the outer borders of the photo-dissociation regions and correspond to $>1.85 \, 10^{-6}$ W m$^{-2}$ sr$^{-1}$ in the MSX 8~$\upmu$m data.}
    \label{fig:IG}
\end{figure}

The final ionised gas column density map is displayed in Figure~\ref{fig:IG}. The angular resolution of the free-free emission measure map from \textit{Planck} is $1\degr$. The final ionised gas column density map was re-binned on the same grid as the MSX 8~$\upmu$m map with a grid spacing of $1"$.

\section{Gamma-ray analysis and results}\label{sec:gamma}

\subsection{Data selection}

We analysed 13.25 years of \F-LAT data from the beginning of the mission on 4 August 2008 to 3 November 2021.
We used the P8R3 data set \citep{Atwood2013, Bruel2018} and selected events in the P8R3\_SOURCE class that is associated with instrument response functions P8R3\_SOURCE\_V3. This event selection has a level of background contamination sufficiently low to study the bright extended emission from Cygnus~X. Furthermore, we restricted the analysis to time intervals in which the LAT configuration and data quality is appropriate for science analysis.

Events are separated in four independent data sets according to their PSF event type, that is the quality of the direction reconstruction. For each event type, we selected events above a minimum energy so that the point spread function (PSF) 68\% containment radius is always better than $0.7\degr$, which roughly corresponds to the characteristic size of the most prominent spatial structures in the gas maps. The minimum energy threshold used is 0.5~GeV. Lowering the minimum energy induced instabilities in the analysis due to bright emission from a few pulsars in the region. To reliably characterise extended emission at energies $< 0.5$~GeV event selection based on the pulsars phases would be necessary, but this is beyond the scope of the current study. The maximum energy is 1~TeV for all event types.

To reduce contamination from the bright gamma-ray emission from the Earth atmosphere, we selected events within a cone from the local zenith with aperture $z_\mathrm{max}$. The value of $z_\mathrm{max}$ was chosen from a visual inspection of the distribution of counts as a function of zenith angle for each event type component in the given energy ranges. Energy and zenith angle selection for the four data sets are summarised in Table~\ref{tab:PSF}.

\begin{table}[htbp]
    \centering
    \caption{The four data sets used in the analysis.}
    \begin{tabular}{ccc}
    \hline\hline
     \rule{0mm}{4mm}Event type & Energy range (GeV) & $z_\mathrm{max}$\\ 
    \hline
     \rule{0mm}{4mm}PSF3  & $0.5$ - $1000$ & $100\degr$\\
     PSF2 & $1$ - $1000$ & $100\degr$ \\
     PSF1 & $3.2$ - $1000$ & $100\degr$ \\
     PSF0 & $5$ - $1000$ & $105\degr$\\
     \hline
    \end{tabular}
    \label{tab:PSF}
\end{table}

\subsection{Region of interest and emission model}\label{sec:anamodel}

A major challenge in the characterisation of extended emission from Cygnus~X is to model the bright interstellar emission from the large-scale population of CRs. Due to the large column densities of the ISM in this region, emission associated with gas is the dominant contribution at GeV energies \citep{Ackermann2012}. Under the assumption that the large-scale CR densities are uniform on the spatial scales of interstellar complexes, we can model the foreground and background intensity associated with interstellar gas $I_\mathrm{gas}$ as a linear combination of the column density maps for the different gas phases and structures along the line of sight:
\begin{equation}
\begin{aligned}
   I_\mathrm{gas} (l,b,E)= q_\mathrm{LIS}(E) \cdot \left[\sum_{\imath=1}^3 \left[A_\imath\, N_{\mathrm{H\,\scriptstyle{I}},\imath} (l,b) + B_\imath \,  W_\mathrm{CO,\imath}(l,b) \right] \right.\\
   + C \,  \tau_\mathrm{DNM}(l,b) \Biggr],
\end{aligned}
\end{equation}
where $q_\mathrm{LIS} (E)$ is the local gas emissivity spectrum, that is the gamma-ray emission rate per hydrogen atom, from \citet{casandjian2015}, derived from LAT data. The summation over $\imath$ describes the combination of the three regions along the line of sight: local arm, Perseus arm, outer arm and beyond. The free parameters $A_\imath$, $B_\imath$, and $C$ account at the same time for variations of the large-scale CR densities across the three regions, and for the \xco ratios and the dust specific opacity $\sigma_{353} = \tau_{353}/$\nhi.
The spectral shape of $q_\mathrm{LIS}$ is fixed throughout the paper.
We know that spectral variations of the emissivity along the line of sight are small towards Cygnus \citep{Ackermann2012} and, in general, towards the outer Galaxy \citep{acero2016}. Conversely, this implies that any spectral deviations from the local interstellar spectrum (LIS) in Cygnus~X are not accounted for by the background model and characterised as part of the cocoon.

An additional diffuse component is given by IC emission from the large-scale population of CR leptons. We accounted for it using the {GALPROP} model $\mathrm{^SY^Z6^R30^T150^C2}$ \citep{IC_model}.
Finally, we need to account for the isotropic gamma-ray background, which is a combination of extra-galactic diffuse gamma-ray emission (probably due to populations of unresolved sources) and of residual contamination by charged CRs. For this component, we used the tabulated spectra provided by the \textit{Fermi}-LAT collaboration and determined from an analysis of LAT data over a large region of the sky\footnote{\texttt{We used files iso\_P8R3\_SOURCE\_V3\_PSFn\_v1.txt}, with \texttt{n} the PSF event type, from \url{https://fermi.gsfc.nasa.gov/ssc/data/access/lat/BackgroundModels.html}.}.

We note that the IC model is also subject to large uncertainties. However, morphological variations over our limited region of interest described below are expected to be small for conventional models. Moreover, uncertainties in the spectrum are mitigated by the fact that the isotropic background spectrum is derived from a fit to the LAT data.

The interstellar emission model, along with the LAT response, sets the choice of the region of interest (ROI) for the analysis. The longitude and latitude extents should be sufficiently large to separate the extended emission of the Cygnus cocoon from the large-scale background, and so that the different components of the background model can be reliably constrained by the data. We chose a ROI with Galactic longitude $73\degr \leq l \leq 87\degr$ and with Galactic latitude $|b| \leq 15\degr$. The longitude interval leaves out complexes associated with Cygnus~OB1 at $l < 73\degr$ and with HB~21 at $l > 87\degr$. The wider coverage in latitude makes it possible to better constrain emission from local \hi, IC scattering, and the isotropic background.

We modeled individual sources within the region based on the most recent catalogue of gamma-ray sources detected by the LAT, 4FGL-DR3 \citep{4FGL-DR3}. All the sources within a square box of $40\degr$ side centred at $l = 80\degr$ and $b = 0\degr$ were included to account for the spill-over due to the PSF.

For two extended sources with potential impact on the characterisation of the cocoon emission, we replaced the 4FGL-DR3 models with dedicated models provided by recent in-depth studies. The SNR $\gamma$~Cygni is modelled according to the results from a joint fit of \textit{Fermi-LAT} and MAGIC data at energies $> 5$~GeV \citep{MAGIC20}. The source is modelled as a disk with a log-parabola spectrum to account for the shell, plus a 2D Gaussian with a power law spectrum to account for an additional component in the north of the shell. The arc component detected by MAGIC is not included since its flux is subdominant at energies $< 1$~TeV. A morphological evolution of the remnant below 5~GeV is very challenging to characterise due to bright emission from PSR~J2021$+$4026 \citep{MAGIC20}. Thus, this possibility is not considered in our study.

The SNR Cygnus Loop is modelled following the analysis by \citet{Tutone21} in the 0.1-100 GeV energy range. We used two templates based on X-ray (ROSAT $0.1 - 2.4$~keV) and UV (GALEX $1771 - 2831$~\AA) data, each of them associated to a log-parabola spectrum. Emission from this source above 100~GeV is expected to be small, and we visually checked in the residuals that there were no excess or deficit of counts at the location of the Cygnus Loop.

\begin{figure}[ht]
    \centering
    \includegraphics[width=1.0\linewidth]{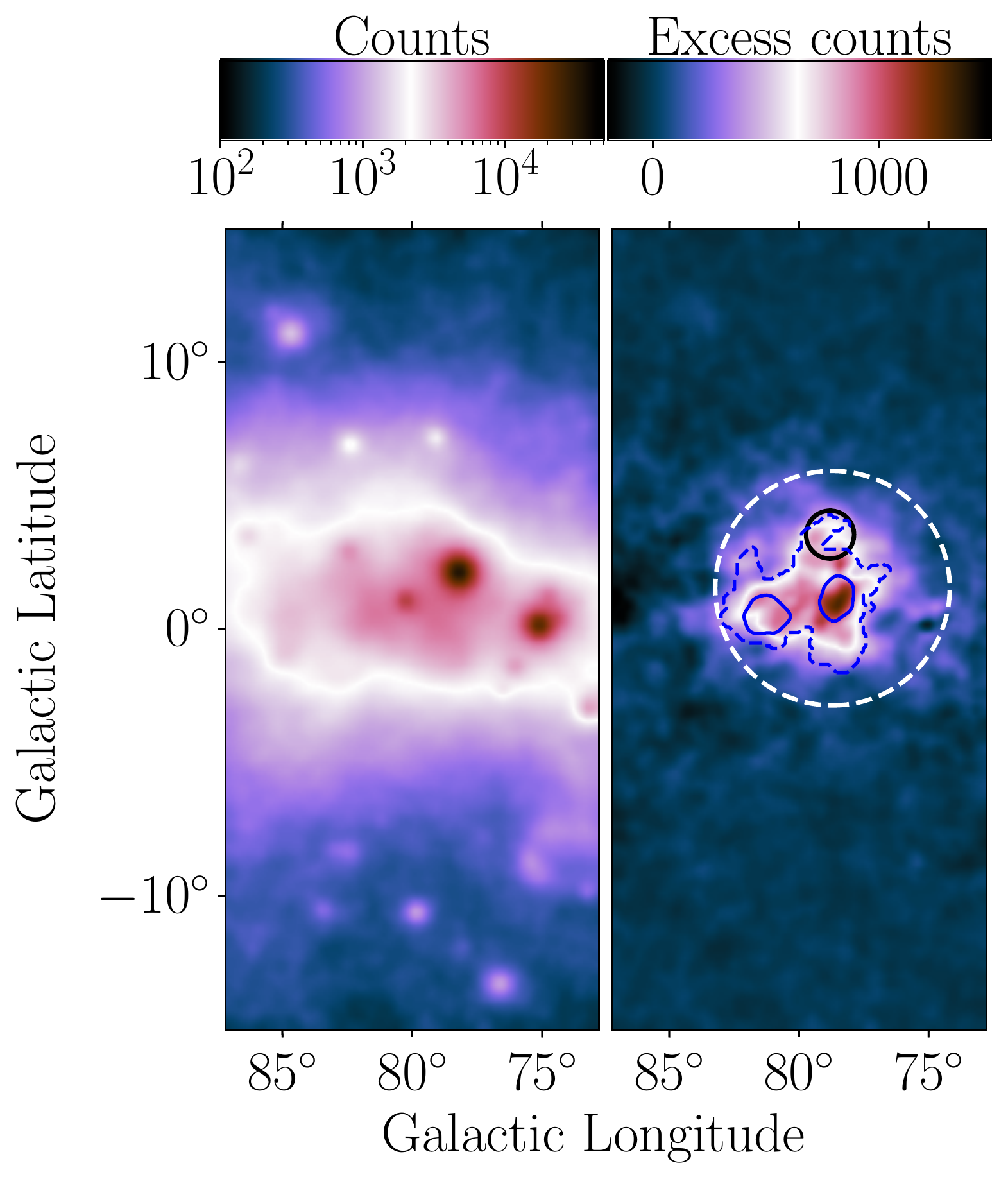}
    \caption{Map of data counts in the full $0.5$~GeV$-1$~TeV energy range (left) and map of excess counts obtained by subtracting from the data counts the best-fit model presented in Section  \ref{sec:final} except for the components associated to the cocoon, namely \halo, \ig, and \sw (right). The dashed contours correspond to the $8~\mu$m emission from MSX data at $1.85 \times 10^{-6}~\mathrm{W~m^{-2}~sr^{-1}}$. The contours correspond to the column density of the ionised gas template at $3.5\times 10^{21}~\mathrm{H~cm^{-2}}$. The circle and the dashed circle correspond to the position and $r_{68}$ of \sw and \halo respectively.}
    \label{fig:DataVsExcess}
\end{figure}

Figure~\ref{fig:DataVsExcess} shows the gamma-ray count map in the region of interest and the excess counts associated with the cocoon obtained by subtracting from the data counts the best-fit model presented in Section  \ref{sec:final}.
The cocoon, that is the excess shown in the right panel of Figure~\ref{fig:DataVsExcess}, was initially modelled as in 4FGL-DR3: a 2D Gaussian with $r_{68} = 3\degr$ \citep{ackermann2011} and a spectrum described by a log-parabola function. The morphological and spectral models were refined later during the analysis.

\subsection{Analysis framework}

The analysis is performed using \texttt{fermitools} v2.0.8 and a modified version of \texttt{Fermipy} v1.0.1 that enables the use of catalogue 4FGL-DR3.
Models are fit to the data via a binned maximum likelihood analysis with Poisson statistics. Events are binned on a grid with 10 bins per decade in energy and on maps with a pixel size of $0.1\degr$ in arrival direction.

Throughout the paper, we compare several models for the region and the source of interest. In the simpler cases we use the likelihood ratio test, that is the test statistic defined as:
\begin{equation}
TS = 2 \left(\ln \mathcal{L} - \ln \mathcal{L}_0 \right),
\end{equation}
where $\mathcal{L}_0$ is the maximum likelihood of a more parsimonious emission model with fewer free parameters (null hypothesis) and $\mathcal{L}$ is the maximum likelihood of the more complex model that we want to test (test hypothesis). In the null hypothesis $TS$ is distributed as a $\chi^2_n$ with a number of degrees of freedom $n$ equal to the difference of degrees of freedom between the two models. This is only valid for nested models, that is if the model in the null hypothesis can be obtained from the model in the test hypothesis by fixing some of its parameters to values in the interior of the allowed range \citep[for instance][]{protassov2002}

For non-nested models, we use the Akaike Information Criterion (AIC). The AIC of a model is defined as:
\begin{equation}
AIC = 2k - 2\ln\mathcal{L},
\end{equation}
with $k$ number of free parameters in the model and $\mathcal{L}$ the maximum likelihood of the model.
The model providing the smallest AIC is taken as the one best representing the data at the smaller cost in terms of free parameters according to information theory \citep[for instance][]{burnham2002}.

Throughout the paper, we use the method described in \citet{Bruel21} to assess the goodness of fit of the different models considered. Practically, we show the deviation of the data with respect to the model in units of significance based on the Poisson statistics using the so-called PS maps.

The analysis starts with a preliminary optimisation of the emission model via a procedure described in Appendix \ref{app:prelopt}. In subsequent steps, unless stated otherwise, we keep free the normalisations of the gas and IC components, as well as the normalisations and spectral parameters of the three pulsars PSR~J2021$+$4026, PSR~J2021$+$3651, and PSR J2032+4127, the two $\gamma$~Cygni extended components, and the two Cygnus Loop extended components. The normalisation of the subdominant isotropic background is fixed after the preliminary iterative optimisation of all the sources in the ROI due to the possible degeneracy with the IC component. The best-fit normalisation obtained is $0.91\pm0.02$ (with variations $\leq 0.01$ for the different spin temperature values).

\subsection{Morphological analysis}\label{sec:morphology}

In this section, we aim at characterising the morphology of the cocoon. As a first step, we optimised the position and extension of the 2D Gaussian model used in \citet{ackermann2011} and the LAT catalogues to describe the extended cocoon emission, following the methodology described in Appendix \ref{app:morph}.

\begin{figure*}[!htbp]
    \centering
    \includegraphics[width=0.78\textwidth]{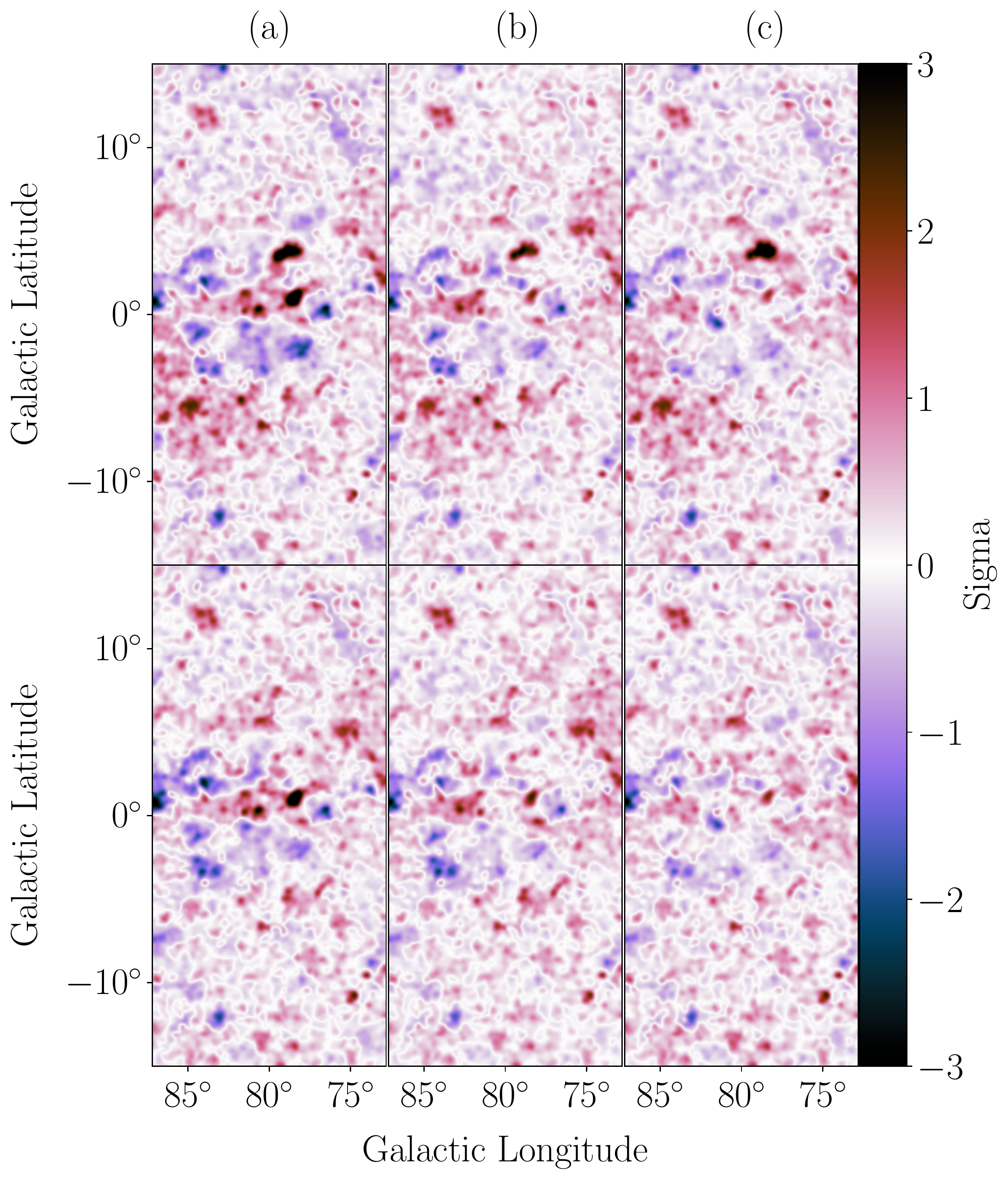}
    \caption{PS maps for different models considered in the morphological analysis. On the top panels we can see the PS maps for different morphological models: a) one extended Gaussian, b) one extended Gaussian plus two smaller Gaussians at the peaks of the ionised gas column density distribution, c) one extended Gaussian plus the ionised gas template. On the bottom panels, we can see the PS maps for the same models with the addition of two Gaussians for the western and off-field excesses in the model (ultimately labelled \sw and \ob). The bin size is $0.1\degr$ and the maps were smoothed for display with a kernel of size $0.13\degr$}
    \label{fig:3cocoonmodels}
\end{figure*}

The corresponding PS map is displayed in Figure~\ref{fig:3cocoonmodels} (panel a, top row). An excess appears in the central region of the cocoon, in part reminiscent of the two main peaks of ionised gas column density within the Cygnus~X cavities (Figure~\ref{fig:IG}). Therefore, we tested the addition of a central component in the cocoon region using two alternative models: either the ionised gas template clipped at the boundaries of the cavities (defined as contours above $1.85~10^{-6}~\mathrm{W~m^{-2}~sr^{-1}}$ emission at 8~$\upmu$m), or two Gaussians with free extensions and positions, initialised at the main peaks in the ionised gas map. All newly added sources on top of the Gaussian model for the extended cocoon component here and elsewhere in this section are modelled using a power-law spectrum.

For the models described above, extended deviations are still apparent (see Figure~\ref{fig:3cocoonmodels} top row). The largest excess appears in the western part of the cocoon at $l \sim 78.8\degr$ and $b \sim 3.7\degr$. It does not overlap with any known sources or structures in the gas maps. A second extended region of positive deviations appears at the edge of our ROI, at $l \sim 84.6\degr$ and $l \sim -5.6\degr$ towards the southern arc of the Cygnus SB as imaged in soft X-rays \citep{cash1980}. 
We added two additional Gaussian components to model those excesses, hereafter referred to as western and off-field excesses. We initialise the Gaussian centres on the excess peaks, and fit their positions and extensions. This results in a significant likelihood improvement ($\Delta\ln\mathcal{L} \sim 200$) for all models of the central cocoon component.

The different models for the cocoon central component are compared in Table~\ref{table:morphology}. The addition of the central cocoon component on top of the 2D Gaussian for the extended one provides a significant improvement in likelihood ($\Delta\ln\mathcal{L} > 240$). Conversely, a model including the ionised gas map without the extended cocoon Gaussian component resulted in a marked degradation of the likelihood ($\Delta\ln\mathcal{L} = -600$).

The model including the ionised gas template for the central cocoon component provides the largest likelihood and the smallest AIC, and therefore it is the one favoured by our analysis. It is strongly preferred over two additional Gaussian components at the peaks in the ionised gas distribution ($\Delta AIC < -124$), which strengthens the evidence for a correlation between part of the gamma-ray signal and the ionised matter distribution\footnote{The $AIC$ criterion may not be fully appropriate in this case due to the information entropy encoded in the geometry of the template and not represented by any fit parameter, but the large improvement of log-likelihood clearly favours the model with the ionised gas template.}.
This conclusion is supported by visual inspection of the deviations in Figure~\ref{fig:3cocoonmodels} (bottom row, panels b and c).

\begin{table*}[ht]
\caption{Comparison of different spatial models}              
\label{table:morphology}      
\centering                                      
\begin{tabular}{l c c c c c}          
\hline\hline                        
Model & $\Delta \ln \mathcal{L}$ & $\Delta AIC$  \\    
\hline                                   
Extended Gaussian & 0 & 0 \\
+Western + Off-field & & \\
\hline
Extended Gaussian + 2 Gaussians (IG peaks) & $316~[246, 316]$ & $-622~[-622, -492]$\\
+ Western + Off-field & & \\
\hline
Extended Gaussian + IG template & $435~[379, 435]$ & $-868~[-868, -746]$ \\
+ Western + Off-field & & \\
\hline                                             
\end{tabular}
\tablefoot{$\ln \mathcal{L}$ and $AIC$ values are provided as differences with respect to the simplest model with only one 2D Gaussian for the extended emission of the cocoon.
The intervals correspond to the minimum and maximum spin temperatures considered. The Western and Off-field components are named later \sw and \ob.}
\end{table*}

We also tested the full ionised gas map, that is not clipped at the boundaries of the cavities, but this yielded a smaller likelihood. The interpretation of this result is not straightforward. The reason may be physical, for example related to confinement of the particles in the cavities. It could also be related to limitations in the analysis, such as systematic biases in the emission measure map extracted from \textit{Planck} data, approximations in the derivation of the ionised gas column density, or degeneracies with other gas templates outside the two main peaks in the map.

Spatial parameters for the multiple overlapping extended sources may be degenerate to some level. To robustly determine their values, we performed an iterative fit of the positions and extensions of all 2D Gaussian sources discussed in this section for the best model. The iterations proceed until the log-likelihood improvement between two iterations is smaller than 1. The iterative fit converged after 6 iterations, with a total improvement in $\ln \mathcal{L}$ of $30$.

Table~\ref{table:morph-pars} provides the best-fit morphological parameters and $TS$ of all the extended components discussed in this section for the case in which the cocoon region is modelled by a broad 2D Gaussian (extended component) plus the ionised gas map (central component), and an additional smaller 2D Gaussian slightly off the emission peak (western component).
The extended emission components are named after their Galactic coordinates as FCES~G$LL.ll\pm B.bb$ (FCES stands for \textit{Fermi} Cygnus Extended Source). To make the paper easier to read, the sources are given a nickname that we use throughout the paper. \halo is the name given to the Gaussian that describes the cocoon extended emission, nicknamed \NNhalo, \ig to the component modelled by the ionised gas map in the cocoon central region, nicknamed \NNig, and \sw to the component corresponding to the excess appearing in the western part of the cocoon, nicknamed \NNsw.  Interestingly, the addition of a central component for the cocoon results in a larger $r_{68}$ for the extended component with respect to previous studies. We remark that the off-field excess, ultimately dubbed \ob and nicknamed \NNob, is best modelled by a Gaussian centred at the edge of the ROI, therefore its characterisation may be inaccurate. A better characterisation of this component is left for further work. The positions and extensions of sources in the cocoon area are shown overlaid to the excess map in the right panel of Figure~\ref{fig:DataVsExcess}.

\begin{table*}[!htbp]
\caption{Best-fit morphological parameters and $TS$ for the extended emission components considered in the morphological analysis.}              
\label{table:morph-pars}      
\centering                                      
\begin{tabular}{l c c c c }          
\hline\hline                        
Name & $l (\degr)$ & $b (\degr)$ & $r_{68} (\degr)$ & $TS$  \\    
\hline                                   
\halo (\NNhalo) & $78.7 \pm 0.1^{+0.0}_{-0.2}$ & $1.56 \pm 0.06^{+0.07}_{-0.02}$ & $4.4 \pm 0.1^{+0.1}_{-0.1}$ & $2751~[2436, 2824]$\\
\ig (\NNig) & ... & ... & ... & $1267~[1267,1301]$\\
\sw (\NNsw) & $78.8 \pm 0.1^{+0.0}_{-0.1}$ & $3.6 \pm 0.1^{+0.0}_{-0.0}$ & $0.9 \pm 0.1^{+0.0}_{-0.0}$ & $93~[93,106]$\\
\ob (\NNob) & $85.0 \pm 0.4^{+0.2}_{-0.2}$ & $-1.8 \pm 0.25^{+0.3}_{-0.2}$ & $6.4 \pm 0.2^{+0.1}_{-0.1}$ & $684~[680,723]$\\
\hline                                             
\end{tabular}
\tablefoot{The first uncertainties are statistical. The second uncertainties and $TS$ variations are systematic from varying the spin temperature.}
\end{table*}

\subsection{Spectral analysis}\label{sec:spectra}

In this section, we aim at characterising the spectral properties of the FCES sources. Based on the best morphological model derived in the previous section, and for each emission component listed in Table~\ref{table:morph-pars}, we tested three spectral models: a simple power law (PL), a log-parabola (LP), and a smooth broken power law (SBPL). The expressions for these models are given in Appendix \ref{app:specmodels}.

The models are compared in Table~\ref{table:spectral_test}. For the components \NNig and \NNsw the best-fit model is the simple PL, with the LP providing a negligible improvement in log-likelihood. The fit of the SBPL for these two components did not converge, presumably due to lack of curvature in the spectrum. Conversely, for \NNhalo and \NNob the models with curvature (LP or SBPL) provide a large improvement in $\ln \mathcal{L}$. From the Akaike criterion, we can conclude that the SBPL is favoured. Spectral parameters for the best-fit models are presented in the next subsection.

The spectral indices of the components \NNig and \NNsw are compatible with each other. If we fix the spectral index of \NNsw to the best-fit value for \NNig, we obtain a decrease in log-likelihood of $1.2~[1.2, 1.8]$ ($1.1~[1.1,1.3]\,\sigma$, where here and in the following variations or uncertainties refer to the different spin temperatures). This demonstrates that the two components have compatible spectral shapes. However, if we model \NNsw or \NNig with the same spectral shape as \NNhalo and a free normalisation, we observe a decrease in log-likelihood of $7.9~[7.9, 9.8]$ ($\Delta AIC = 19.8~[19.8, 23.6]$) and $70.6~[66.6, 70.6]$ ($\Delta AIC = 145.2~[137.2, 145.2]$), respectively. So both \NNig and \NNsw have spectra incompatible with the spectrum of \NNhalo, this time suggesting a different origin of the gamma-ray emission. 

\begin{figure*}[!htbp]
    \centering
    \includegraphics[width=1.\textwidth]{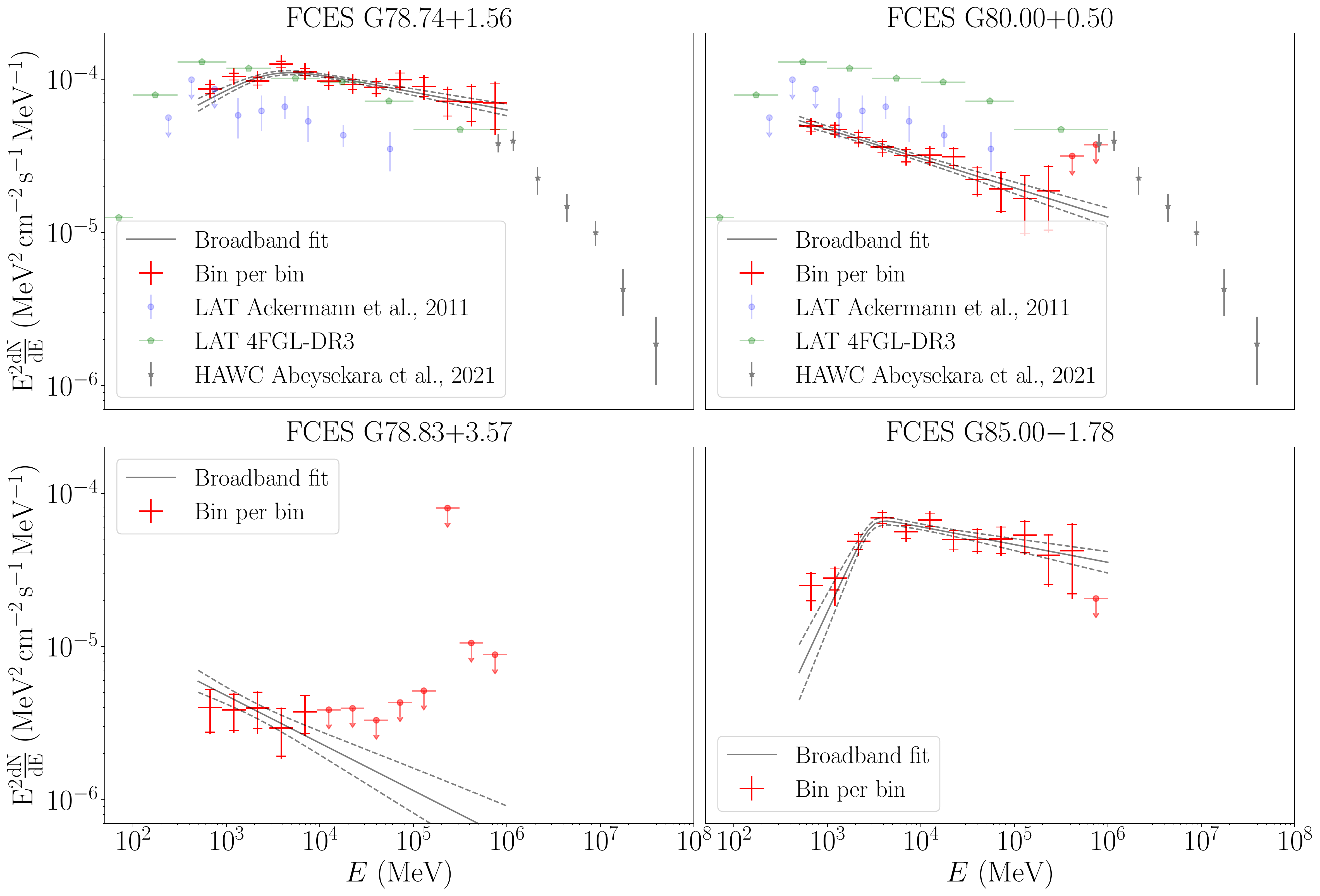}
    \caption{Spectral energy distribution of the four extended components studied. In the top panel, we show for reference earlier determinations of the cocoon spectrum. The statistical uncertainties are displayed within the error caps and the full error bar is the quadratic sum of the statistical uncertainties, the uncertainties related to the different spin temperatures, and the uncertainties related to the effective area of the telescope.}
    \label{fig:SEDCocoon}
\end{figure*}

We then computed the spectral energy distribution (SED) of the four sources. To this end we performed independent analyses over 4 energy bins per decade between 500~MeV and 1~TeV. For this part of the analysis all spectral-shape parameters are fixed and only normalisations are allowed to vary. However, the normalisations of the gas maps are fixed to the best-fit values obtained for the entire energy range to preserve the local emissivity shape. The IC normalisation is also fixed to the best-fit value obtained for the entire energy range due to the large degeneracy with the extended components. Finally, the normalisations of all pulsars are fixed above $5$~GeV because their emission fades off rapidly. The results are shown in Figure~\ref{fig:SEDCocoon}. For flux densities (and all derived quantities later) we  include in the systematic uncertainties those from the effective area of the LAT, combined in quadrature with those from the spin temperature choice.

The source with the highest flux is \NNhalo, followed by \NNig. The spectrum of \NNhalo extends to higher energies and connects to the cocoon spectrum measured by HAWC \citep{cocoon_hawc}, confirming earlier indications of a spectral break between the GeV and the TeV energy ranges. Our spectrum for \NNhalo is similar at energies $> 1 $~GeV to the cocoon SED in catalogue 4FGL-DR3. On the contrary, our SED lies above the one presented in \citet{ackermann2011}, which is closer to our SED for \NNig. Presumably, the SED determination in \citet{ackermann2011} was biased towards the central component, while the extended component captured by \NNhalo in our analysis was difficult to detect at that time due to a reduced amount of data (2 years versus more than 13 years here) and a less advanced event reconstruction scheme.

\begin{table*}[!htbp]
\caption{Statistical comparison of different spectral models for the extended sources in Cygnus X}              
\label{table:spectral_test}      
\centering                                      
\begin{tabular}{l c c c}
\hline\hline
Component & $\Delta \ln \mathcal{L}_\mathrm{LP-PL}$ & $\Delta \ln \mathcal{L}_\mathrm{SBPL-PL}$ & $\Delta AIC_\mathrm{SBPL-LP}$ \\
\hline
\halo (\NNhalo) & $33~[27, 35]$ & $42~[36, 43]$ & $-14~[-14, -14]$ \\
\ig (\NNig) & $1~[1, 1]$ & ... & ... \\
\sw (\NNsw) & $1~[1, 2]$ & ... & ... \\
\ob (\NNob) & $26~[24, 27]$ & $38~[35, 38]$ & $-22~[-22, -20]$ \\
\hline
\end{tabular}
\tablefoot{The columns show the log-likelihood differences between a log-parabola and a power-law model, $\Delta \ln \mathcal{L}_\mathrm{LP-PL}$, or between a smooth broken-power-law and a power-law model, $\Delta \ln \mathcal{L}_\mathrm{SBPL-PL}$, and the Akaike information criterion difference between a smooth broken-power-law and a log-parabola model, $\Delta AIC_\mathrm{SBPL-LP}$. The intervals correspond to the minimum and maximum values obtained from variation of the spin temperature.}
\end{table*}

\subsection{Final global fit}
\label{sec:final}

After the selection of the best spectral models we performed a final optimisation of the ROI, including free normalisation and spectral-shape parameters for the FCES sources.
The final spectral parameters of the FCES sources are displayed in Table~\ref{table:finalspectralpars}.

\begin{table*}[!htbp]
\caption{Best-fit values after the final optimisation of the model.}              
\label{table:finalspectralpars}      
\centering                                      
\begin{tabular}{l c c c c c c c c}          
\hline\hline                        
Component & Model & \multicolumn{4}{c}{Spectral parameters} \\ 
 & & $N_0$ (cm$^{-2}$ s$^{-1}$ MeV$^{-1}$) & $\gamma$ or $\gamma_1$ & $\gamma_2$ & $E_b$ (GeV) \\
 
\hline                                   
\halo (\NNhalo) & SBPL & $8.6\pm0.6^{+0.2}_{-0.9} \times 10^{-11}$ & $1.67\pm0.05^{+0.02}_{-0.01}$ & $2.12\pm0.02^{+0.00}_{-0.01}$ & $3.0\pm0.6^{+0.0}_{-0.2}$ \\
 
\ig (\NNig) & PL & $4.7 \pm 0.2^{+0.0}_{-0.0} \times 10^{-11}$ & $2.19\pm 0.03^{+0.00}_{-0.01}$ & ... & ... \\

\sw (\NNsw) & PL & $4.8 \pm 0.6^{+0.3}_{-0.0} \times 10^{-12}$ & $2.3\pm 0.1^{+0.0}_{-0.0}$ & ... & ... \\

\ob (\NNob) & SBPL & $1.7 \pm 0.5^{+0.7}_{-0.0} \times 10^{-11}$ & $0.7\pm0.3^{+0.2}_{-0.0}$ & $2.12\pm0.04^{+0.00}_{-0.00}$ & $3.0\pm0.4^{+0.0}_{-0.2}$\\
\hline                                             
\end{tabular}
\tablefoot{The first uncertainties are statistical and the second uncertainties are systematic and result from varying the spin temperature. $N_0$ is given at the reference energy of $1~$GeV.}
\end{table*}

\begin{figure}[!htbp]
    \centering
    \includegraphics[width=0.45\textwidth]{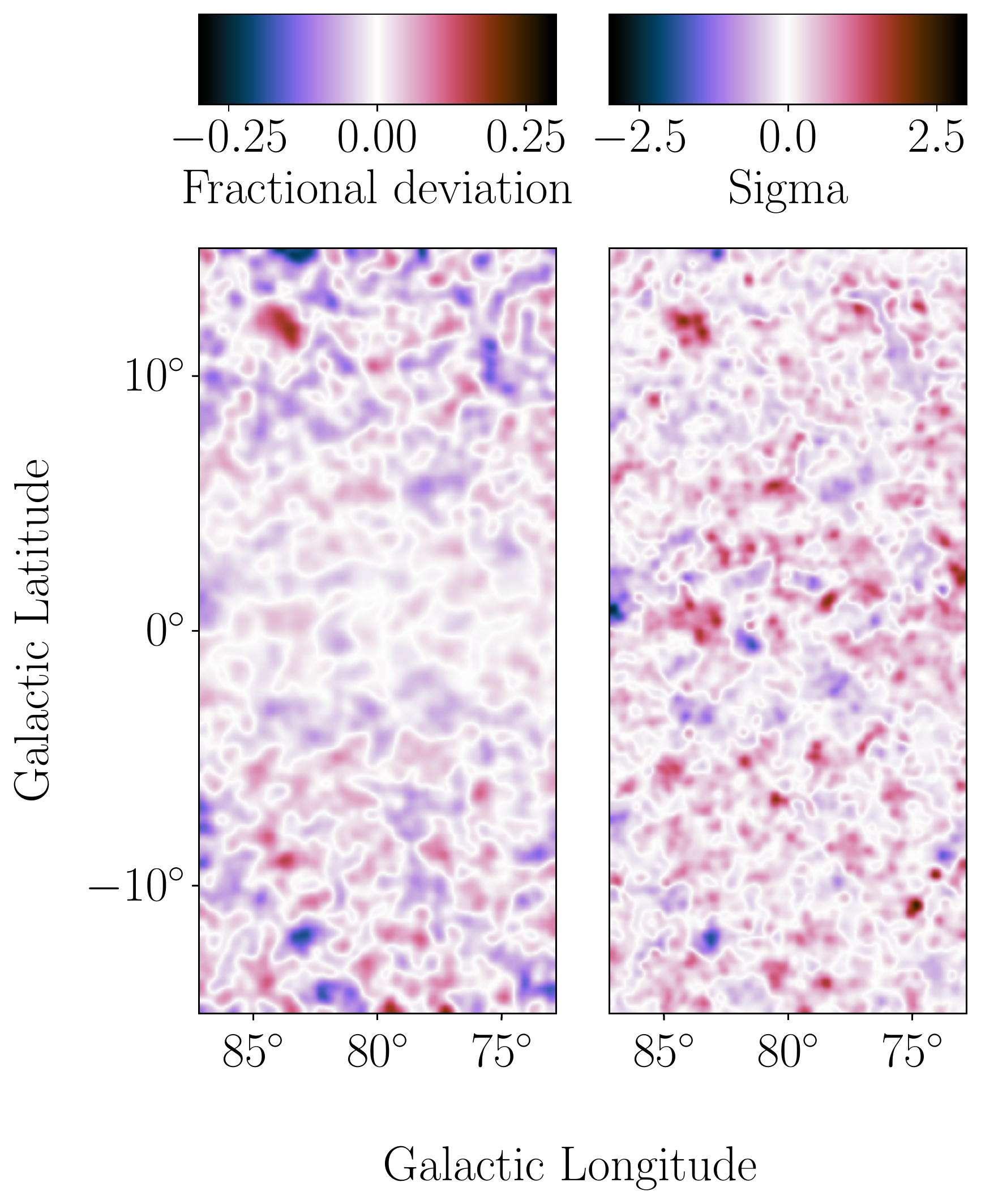}
    \caption{Final deviation maps for the best-fit model over the entire energy range, on the left as fractional deviation, and on the right using the PS map. The bin size is $0.1\degr$ and the size of the smoothing kernel $0.13\degr$.}
    \label{fig:bestmodel}
\end{figure}

Figure~\ref{fig:bestmodel} illustrates the quality of the ROI model after all the optimisation steps. The PS map show that we obtained a model of the ROI with no deviation above $3\sigma$ and the fractional deviation in the bottom left panel show no deviation above $10\%$ in the central part of the ROI.  This model serves as a reference for the following spectro-morphological analysis. The largest deviation with a PS value close to $3\sigma$ lies at $l \sim 84\degr,\, b\sim 12\degr$ \citep[at $1\degr$ from the Geminga-like pulsar PSR~J1957$+$5033; see][]{saz-parkinson2010}. The study of this excess is left for another work. 

\begin{figure}[!htbp]
    \centering
    \includegraphics[width=0.4\textwidth]{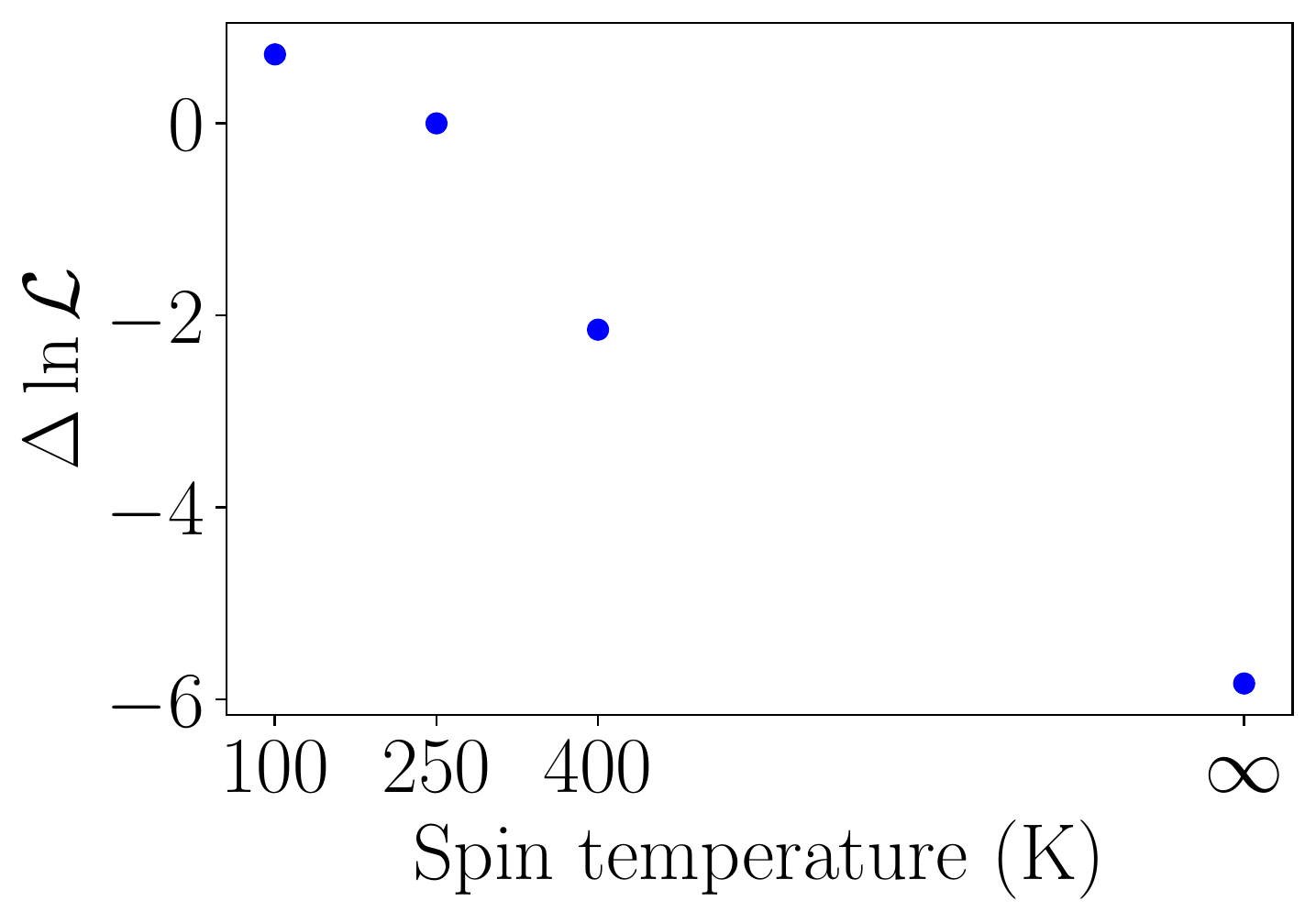}
    \caption{Difference in log-likelihood in fits of the final model for different spin temperatures.}
    \label{fig:diffTspinLoglike}
\end{figure}

Figure~\ref{fig:diffTspinLoglike} shows the final likelihood values for the different spin temperature values considered. The largest likelihood is obtained for a spin temperature of 100~K, but with a difference in log-likelihood $< 1$ with respect to the reference value of 250 K. The largest difference $\Delta \ln \mathcal{L} \sim 6$ is found for the optically thin case.

After the final global fit the normalisation of the \hi emissivity in the local arm is $0.97\pm0.01\,^{+0.06}_{-0.17}$. The normalisation is in reasonable agreement with the average value for the local neighbourhood from \citet{casandjian2015}. Under the hypothesis that the same CR population interacts with atomic, molecular and dark gas in the local arm, we can use the normalisations of the gas maps to infer the \xco 
factor, which yields $(0.75 \pm 0.06\, ^{+0.15}_{-0.02}) \times 10^{20} \mathrm{\,H\,cm^{-2}(K\,km\,s^{-1})^{-1}}$. This is a factor of $\sim$2 lower than results from the earlier analysis of Cygnus in \citet{Ackermann2012}, but close to gamma-ray estimates from nearby CO clouds \citep[for instance][]{remy2017}, which strengthens the hypothesis that \xco variations found between local high-latitude clouds and the local arm may be highly sensitive to the separation of DNM and CO bright molecular cloud in the construction of gas maps and gamma-ray analyses. Other effects related to the increasing difficulty to separate the gas phases at larger distances may also be at play. Our analysis exploiting the PSF event types provides an improvement in this respect over the work in \citet{Ackermann2012}.

The determination of a conversion factor for the DNM tracer is less obvious due to the lack of knowledge on the distribution along the line of sight. However, the morphology of the DNM in Figure~\ref{fig:DNM} closely resembles the structures in the local arm and Cygnus complex in Figure~\ref{fig:gasmaps}. Therefore, for simplicity we assume that all the DNM is in the closest region. Based on this assumption, we can follow the same procedure used for \xco and infer a DNM dust specific opacity $\sigma_{353} = (1.370\pm0.05\,^{+0.059}_{-0.170}) \times 10^{-26}$~cm$^{2}$~H$^{-1}$, also close to gamma-ray results for nearby clouds \citep{remy2017}.

\begin{figure}[!htbp]
    \centering
    \includegraphics[width=1.\linewidth]{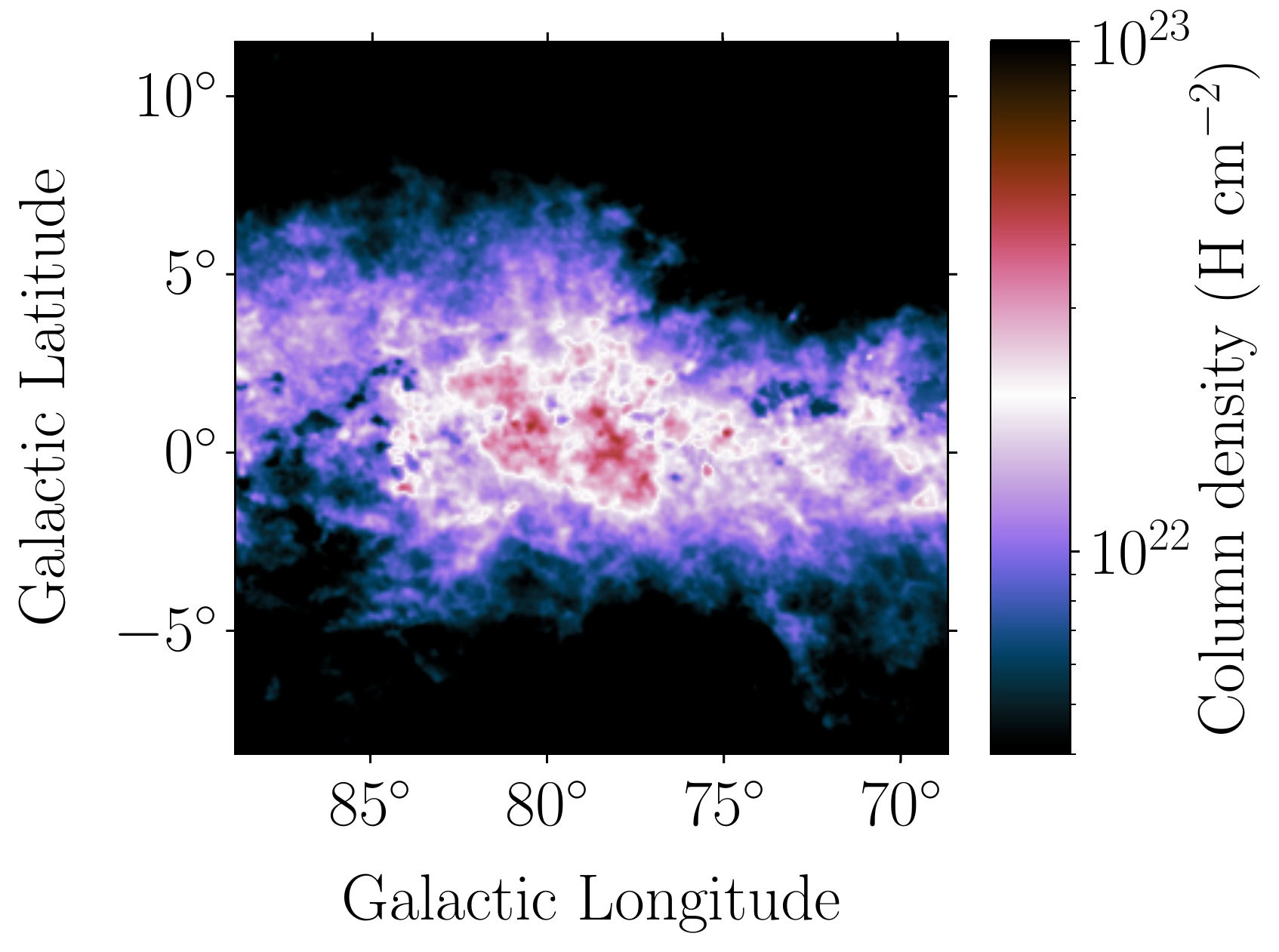}
    \caption{Neutral gas column density in the local arm and Cygnus region, obtained by summing \nhi, \nhd and an estimate of the column density for the DNM, with conversion factors calibrated on the gamma-ray analysis.}
    \label{fig:totcoldens}
\end{figure}

We use these coefficients to build a total column density map of neutral gas in the local arm and the Cygnus complex, which is shown in the left panel of Figure~\ref{fig:totcoldens} and is used for the interpretation of the results in the Section  \ref{sec:discussion}. With respect to the reference spin temperature of 250~K, the total column density of neutral gas increases by $\sim 16\%$ for a spin temperature of $100$~K and decreases by $\sim 5\%$ for the optically thin case.

\subsection{Spectro-morphological analysis}\label{sec:spectromorpho}

\subsubsection{Extension and position versus energy}

We first tested if the best-fit spatial model of the \NNhalo and \NNsw components changes as a function of energy. The component \NNob is left aside in the spectro-morphological analysis because it is displaced regarding the Cygnus X region which we aim to study in this paper, and it lies on the border of the ROI, and therefore its characterisation may not be optimal.

We fitted the extension and position of \NNhalo and \NNsw in five energy bands: 0.5 to 1.6 GeV, 1.6 to 5 GeV, 5 to 16 GeV, 16 to 50 GeV, and 50 to 1000 GeV. As shown in Section ~\ref{sec:spectra}, \NNsw has a softer spectrum: above $\sim5$ GeV the source flux becomes very low and its TS is below $25$. Therefore, fitting the position and extension of the source above $\sim5$ GeV is not possible.
In this section, the diffuse components, that is the gas maps and the IC component, are fixed, while the two components of Cygnus Loop are fixed above $5$~GeV because of their very steep spectrum. \NNob is fixed due to its off-centre position.

\begin{figure*}[!hbtp]
    \centering
    \includegraphics[width=1\linewidth]{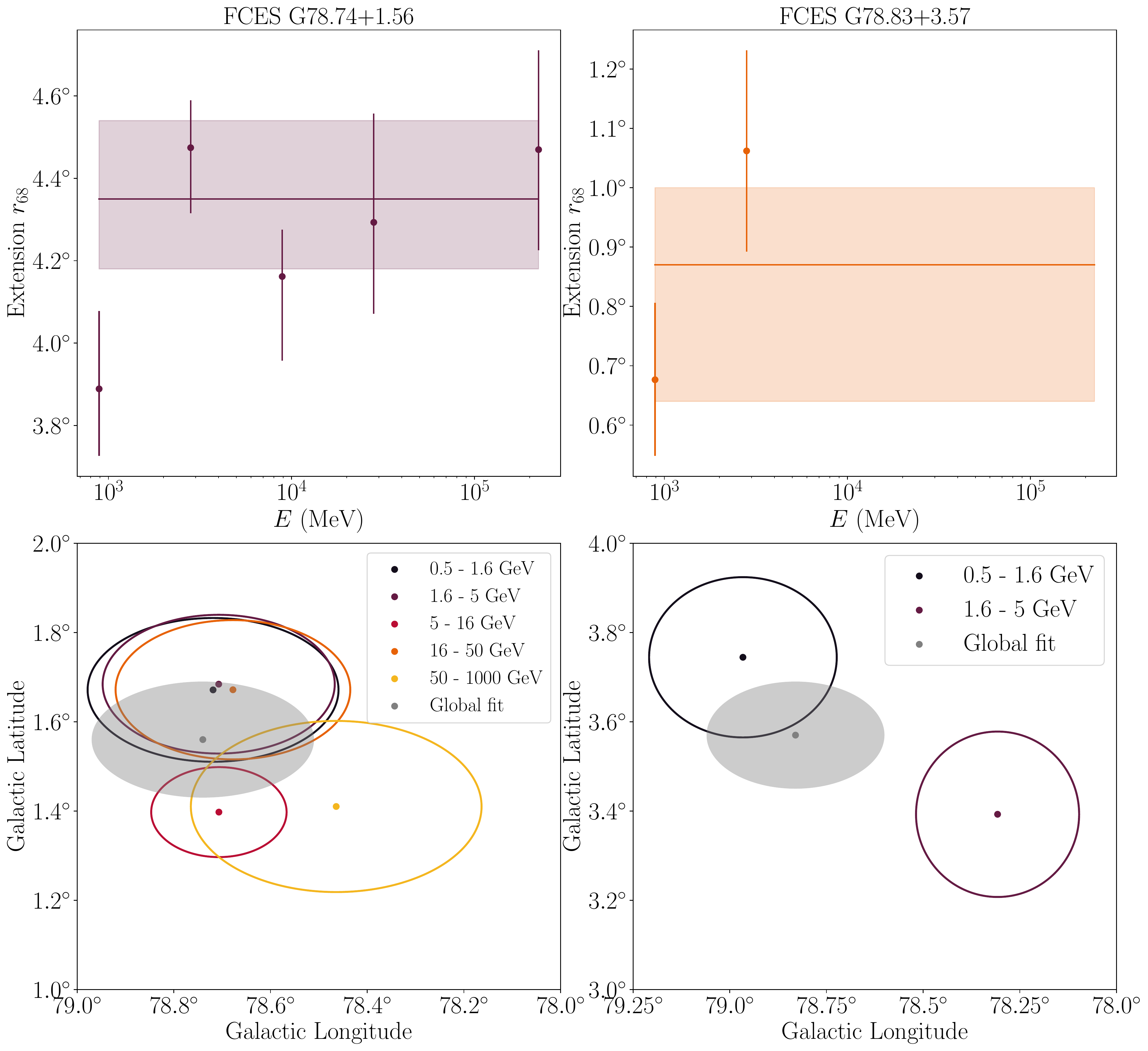}
    \caption{Extension (top) and position (bottom) for \halo (left) and \sw (right) as a function of energy. In the top panels the bands show the values in the global fit over the entire energy range}. In the bottom panels the grey areas show the results in the entire energy range. All uncertainties are provided at 1$\sigma$ level.
    \label{fig:ExtensionVsEnergy}
\end{figure*}

The results are shown in Figure~\ref{fig:ExtensionVsEnergy}. The top panel  shows the extension as a function of energy for both sources. There is no indication of an evolution of the extension as a function of energy, and the values in different energy bands are compatible with that obtained in the broadband analysis. The lower panels show the best-fit centroid positions for the two components. For \NNhalo, all positions are compatible with each other and the broadband fit within $1\sigma$. For \NNsw, we can see a hint of evolution of the position in the first two bins, but the two values are compatible within $2\sigma$ with the value from the broadband fit over the full energy range.

\subsubsection{Spectral variations across the extended sources}\label{sec:rings}

In this section, we search for spectral variations across the emitting regions for \NNhalo and \NNig. We started by examining \NNhalo. To this end, we replaced the Gaussian model with a combination of rings and segments. We tested several combinations but here we describe the profile obtained with a combination of: a central disk of radius $0.7\degr$; five rings of external radius $1.7\degr$, $2.7\degr$, $3.7\degr$, $4.7\degr$ and $6\degr$, that can be decomposed into four segments spanning $90\degr$ in azimuth; and two large rings of external radius $7.4\degr$ and $8.9\degr$.
This somewhat arbitrary setup was chosen to ensure a minimal TS (at least 25) in every segment and ring (see Figure~\ref{fig:TSvar} in Appendix \ref{app:TSvar}). Eventually, however, the wider ring has a low TS ($\sim$10) therefore its parameters have to be interpreted with some caution.

All the components are modelled using a LP spectrum with parameters initiated at the best-fit values found in Section ~\ref{sec:spectra}. The LP model was chosen instead of the SBPL model for this part of the analysis because it yields more stable results when fitting several free components at once.
For this section the two components of Cygnus Loop are fixed due to their off-centre position. \NNob is also fixed due to its off-centre position and proximity with the border of the ROI.

We also tested a combined description of \NNhalo and \NNig via rings and segments by removing the ionised gas template from the emission model, but the highly structured central part was poorly described by the latter model. Thus, we decided to proceed with the ionised gas map in the model, and we used the combination of segments and rings to only represent the source \NNhalo. The spectral shape and normalisation is left free for \NNig.
The decomposition of \NNhalo proceeded through a few subsequent steps.
\begin{description}
\item[A] We replaced the Gaussian by the aforementioned combination of seven rings and a central disk. Only the normalisation of each template was free, while the spectral parameters ($\alpha$ and $\beta$, see Appendix~\ref{app:specmodels} for a definition) were fixed to the initial values from the analysis in Section ~\ref{sec:spectra}.
\item[B] The parameters $\alpha$ and $\beta$ for the disk and rings were free.
\item[C] The innermost five rings (beyond the central disk) were decomposed into four azimuthal segments. Only the normalisations were free and the spectral parameters were fixed to the values obtained in the step B.
\item[D] All spectral parameters were free.
\end{description}

For step C, a few different orientations for the segments were tested, and we present the results for the one yielding the best likelihood. The values of $\Delta \ln \mathcal{L}$ and $\Delta AIC$ for the four steps are provided in Table~\ref{table:ringsevolution}.

\begin{table}[!hbtp]
\caption{Variations of $\ln \mathcal{L}$ and $AIC$ for the decomposition of \halo in rings and segments.}
\label{table:ringsevolution}      
\centering                                      
\begin{tabular}{l c c }          
\hline\hline                        
Step & $\Delta \ln \mathcal{L}$ & $\Delta AIC$ \\    
\hline                                   
A & $-9~[-13, -9]$ & $32~[32, 40]$ \\
B & $7~[7, 14]$ & $18~[4, 18]$ \\
C & $58~[47, 58]$ & $-86~[-86, -64]$ \\
D & $35~[35, 42]$ & $-10~[-24, -10]$ \\
\hline                                             
\end{tabular}
\tablefoot{Values are provided as difference with respect to the previous step, and, for step A, with respect to the global analysis presented in Section ~\ref{sec:spectra}. See the text for details.}
\end{table}

The degradation in step A is due to the approximation of representing a 2D Gaussian with concentric rings and a central disk. However, this is not a cause for concern as the decrease in log-likelihood is small. Step B does not provide an improvement in the description of the emitting region, that is there are no significant variations of the spectrum as a function of distance from the centre. Conversely, we find a model improvement in step C, which demonstrates that the emission is not azimuthally symmetric in intensity. The likelihood improvement is equally shared by all the rings concerned, and it is not surprising given the diversity of regions inside and outside the plane spanned by each broad ring. A further improvement in the likelihood is obtained in step D, showing also the presence of azimuthal spectral variations, mostly driven by the two innermost rings and the central disk with an improvement in log-likelihood of $24$ ($\Delta\mathrm{AIC}=-15$) when only those components have the spectral shape free.

However, the azimuthal variations in the first two rings could be explained by spectral variations across \NNig. To check this hypothesis, we sliced the ionised gas template vertically at $l=79.8\degr$ to separate the two main lobes of ionised gas and repeated the step D.
This results in an improvement of the log-likelihood smaller than one, meaning that no spectral variation is detectable between the two sides of the source \NNig.

We conclude that the best model is the one combining the ionised gas template to describe \NNig and with \NNhalo decomposed and fitted as in step D. This is used as a basis for the interpretation of the results in the next section, where the emission profiles extracted from the data is also shown. Some additional plots illustrating the results are provided in Appendix~\ref{app:TSvar}.

\section{Discussion}\label{sec:discussion}

\subsection{The cocoon and its landscape}
\label{sec:discussion:landscape}

Our analysis shows that the Cygnus cocoon in the LAT energy band is best described by at least two spatial components with different spectra: a central component, \NNig, with a power law spectrum of index $2.19 \pm 0.03^{+0.00}_{-0.01}$, and an extended component, \NNhalo, with a smooth broken power law spectrum with indices $1.67 \pm 0.05^{+0.02}_{-0.01}$ below $3.0 \pm 0.6^{+0.0}_{-0.2}$~GeV and $2.12 \pm 0.02^{+0.00}_{-0.01}$ above. A third newly discovered extended emission component, \NNsw, overlaps in projection with the cocoon and has a spectrum compatible to the one of the central component, a power law with index $2.3 \pm 0.1$.

\begin{figure*}
    \centering
    \includegraphics[width=\textwidth]{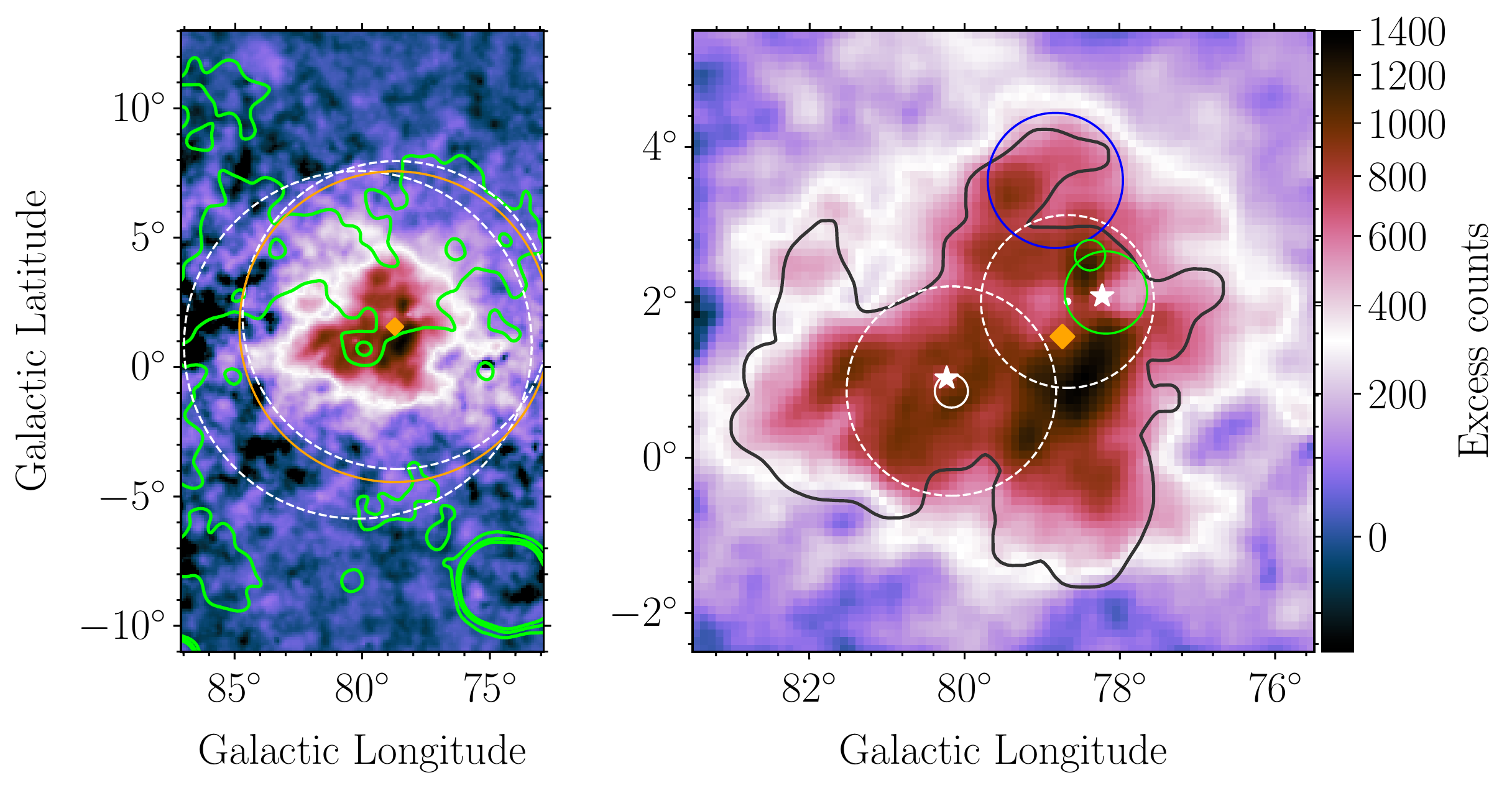}
    \caption{Excess counts corresponding to the three gamma-ray sources associated or potentially related to the cocoon, namely \halo, \ig, and \sw (zoomed in from the left panel in Figure~\ref{fig:DataVsExcess}). Left: extended region. Green contours correspond to the X-ray emission from the ROSAT all-sky survey in the 0.4~keV to 2.4~keV band. The orange circle shows the outer radius of the third to last annulus included in the analysis. The dashed circles show the estimated sizes of the wind bubbles for Cyg~OB2 (lower left) and NGC~6910 (upper right). See text for details. Right: zoom in the central region. Black contours correspond to the 8 $\upmu$m emission from MSX data at $1.85 \times 10^{-6}$ W m$^{-2}$ sr$^{-1}$. The stars show the positions of  PSR J2032$+$4127 (lower left) and PSR J2021$+$4026 (upper right). The green circles show the radius/68\% containment radius of the two emission components associated with the $\gamma$~Cygni SNR (subtracted from the map, see Section ~\ref{sec:anamodel} for details). The blue circle shows the 68\% containment radius of \sw. The continuous circles show the 50\% containment radius for members of the Cyg~OB2 association \citep{berlanas2019} and of the NGC~6910 cluster \citep[][NGC~6910 has a 50\% containment radius of 8.9\arcsec which appears as a dot on this image]{cantat-gaudin2020}. The dashed circles show the estimated sizes of the cluster wind termination shock for Cyg~OB2 (lower-left) and NGC~6910 (upper right). In both panels the orange diamond shows the centre of \halo.}
    \label{fig:morph_MWL}
\end{figure*}

Figure~\ref{fig:morph_MWL} shows the excess counts corresponding to the three gamma-ray sources associated or potentially related to the cocoon, that is total counts minus the best-fit model for all components except \NNhalo, \NNig, and \NNsw (zoomed in from Figure~\ref{fig:DataVsExcess}). The brightest emission in the central region of the cocoon lies in the cavities bounded by the photo-dissociation regions traced by 8 $\upmu$m emission (right panel), as found by \citet{ackermann2011}, and the majority of it is traced by our ionised gas template and associated to source \NNig. The extended cocoon component, \NNhalo, overlaps with the northern rim of the X-ray structure known as Cygnus SB \citep{cash1980}, which may be associated with star-forming regions in Cygnus X \citep{uyaniker01}, although recent data may suggest that the entire X-ray structure is rather a hypernova remnant at a distance of 1.1-1.4 kpc \citep{bluem2020}. Last, source \NNsw is situated along a bright arc of 8 $\upmu$m emission, but does not coincide with any over-densities in neutral or ionised gas densities (see Figs.~\ref{fig:gasmaps}, \ref{fig:DNM}, and \ref{fig:IG}). Its centroid lies at approximately 1\degr\ from the $\gamma$~Cygni SNR, that, if we assume a distance to the Earth of 1.7~kpc, corresponds to a physical distance of $\sim$30~pc. 

\begin{figure}[!htbp]
    \centering
    \includegraphics[width=0.5\textwidth]{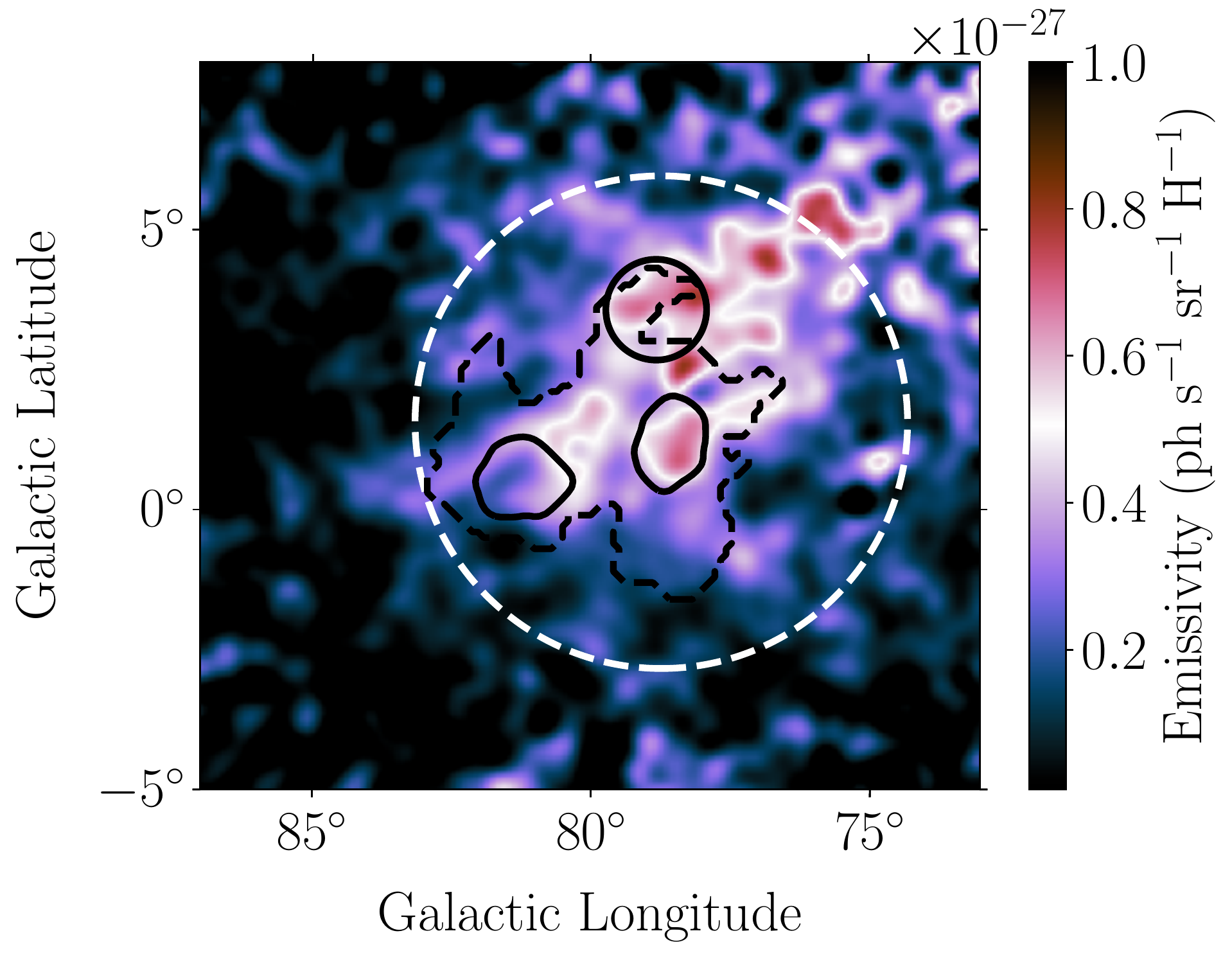}
    \caption{Emissivity map calculated from the excess counts associated to the cocoon in Figure~\ref{fig:DataVsExcess} (right panel). The dashed contours correspond to the $8~\mu$m emission from MSX data at $1.85 \times 10^{-6}~\mathrm{W~m^{-2}~sr^{-1}}$. The contours correspond to the peak column density of the ionised gas template at $3.5\times 10^{21}~\mathrm{H~cm^{-2}}$. The circle and the dashed circle correspond to the position and $r_{68}$ of \sw and \halo respectively.}
    \label{fig:EmissExcess}
\end{figure}

Under the hypothesis that the observed gamma-ray emission is of hadronic origin we can convert the excess map into an emissivity map. To this aim we divided the excess cube in the analysis energy bins by the exposure cube and the total, neutral plus ionised, gas column density map. The latter quantity is an upper limit to the relevant gas column densities because gas could be distributed over a larger distance along the line of sight compared to the volume probed by the particles in the cocoon. However, we expect most of the gas in this region to be concentrated around the star-forming complex in Cygnus X, and we do not have an alternative simple prescription to estimate the foreground and background column densities to be subtracted. The results are displayed in Figure \ref{fig:EmissExcess}.

On one hand, we can see an emissivity peak in the cocoon central area coincident with the peaks in the ionised gas distribution (modelled by \NNig in our analysis) with broad wings extending to several degrees from the centre (\NNhalo). On the other hand, we see a marked peak at the position of \NNsw and around the $\gamma$~Cygni SNR and NGC~6910 stellar cluster. Although position and spectral similarity to \NNig suggest that this source is related to the cocoon, the interpretation is not obvious. \NNsw may be related to gas missing in our model, or else to a nearby source or some peculiar transport configuration that results in an accumulation of particles in this region. In the following for simplicity we concentrate on the interpretation of the two brightest sources in the cocoon area, namely \NNig and \NNhalo.

The striking spatial coincidence of the brightest part of the gamma-ray signal and the contours of the cavity, and to a lesser extent the resemblance with the extended X-ray emission structure, have suggested that both phenomena may have a common origin: the abundant massive-star population of the region. The most prominent stellar clusters in the regions, the Cyg OB2 association and NGC~6910 cluster, are natural candidates, powerful enough to accelerate particles able to produce non-thermal emission at the observed level.

We evaluated the properties of these two objects, following what was done in \citet{ackermann2011}. For Cygnus OB2, we considered 78 O stars \citep{berlanas2020} and a power law mass function of index 1.09 \citep{wright2010}. For NGC~6910 we assumed a power law mass function of index 0.74 \citep{kaur2020} normalised according to Figure~9 of their paper. We evaluated mass loss rates, cluster wind terminal velocities, and mechanical power of the winds by separating stars in four groups, namely O5 to O3, O9 to O5, B5 to B0, and B8 to B5. The sample is limited to stars heavier than B8 due to the validity range for the reference mass-loss rate model adopted. We assumed standard properties of O stars from \citet{martins2005} and for B stars from \citet{cox2000}. We used the parametric wind model by \citet{vink2000}. This yields a mass loss rate of $5.1 \times 10^{-4}$~$M_\sun$~yr$^{-1}$ for Cyg~OB2 and of $2.7 \times 10^{-4}$~$M_\sun$~yr$^{-1}$ for NGC~6910. The mechanical power of the winds is evaluated to $8 \times 10^{38}$~erg~s$^{-1}$ for Cyg~OB2 and $4 \times 10^{38}$~erg~s$^{-1}$ for NGC~6910. The collective wind terminal velocity therefore is $\sim$2200~km~s$^{-1}$ for both clusters. We show in the next section that such powers are sufficient to account for the observed signal in some scenarios.

We can estimate the physical and angular sizes of the cluster wind termination shock and shocked wind bubble using the formulae in \citet{morlino2021}, which follow the simple models in \citet{weaver1977,gupta2018}. If we assume ages of 5~Myr \citep{berlanas2020, kaur2020} for both clusters and interstellar gas densities of 5 H~cm$^{-3}$ we obtain a size of the wind termination shock of 40~pc for Cyg~OB2, and of 33~pc for NGC~6910. The total size of the wind bubble is 200~pc for Cyg~OB2, and 180~pc for NGC~6910. Figure~\ref{fig:morph_MWL} shows that the sizes of the termination shocks are comparable to that of the central emission component, with \NNsw being located at the edge of the termination shock from NGC~6910, while the sizes of the wind bubbles compare well to that of the cocoon extended emission component.

While these results tend to lend support to the idea that Cyg~OB2 and NGC~6910 may be the sources ultimately responsible for the observed gamma-ray emission, we emphasise that the modelling of the winds and bubbles is without any doubts oversimplified. Several effects can be expected to affect the results and weaken the similarity of the gamma-ray emission and expected SB signatures. Generally, the classical theory from \citet{weaver1977} is known to underestimate the radiative losses, hence overestimate the size of the bubble. Hydrodynamical simulations reveal enhanced radiative losses due to instabilities at the interfaces. This results in a bubble being $\sim40$\% smaller than predicted by the classical analytical solution for well-developed bubbles, in agreement with observations \citep{Krause:2014}. At earlier stages, before SB breakout from the parent molecular cloud, the dense and fractal medium surrounding the cluster drives turbulent mixing and efficient cooling, resulting in a reduction of $\sim$30\% in bubble size for parameters relevant to Cyg~OB2 and NGC~6910 \citep{lancaster2021}. More fundamentally, the simple bubble model from \citet{weaver1977,gupta2018,morlino2021} may not be straightforwardly applied to Cyg~OB2, which is not a compact cluster but instead presents multiple substructures with a 50\% containment radius of stellar members spanning 0.2\degr, that is 5 pc at a distance of 1.7~kpc \citep{berlanas2019}. Furthermore, as illustrated in Figure~\ref{fig:morph_MWL}, the bubbles from the two stellar clusters may have interacted and it is not clear how this would have impacted the development of the whole region and whether this should have left specific signs that we should now see.

Therefore, the connection of the observed gamma-ray signal with gas structures imprinted by the development of a SB is far from obvious. Actually, the separation of the cocoon emission into \NNhalo and \NNig, as well as the correlation of the innermost bright signal with ionised gas, can be interpreted in a way that weakens the link between the gamma-ray emission and the cavity delineated by photo-dissociation regions. Indeed, the emission could arise from the interaction of freshly accelerated CRs with the ionised gas inside the cavity, the latter playing no role.
These CRs extend much beyond the limits of the cavity, as was already clearly observed in \citet{ackermann2011}, and their interactions with the ambient medium give rise to the source \NNhalo.

The origin of the CRs powering the cocoon emission can therefore be unrelated to the Cyg OB2 association and NGC~6910 cluster (in the sense that they play no role as a whole, but they can harbour or have harboured the actual source). The central region actually contains a handful of extremely energetic objects and potential particle sources.

We can list: the $\gamma$~Cygni SNR (G78.2$+$2.1) with a probable distance of 1.7 to 2.6~kpc from association with the $\gamma$~Cygni nebula \citep{leahy2013} and dynamic properties estimated in \citet{leahy2020} as ejecta mass of $5\,M_\odot$, age of $9.4^{+2.3}_{-1.6}$~kyr, and supernova (SN) energy of $6.3^{+5.8}_{-3.7} \times 10^{50}$~erg; the $\gamma$~Cygni pulsar, PSR~J2021$+$4026, associated with the $\gamma$~Cygni SNR and with a spin-down power of $1.2 \times 10^{35}$~erg~s$^{-1}$ \citep{ray2011}; PSR J2032$+$4127, a pulsar in a highly eccentric binary system with a Be-type star \citep{lyne2015}, probably part of Cyg~OB2, with an orbital period of 45-50~yr, a spin-down power of $1.5 \times 10^{35}$~erg~s$^{-1}$, and a characteristic age of $\sim$200~kyr \citep{ho2017}.

Furthermore, the centroid of the emission from \NNhalo and the peak in the emissivity map do not coincide with any of the potential particle accelerators, stellar clusters or others (Figure~\ref{fig:morph_MWL} and~\ref{fig:EmissExcess}). Therefore, we tried to account for our observations in a generic way, with a simple diffusion model based on an unspecified source and not exclusively relevant to massive star clusters and their associated SBs. We introduce in the following sections the model framework used and the parameter setups yielding satisfactory fits to the data.

\subsection{A simple diffusion-loss framework for the cocoon}
\label{sec:discussion:model}

Given the layout of the emission exposed in the previous subsection, with significantly extended radiation from a region reaching well beyond the vicinity of potential sources, it seems reasonable to consider that gamma rays are produced by particles that were released by one or several sources some time ago and were transported in the surrounding medium since then. In this section, we aim to provide a quantitative assessment of this idea. 

We interpret the observations in the framework of a one-zone diffusion-loss transport model where particles are continuously injected at a point in space for some duration and then experience diffusive transport in a uniform and isotropic medium. This is very likely an overly simplistic description of the processes at stake because there may be multiple sources, not all of them can be assumed to be of negligible size, the medium is probably not uniform over the few hundreds of parsecs probed by the emission, and there may be other transport processes than diffusion. Yet, our goal is to draw a few key inferences from the observables and we defer more advanced modelling efforts to subsequent publications. Moreover, we show later that such a modelling with a very limited number of free parameters can yield a fairly good representation of the observables.

The full formalism of the model framework is provided in Appendix \ref{app:framework}. Ultimately, the main parameters of the model are: injection luminosity $Q_0$, power law injection spectrum slope $\alpha$, characteristic injection duration $t_{\rm inj}$, diffusion duration $t_{\rm diff}$, and diffusion coefficient normalisation $D_0$. We explored a large parameter space for these four parameters and fitted the predictions to the results of the gamma-ray analysis.

The diffuse emission from the Cygnus cocoon is very extended, with an angular size of $4.4\degr \pm 0.1\degr$ for \NNhalo that translates into a $\sim130$\,pc length at a distance of 1.7\,kpc.
A more compact and central emission component \NNig is correlated with the distribution of ionised gas within a radius of about 50\,pc; the spectrum of \NNhalo is flat and that of \NNig, although significantly softer, is also pretty hard compared to interstellar emission on larger scales. 

Given these observables, we proceeded to educated guesses for the main parameters of the model, considering first the case of a hadronic scenario. The typical extent of the emission provides a constraint on the diffusion length, that is on the product of diffusion coefficient and diffusion time: 
\begin{align}
&r_d = \sqrt{4Dt_{\rm diff}} \gtrsim 100\mathrm{\,pc}, \\
&D = D_{\rm ism}(10\mathrm{\,GeV}) = 10^{29}\mathrm{\,cm}^2\,\mathrm{\,s}^{-1} \Rightarrow t_{\rm diff} \simeq 7.5\mathrm{\,kyr}, \\
&D = D_{\rm supp}(10\mathrm{\,GeV}) = 10^{27}\mathrm{\,cm}^2\,\mathrm{\,s}^{-1} \Rightarrow t_{\rm diff} \simeq 0.75\mathrm{\,Myr}.
\end{align}
If diffusion has the average properties inferred for transport over large scales in the Galaxy \citep{Trotta:2011}, defined by the coefficient $D_{\mathrm{ism}}$, particles need less than 10\,kyr to fill a volume that would account for the extent of the observed emission at a distance of 1.7\,kpc. Conversely, if diffusion is for some reason strongly suppressed by one to two orders of magnitudes as inferred for a variety of sources including star-forming regions \citep{aharonian2019,Abeysekara:2017b,Abramowski:2015}, and is characterised by coefficient $D_{\mathrm{supp}}$, then about 1\,Myr is needed.

The emissivity enhancement inferred for the cocoon is comparable to the local emissivity within a factor of two to three depending on the energy range, such that the CR energy density $u_{\rm CR}$ in the region is similar to the one in the solar neighbourhood, $u_{\rm CR,local}$. This makes it possible to constrain the properties of particle injection, namely its power $L_{\rm inj}$ and typical duration  $t_{\rm inj}$:
\begin{align}
& \frac{4\pi}{3}r_d^3 \times u_{\rm CR} \simeq \frac{1}{2}L_{\rm inj}t_{\rm inj}, \\
& u_{\rm CR} \simeq u_{\rm CR,local} \simeq 1\mathrm{\,eV}\,\mathrm{\,cm}^{-3}, \\
& L_{\rm inj}t_{\rm inj} \simeq 4 \times 10^{50}\mathrm{\,erg}, \\
& L_{\rm inj} = 10^{38}\mathrm{\,erg}\,\mathrm{\,s}^{-1} \Rightarrow t_{\rm inj} \simeq 0.1\mathrm{\,Myr}.
\end{align}
As computed in the previous subsection, the mechanical luminosity of the most prominent star clusters in Cygnus is in the range $4-8 \times 10^{38}$~erg~s$^{-1}$. Such a power source can deliver particle injection at a level of $10^{38}$~erg~s$^{-1}$, pending efficient particle acceleration with a yield of $\sim10-30$\% (by some unspecified mechanism at this stage). In that case, the inferred CR density enhancement can be attained if injection lasts over about 100\,kyr (and particles accumulate in the volume, see the discussion in the next paragraph). If particle acceleration is less efficient or the source is less powerful, by about an order or magnitude, then injection has to proceed on Myr time scales. Alternatively, a supernova producing $10^{50}$\,erg of accelerated particles and releasing the majority of them over $\sim3-10$\,kyr would provide an injection power of $3-10 \times 10^{38}$~erg~s$^{-1}$ and thus allow short-lived injection.

Scenarios with $t_{\rm inj}$ much smaller than $t_{\rm diff}$ are not viable because particles spread out and leave the volume too rapidly, which results in too flat intensity profiles and too steep spectra (because energy-dependent diffusion depletes the particle population at the high end of the spectrum). So $t_{\rm inj}$ has to be comparable to or greater than $t_{\rm diff}$, with the additional constraint that sufficient energy should be released within a time $t_{\rm diff}$ to match the observed level of emission. In practice, this means that: for average interstellar diffusion, the region is filled over a $\sim10$\,kyr timescale, thus requiring a strong enough source with injection power $\sim10^{39}$~erg~s$^{-1}$ typical of a SN; alternatively, weaker sources such as the star clusters in Cygnus with injection power $\sim10^{37-38}$~erg~s$^{-1}$ require moderate to strong diffusion suppression and transport occurring over hundreds to thousands of kyr.


These considerations remain mostly valid in the case of a leptonic scenario. The main difference with a hadronic scenario is the importance of energy losses, mostly from synchrotron radiation and inverse-Compton scattering. Yet, for an interstellar magnetic field strength $B=3$\,$\upmu$G and optical and infrared interstellar radiation fields with total energy density of about 1\,eV\,cm$^{-3}$, such as those predicted in the large-scale model of \citet{Popescu:2017} at the Galactic position of the cocoon, particles with energy below $300-400$\,GeV have a cooling time of 1\,Myr or more. The transport of electrons over the distances and time scales considered above may therefore be little affected by energy losses in many scenarios. The actual situation is however far more complex because radiation densities in the innermost region of the cocoon are much stronger than the large-scale interstellar average\cmmnt{(\citep{Orlando2007})}, by an order of magnitude, which could significantly affect the spectral and morphological properties of the emission from the population of propagated electrons. Unfortunately, the model framework that we used for this work cannot handle inhomogeneous energy losses.


\subsection{Possible diffusion scenarios for the cocoon}
\label{sec:discussion:results}

\begin{table*}[!hbtp]
\caption{Summary of the different diffusion-loss model setups}
\label{table:diffmodel:params} 
\centering
\begin{tabular}{l c c c c c c c |}
\hline
& H1 & H2 & H3 & H4 & L1 & L2 \\
\hline
& hadronic & hadronic & hadronic & hadronic & leptonic & leptonic \\
$t_{\rm inj}$ (yr) & $10^8$ & $10^8$ & $3 \times10^3$ & $3 \times10^4$ & $10^8$ & $10^8$ \\
$t_{\rm diff}$ (yr) & $3 \times 10^6$ & $3 \times 10^5$ & $10^4$ & $10^5$ & $3 \times 10^6$  & $10^6$  \\
$D_0$ (cm$^2$\,s$^{-1}$) & $10^{27}$ & $10^{28}$ & $10^{29}$ & $10^{28}$ & $10^{27}$ & $10^{28}$  \\
$L_{\rm inj}$ (erg/s) & $3.6 \times 10^{36}$ & $3.2 \times 10^{37}$ & $2.8 \times 10^{39}$ & $2.8 \times 10^{38}$ & $6.8 \times 10^{35}$ & $2.4 \times 10^{36}$ \\
$\alpha$ & 2.0 & 2.0 & 2.0 & 2.0 & 2.0 & 2.0 \\
\hline
\hline
\end{tabular}
\end{table*}

We tested the hypothesis that components \NNhalo and \NNig are produced by a single population of non-thermal particles. In that context, \NNig is gas-related emission (pion decay in hadronic scenarios, and Bremsstrahlung in leptonic scenarios) from the innermost $\sim50$\,pc region of the cocoon, where a significant amount of ionised gas is present as evidenced by free-free emission, while \NNhalo is additional emission on top and beyond \NNig that is not necessarily gas-related (it can be a mix of inverse-Compton and Bremsstrahlung in leptonic scenarios). In the following, we relate \NNig and \NNhalo to so-called central and extended regions in our model, respectively.

For computational reasons, we did not perform an overall optimisation of all model parameters and instead investigated a limited number of scenarios selected from the above guess for viable parameter values. For each parameter setup, the comparison of model predictions and gamma-ray analysis results goes through the following steps that are expected to guarantee a maximum consistency. 
\begin{description}
\item[Central and extended component separation:] for a given run of the model, gas-related emission from the innermost region within 50\,pc in radius of the injection point is handled separately. All related quantities (particle density, gas column density, emission intensity,...) are not included in the properties of the complementary extended region. For instance, gas-related emission intensity along a line of sight that passes through the central region seen in projection is split into a central contribution and the remaining extended contribution. Although the separation is clear-cut in the model, we cannot exclude that there is some cross-talk between overlapping components in the data analysis.

\item[Gas column density correction:]: the model-predicted emission for the extended component was divided into rings, with the angular binning used in the spectro-morphological analysis, and gas-related emission in each ring was rescaled by the ratio of the actual average gas column density in the ring to that corresponding to the default uniform density assumption of the model. 
The same rescaling is also applied to the central component, treated as a single region with average properties.

\item[Fitting of total emission spectra:] the total emission spectrum of the extended component is fitted to the observed spectrum for \NNhalo, which yields the injection luminosity for the whole particle population. Then, the total emission spectrum of the central component, re-scaled by the fitted injection luminosity, is further fitted to the observed spectrum for \NNig. This second fit is meant to correct for the uncertain average column density for the ionised gas in the central region. Both fits are performed via $\chi^2$ minimisation, from significant spectral points and using statistical uncertainties only.
\end{description}


There is, however, a subtlety regarding how emission from the ionised gas should be handled. Atomic and molecular gas in the cocoon region enter twice in the analysis: in the fit to the \F-LAT data over a large ROI, where they trace emission from the background population of CRs, and in the interpretation of the extended emission from \NNhalo and \NNig, where they are associated to an additional population of particles. Conversely, the ionised gas template enters just once, and the associated emission may therefore comprise contributions from background CRs and from an additional population of particles. In practice, what is fitted to the spectrum of \NNig is the spectrum of the central component of the model, possibly augmented by the spectrum of the emission from the ionised gas for a local emissivity (because the background CR population in the cocoon region can be considered close to the local one; see Section  \ref{sec:final}). We tested both options and, as illustrated below, it turns out that not including a contribution from background CRs provides much better fits to the data. 

\subsubsection{Hadronic scenarios}
\label{sec:discussion:results:hadro}

To begin with, we present the result of complete calculations for hadronic scenarios. Following the above discussion of the most likely diffusion-loss model setups given the observables at hand, we present the results of four scenarios dubbed H1, H2, H3 and H4, the parameter sets of which are presented in Table \ref{table:diffmodel:params}. The H1 and H2 scenarios feature constant injection of a hard spectrum of protons and mainly differ by the diffusion time (300\,kyr or 3\,Myr) and level of diffusion suppression (by a factor 10 or 100 with respect to the interstellar average). The H3 and H4 scenarios corresponds to shorter-lived injection over 3 or 30~kyr and transport over 10 or 100~kyr in a medium with no or moderate diffusion suppression (by a factor 10 at most with respect to the interstellar average). Scenario H4 actually corresponds to a scaled version of scenario H3 (multiplying injection and diffusion times and dividing diffusion normalisation and injection power by the same amount, that is ten), such that both setups are completely identical in terms of predictions.

\begin{figure}[!hbtp]
\centering
\includegraphics[width=0.9\columnwidth]{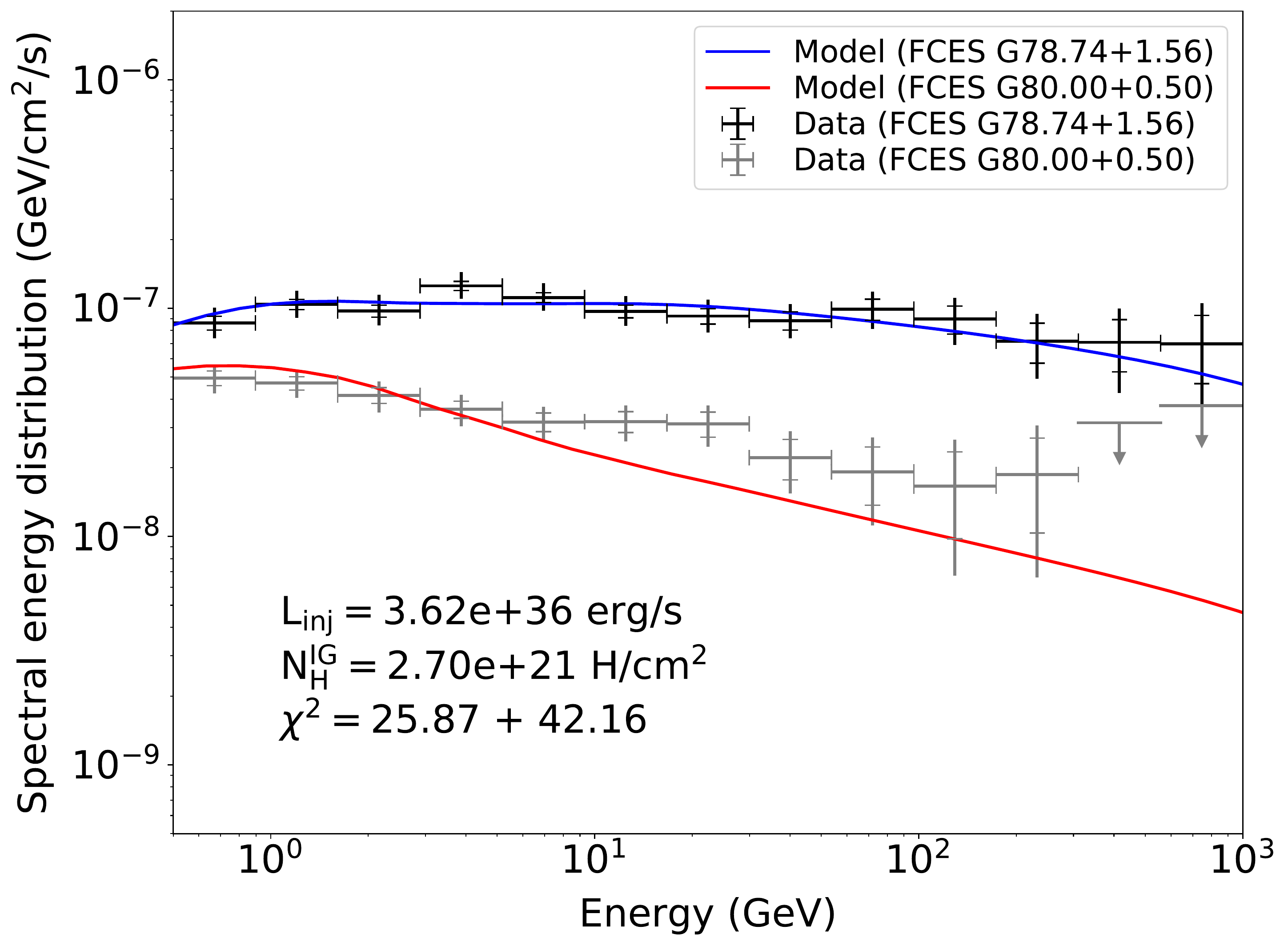} \\
\includegraphics[width=0.9\columnwidth]{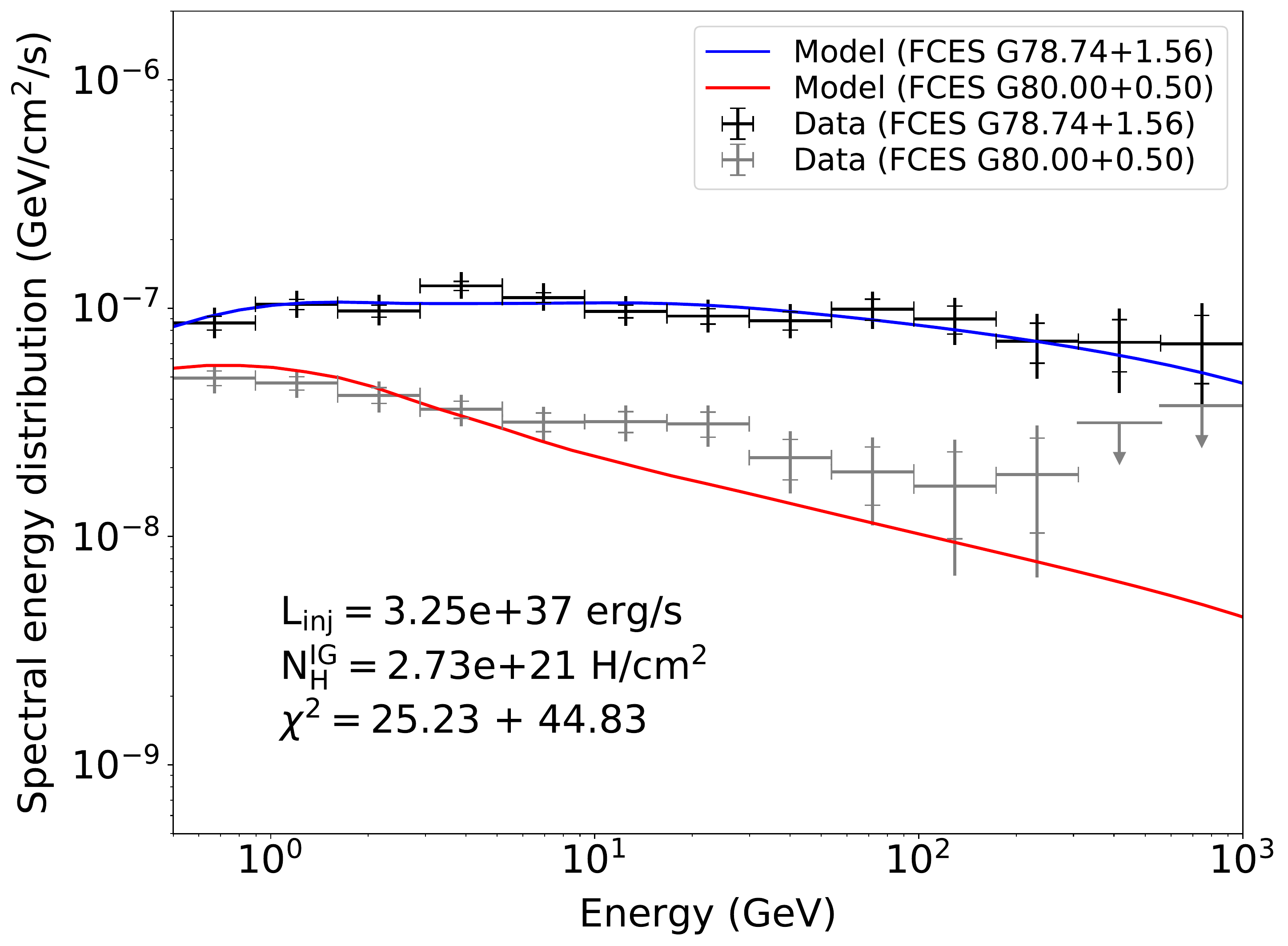}
\caption{Fitted model spectra for the \halo and \ig emission components, for model setup H1 (top) and H2 (bottom). A fit of the model to the spectrum of \halo is done first, with the injection luminosity as fitting parameter, followed by a second fit to the spectrum of \ig, with ionised gas column density as fitting parameter. This last fit includes a contribution to the emission from a background population of CRs (see text). The full extent of the error bars corresponds to the quadratic sum of the statistical and systematic uncertainties, while the caps mark the contribution of the statistical uncertainty only. The first $\chi^2$ corresponds to the fit to \halo and the second one to \ig.}
\label{fig:diffmodel:hadro:spec}
\end{figure}

Figure~\ref{fig:diffmodel:hadro:spec} shows the total fitted spectra for \NNhalo and \NNig for model setups H1 and H2. The predicted shape for the \NNhalo spectrum is in good agreement with the data, while that for the \NNig spectrum is too steep. A steeper predicted spectrum for \NNig is obtained because higher-energy particles leave the innermost regions more rapidly than lower-energy particles, and also because it contains a contribution from the background CR population that has a steeper spectrum. 

\begin{figure}[!hbtp]
\centering
\includegraphics[width=0.9\columnwidth]{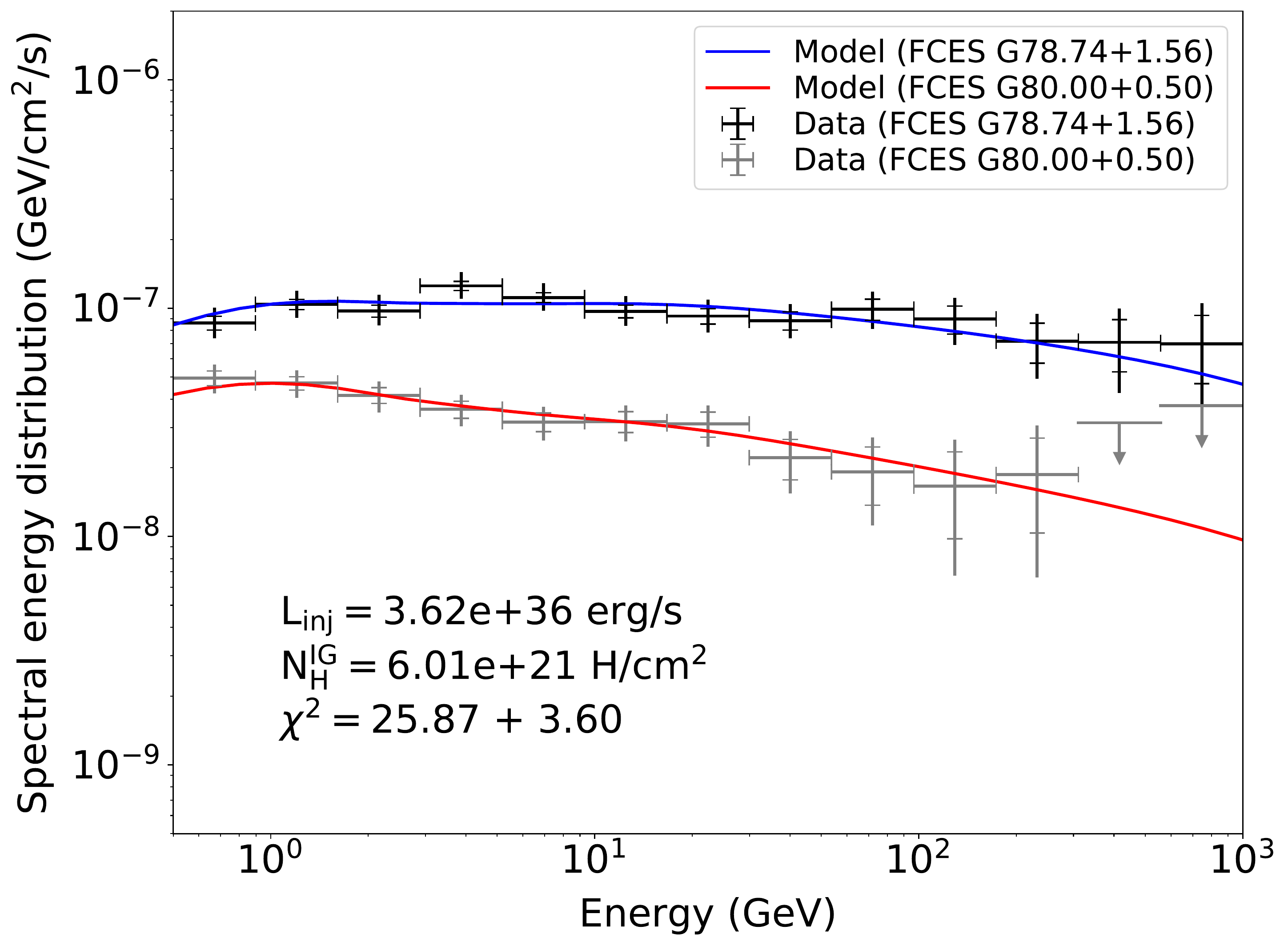} \\
\includegraphics[width=0.9\columnwidth]{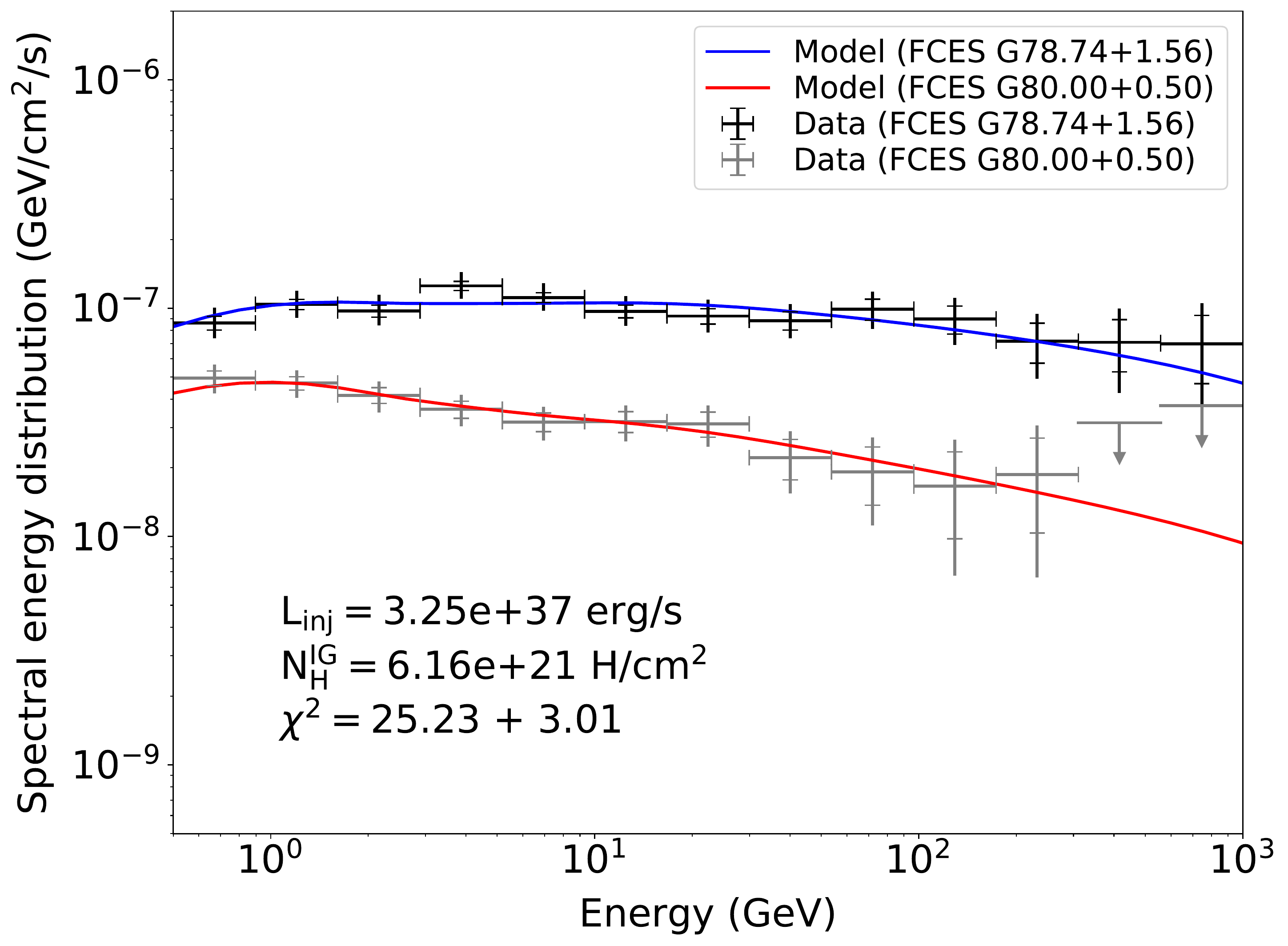} \\
\includegraphics[width=0.9\columnwidth]{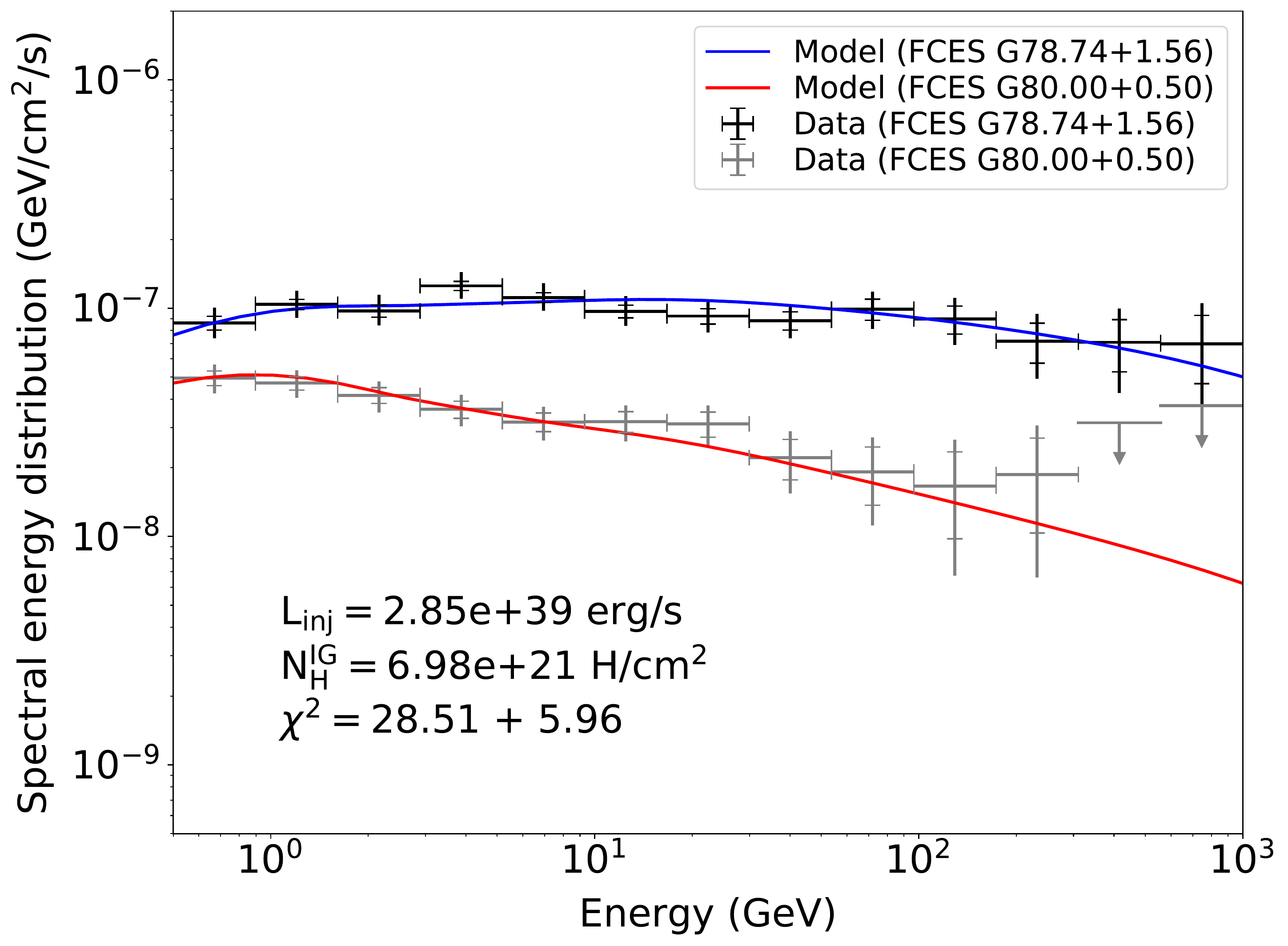}
\caption{The top two panels are the same as in Figure \ref{fig:diffmodel:hadro:spec}, without a contribution to the emission from a background population of CRs in the fit to the spectrum of \ig. The bottom panel is the corresponding figure for scenario H3. The first $\chi^2$ corresponds to the fit to \halo and the second one to \ig.}
\label{fig:diffmodel:hadro:spec-nolocal}
\end{figure}

Interestingly, a much better fit to the spectrum of \NNig is obtained when not adding a local emissivity contribution to the model spectrum for the central region, at the expense of higher fitted column density for the ionised gas. This is illustrated in the top two panels of Figure \ref{fig:diffmodel:hadro:spec-nolocal}. Emission from the ionised gas in the innermost regions of the cocoon would then arise only from CRs produced in Cygnus. Background pre-existing CRs may have been evacuated in the past during the SB growth, for instance by advection in the stellar winds. Alternatively, the spectral signature of these pre-existing CRs may have been absorbed by another component in the fit to the LAT observations (for instance by the molecular gas or DNM templates that have a high degree of correlation with ionised gas in the central region). Since the spectral fit is so much better when not including the contribution from background CRs for \NNig (this is true also in leptonic scenarios), we present in the following only results produced with this approach. We note, however, that this has almost no influence on the intensity and emissivity profiles presented thereafter. 

\begin{figure*}[!hbtp]
\centering
    \subfloat{\includegraphics[width=0.46\textwidth]{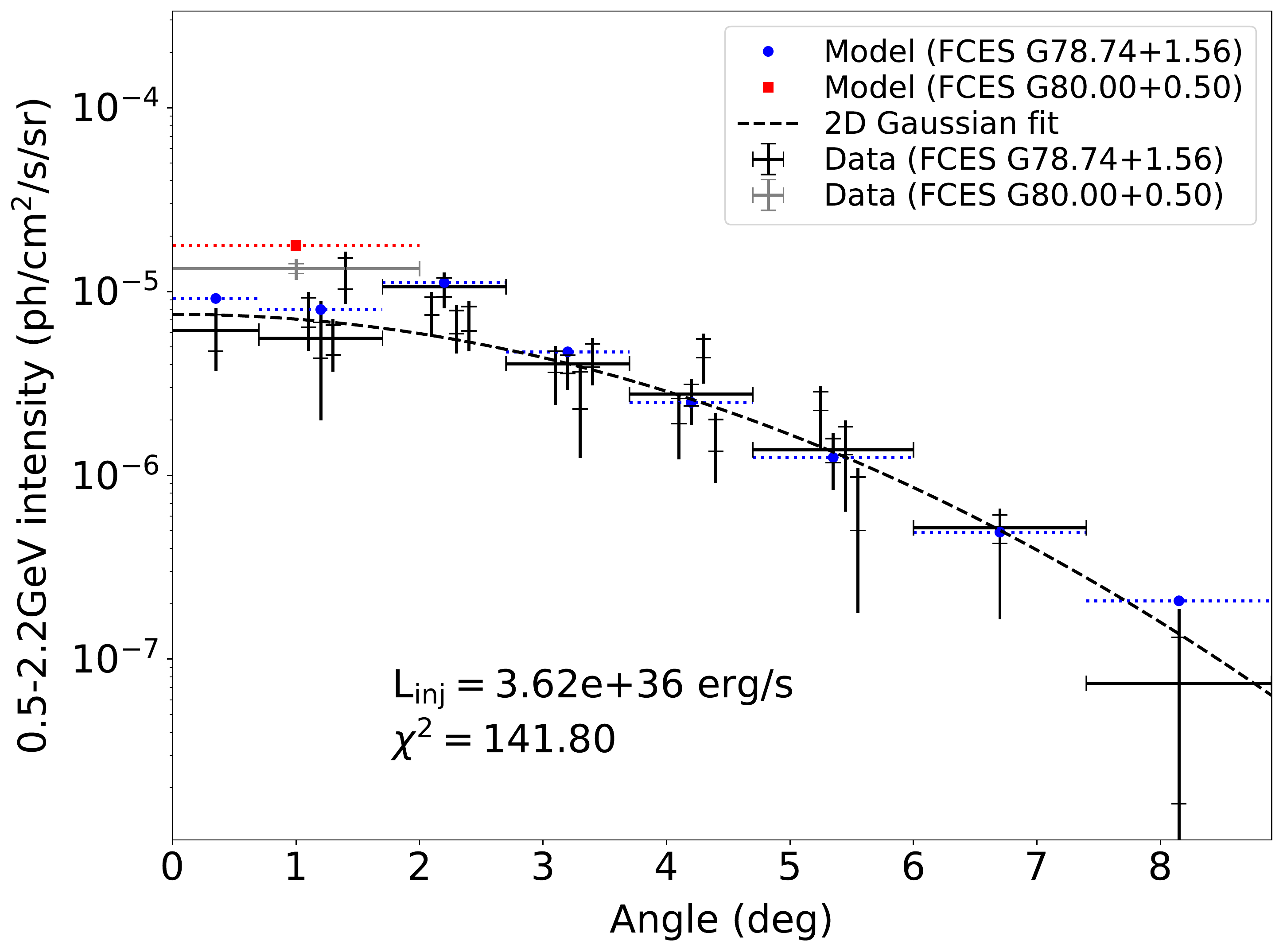}}
    \qquad
    \subfloat{\includegraphics[width=0.46\textwidth]{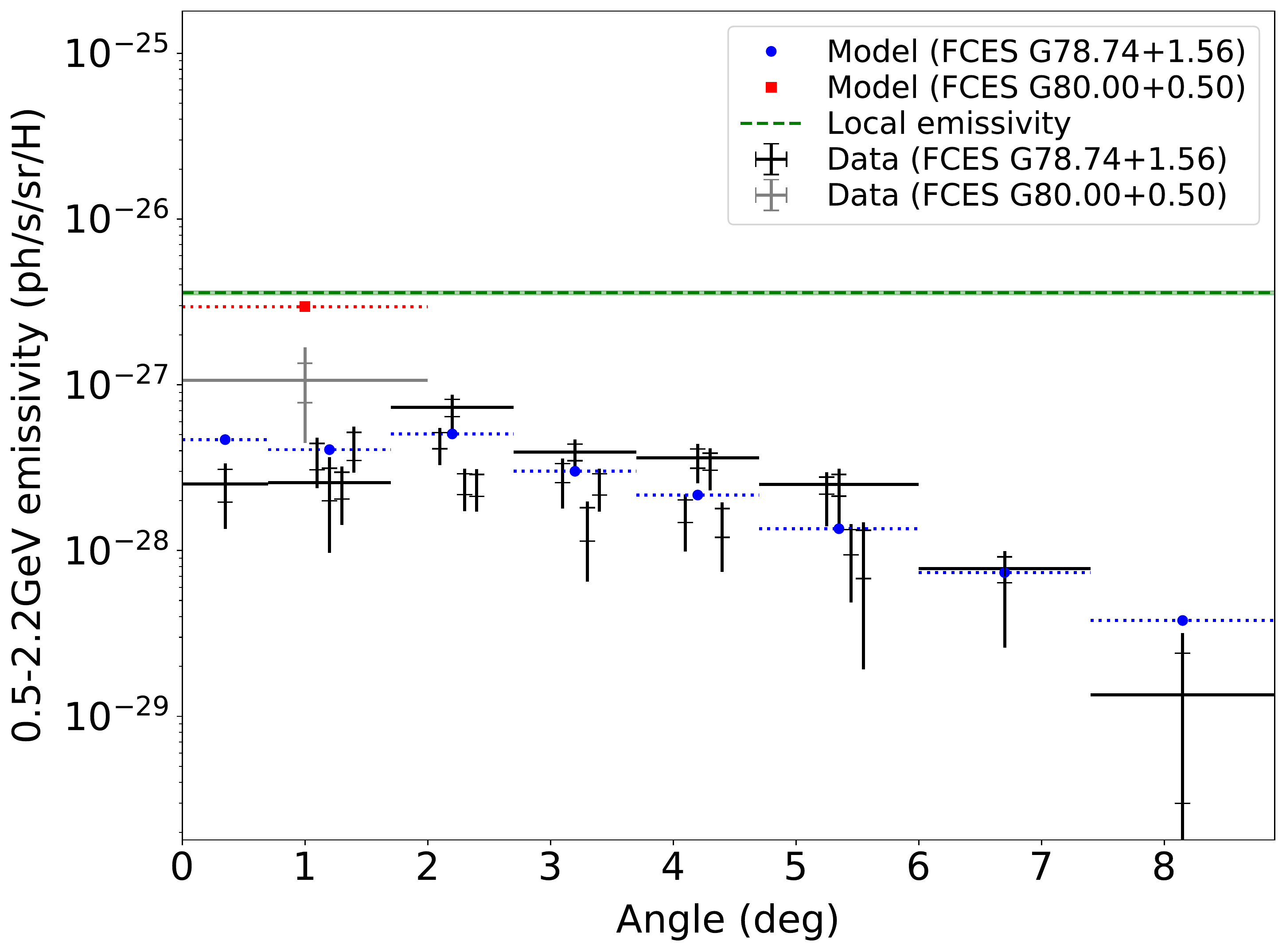}}
    \qquad
    \subfloat{\includegraphics[width=0.46\textwidth]{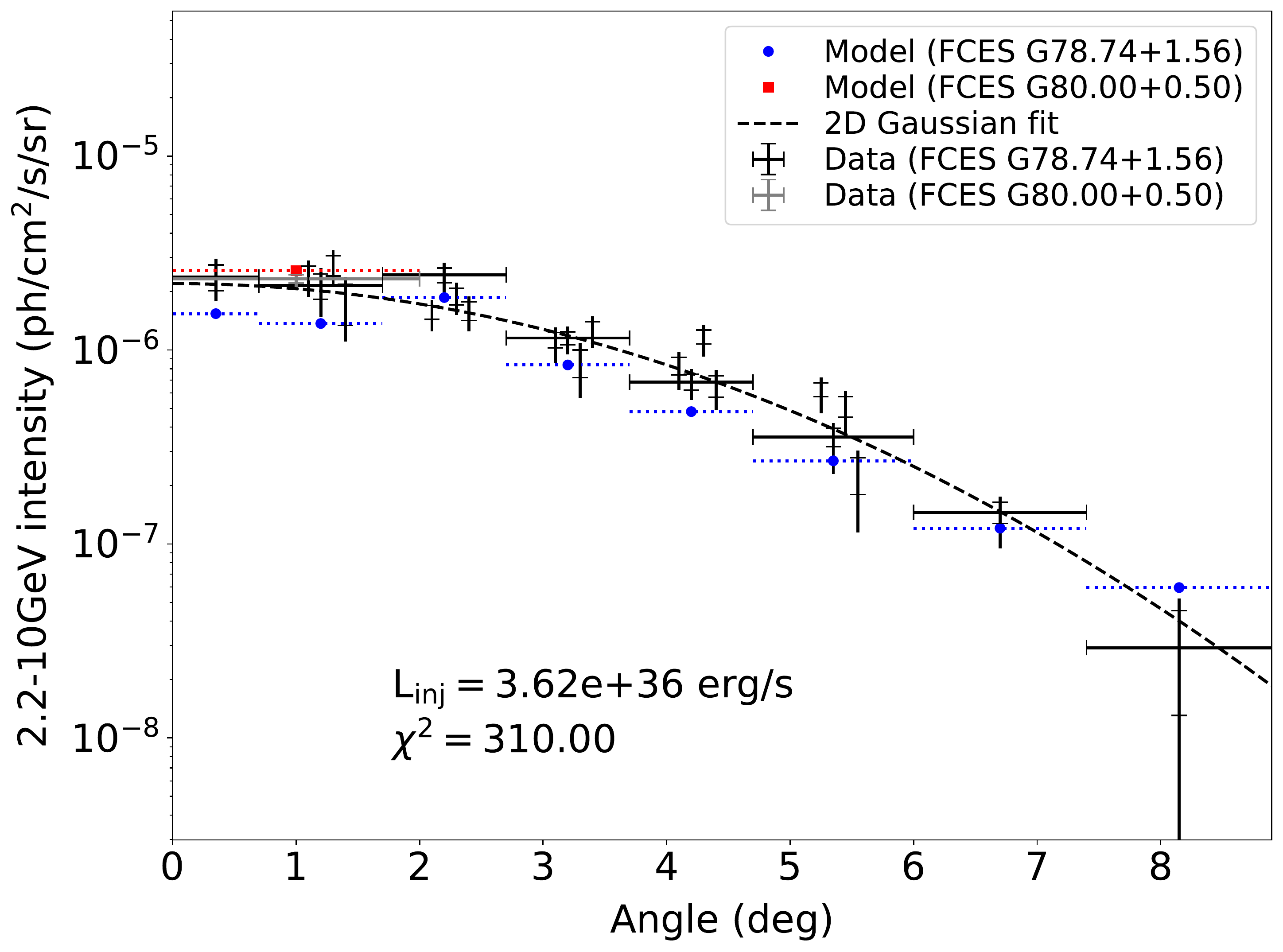}}
    \qquad
    \subfloat{\includegraphics[width=0.46\textwidth]{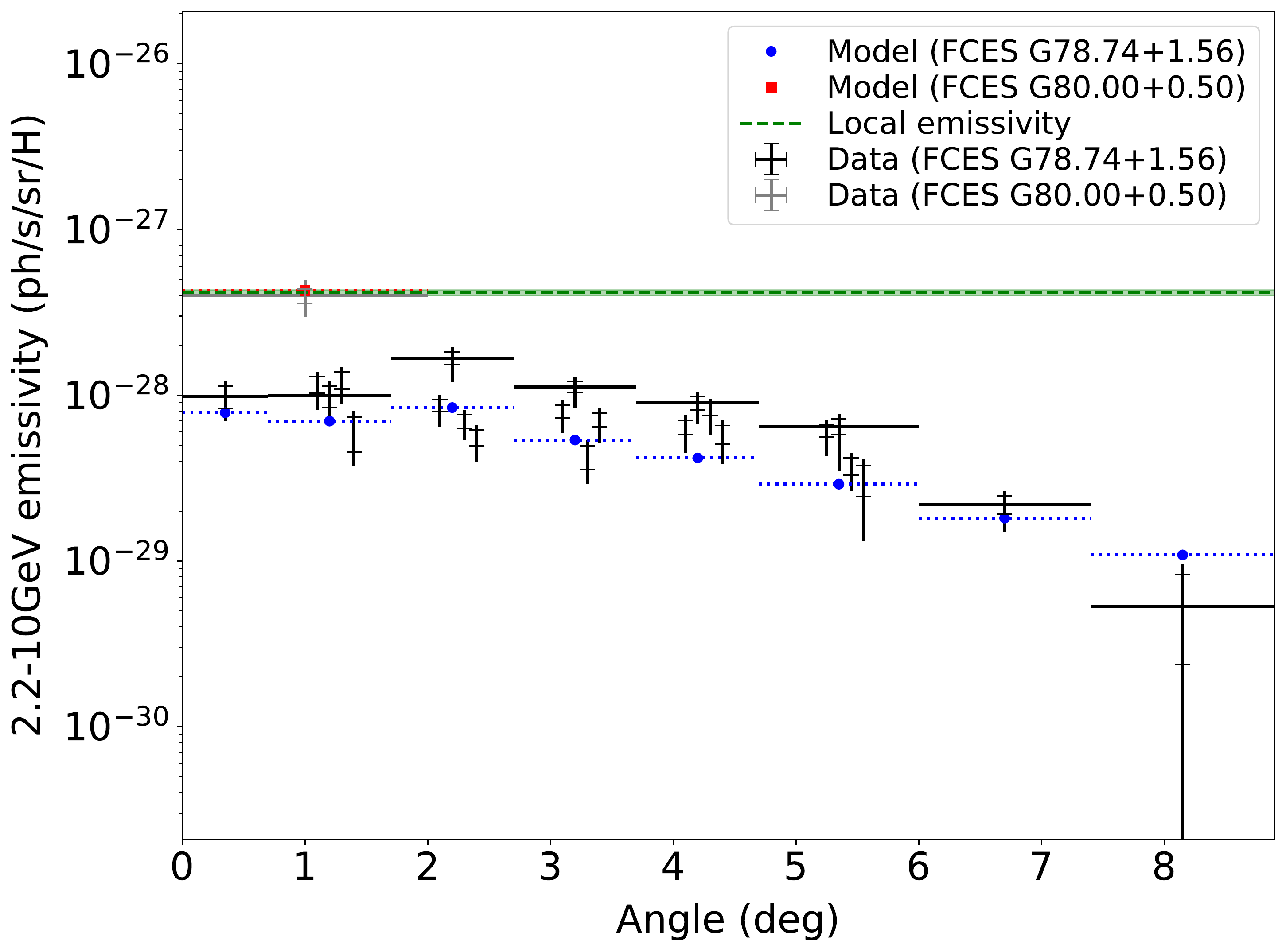}}
    \qquad
    \subfloat{\includegraphics[width=0.46\textwidth]{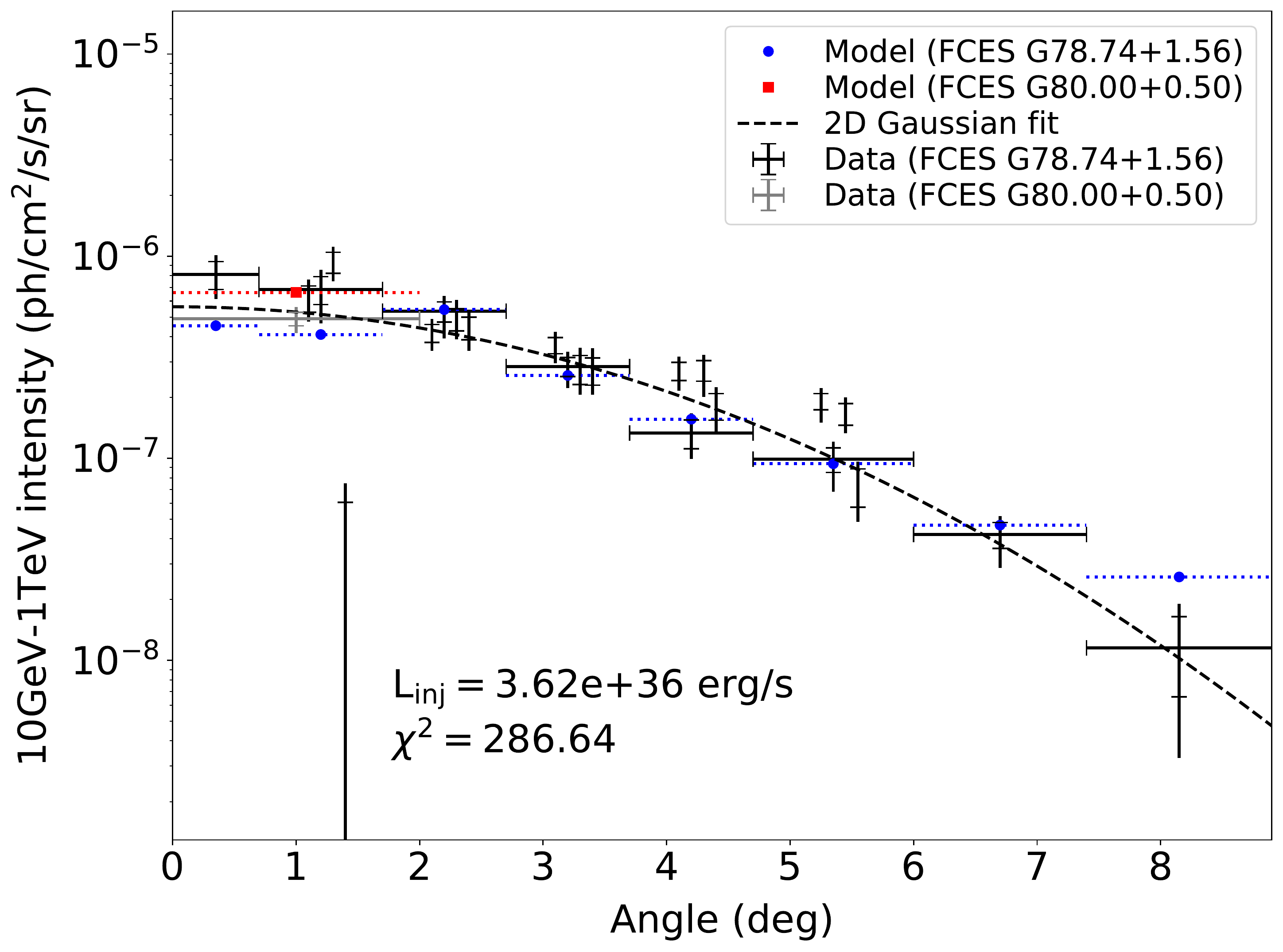}}
    \qquad
    \subfloat{\includegraphics[width=0.46\textwidth]{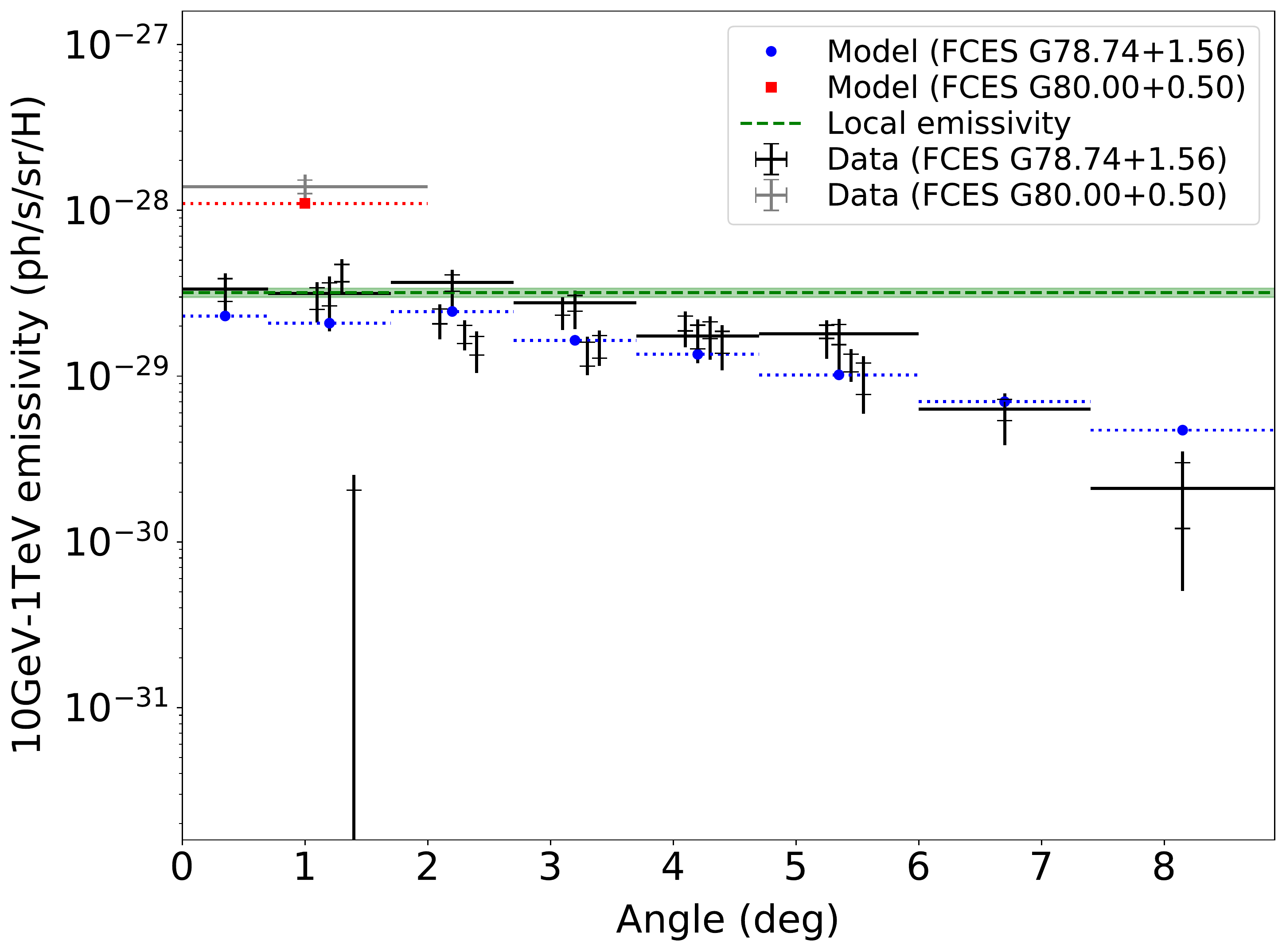}}
    \caption{Intensity and emissivity radial profiles in three different gamma-ray energy bands for the \halo and \ig emission components, compared to predictions for model setup H1. In the intensity plots, the intensity distribution corresponding to the best-fit two-dimensional Gaussian model is displayed for comparison as a dotted line. In the emissivity plots, the local emissivity and its uncertainty in each energy range are displayed for comparison as a dotted line and a shaded band. The data points correspond to a decomposition of the emission into the ionised gas template, a central disk, two outer rings, and five intermediate rings split azimuthally into four segments. For the latter, we displayed the corresponding angular range only for one segment in each ring and introduced a small horizontal shift of the others, for readability. The full extent of the error bars corresponds to the quadratic sum of the statistical and systematic uncertainties, while the caps mark the contribution of the statistical uncertainty only.}
\label{fig:diffmodel:hadro:intemiss:H1}
\end{figure*}

For model setup H1 (resp. H2), the fit implies a proton injection luminosity of $3.6 \times 10^{36}$~erg~s$^{-1}$ (resp. $3.2 \times 10^{37}$~erg~s$^{-1}$), which would correspond to proton injection efficiencies $<0.5-1$\% (resp. $<5-10$\%) for the Cyg OB2 or NGC 6910 clusters. Such low efficiencies are consistent with the assumed flat injection spectra with $\alpha=2.0$, at least in the framework of diffusive shock acceleration. For model setup H3 (resp. H4), the fits implies a much higher proton injection luminosities of $2.8 \times 10^{39}$~erg~s$^{-1}$ (resp. $2.8 \times 10^{38}$~erg~s$^{-1}$). This would correspond either to a high particle acceleration efficiency of $\sim20$\% in an SN with a $10^{51}$\,erg explosion kinetic energy, with subsequent release of accelerated particles over a timescale of 3\,kyr (resp. 30\,kyr), or to a lower acceleration efficiency in an SN more energetic than in the canonical picture. The first option, with relatively high efficiency, may be conflicting with our assumption of a flat injection spectrum.e

Figure \ref{fig:diffmodel:hadro:intemiss:H1} displays the predicted intensity and emissivity profiles for both central and extended model components in scenario H1, compared to the values inferred from the spectro-morphological analysis in segments, in three different energy bands: 0.5-2\,GeV, 2-10\,GeV, and 10\,GeV-1\,TeV. The comparison for other scenarios is shown in Appendix~\ref{app:scenarios}. To compare the different model setups, we provide in each panel the $\chi^2$, computed from all significant intensity data points using statistical uncertainties only. There is, however, no formal fit of the model to the measured intensity profiles, and this $\chi^2$ is just a figure of merit to characterise each model setup. The agreement is overall quite good from the centre up to beyond 8\degr, especially considering the simplicity of the model and the limited number of parameters. Most measurements are within a factor two of the predictions. When subtracting the contribution from the central model component, relatively flat emissivity profiles are obtained for the extended model component, in agreement with the trend inferred from the data analysis. The results lend support to the idea that \NNhalo and \NNig are produced by the same population of particles. All four model setups are overall equally good at accounting for the data, despite widely different parameter sets.

\subsubsection{Leptonic scenarios}
\label{sec:discussion:results:lepto}

We now present the result of complete calculations for leptonic scenarios, in which the emission can be produced by non-thermal Bremsstrahlung and inverse-Compton scattering. We considered two scenarios dubbed L1 and L2, the parameter sets of which are presented in Table \ref{table:diffmodel:params}. Both scenarios feature constant injection of a hard spectrum of electrons and mainly differ by the diffusion time (1 or 3\,Myr) and level of diffusion suppression (by a factor 10 or 100 with respect to the interstellar average). The magnetic field is assumed to have a strength $B=3$\,$\upmu$G, and the interstellar radiation field model is taken from the large-scale model of \citet{Popescu:2017} at the Galactic position of the cocoon (at a 1.7\,kpc distance from us). We tested the effect of a stronger interstellar radiation field model, such as the one used in the original Cygnus cocoon paper \citep{ackermann2011}, and obtained a much poorer fit to our measurements. This stems mostly from the shorter propagation range and different relative contributions of Bremsstrahlung and inverse-Compton to the emission. This scenario is however extreme since it enforces very strong inverse-Compton losses over an extended volume, whereas in reality enhanced radiation fields are expected only in the innermost regions of the cocoon. This is a caveat of the model framework used in this work, which cannot handle inhomogeneous environments. We also tested a leptonic version of scenario H3, but that yields a poor fit to the data.

\begin{figure}[!]
\centering
\includegraphics[width=0.9\columnwidth]{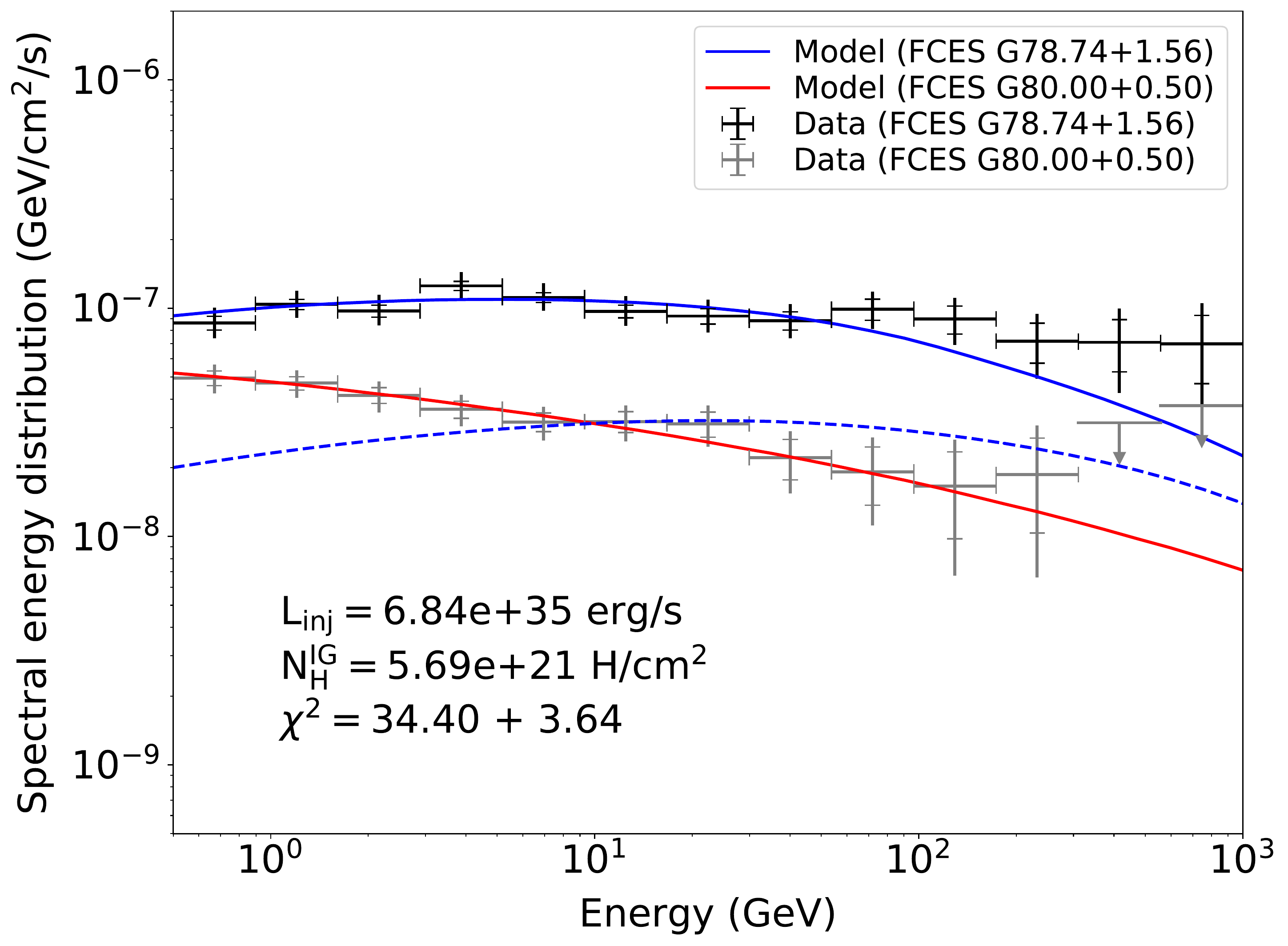} \\
\includegraphics[width=0.9\columnwidth]{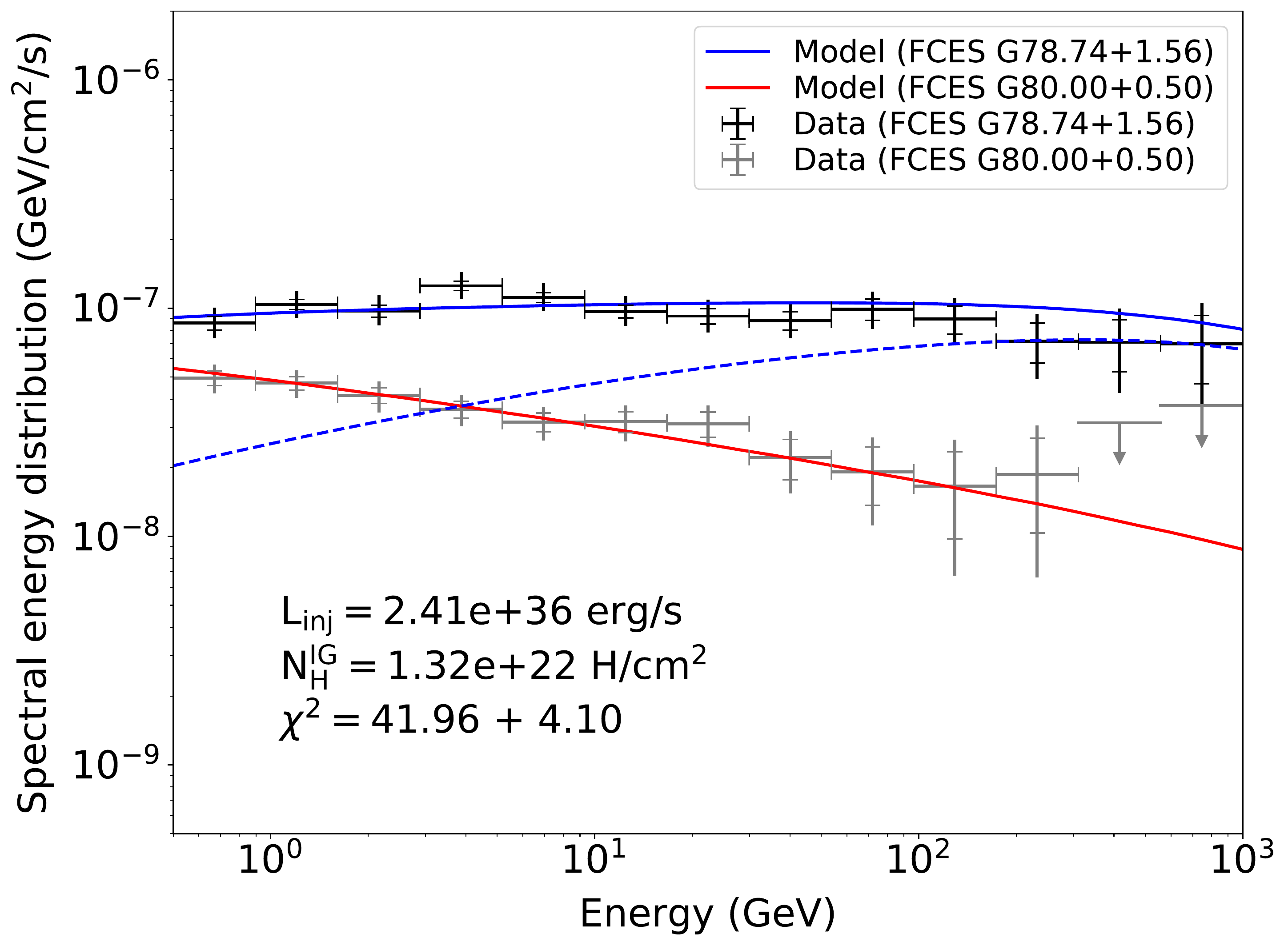}
\caption{Fitted model spectra for the \halo and \ig emission components, for model setup L1 (top) and L2 (bottom). A fit of the model to the spectrum of \halo is done first, with the injection luminosity as fitting parameter, followed by a second fit of the Bremsstrahlung emission only to the spectrum of \ig, with ionised gas column density as fitting parameter. This last fit does not include a contribution to the emission from a background population of CRs (see text). The dashed blue line is the inverse-Compton contribution in the emission model for \halo. The first $\chi^2$ corresponds to the fit to \halo and the second one to \ig.}
\label{fig:diffmodel:lepto:spec}
\end{figure}

The fits of the model to the spectra of \NNhalo and \NNig are displayed in Figure~\ref{fig:diffmodel:lepto:spec}. They are overall pretty satisfactory and yield $\chi^2$ similar to those obtained with the hadronic models, although slightly higher. The diffusion time range in leptonic scenarios seems rather constrained: small ages $\lesssim1$\,Myr tend to produce too hard spectra for the extended component, while ages $\gtrsim3$\,Myr result in too steep spectra. The injection luminosities resulting from these fits are $6.8 \times 10^{35}$ and $2.4 \times 10^{36}$\, erg~s$^{-1}$ for the L1 and L2 scenarios, respectively. This translates into electron injection efficiencies at the sub-percent level at most if the mechanical luminosity from Cyg OB2 and NGC 6910 is the power source for particle acceleration. We checked that the corresponding synchrotron emission does not exceed the radio constraints presented in \citet{mizuno2015}. 

\begin{figure}[!p]
\centering
\includegraphics[width=0.9\columnwidth]{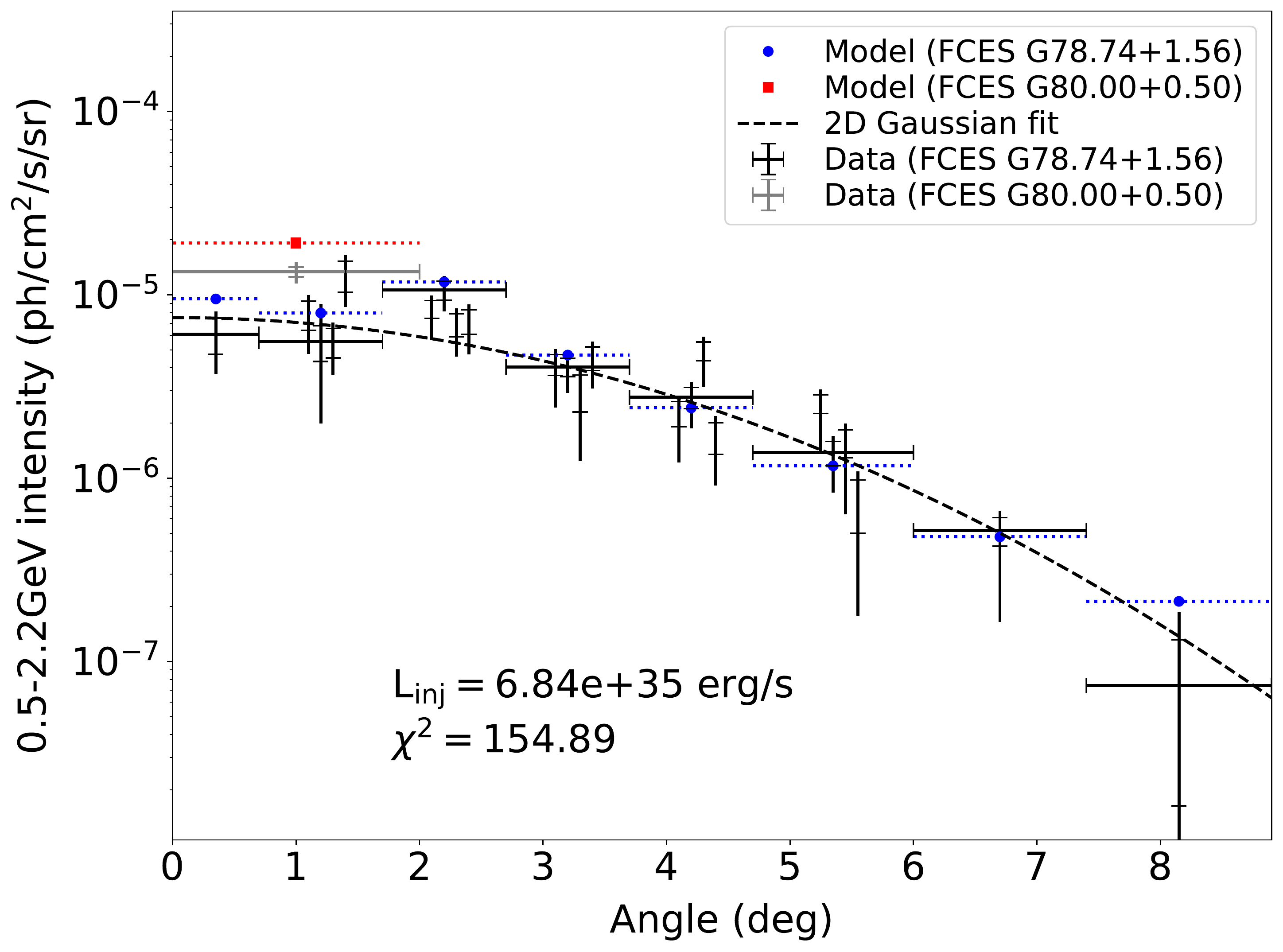} \\
\includegraphics[width=0.9\columnwidth]{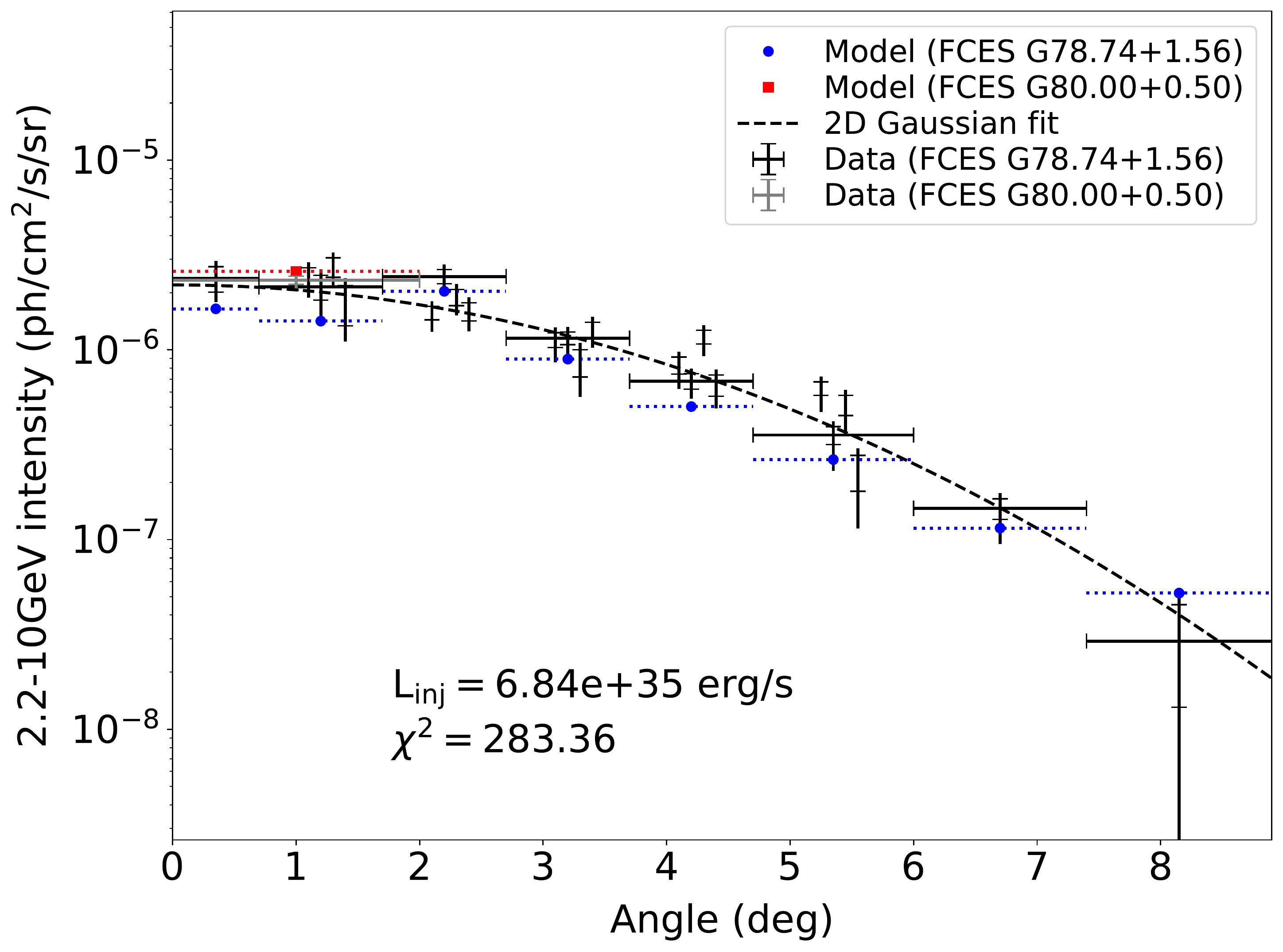} \\
\includegraphics[width=0.9\columnwidth]{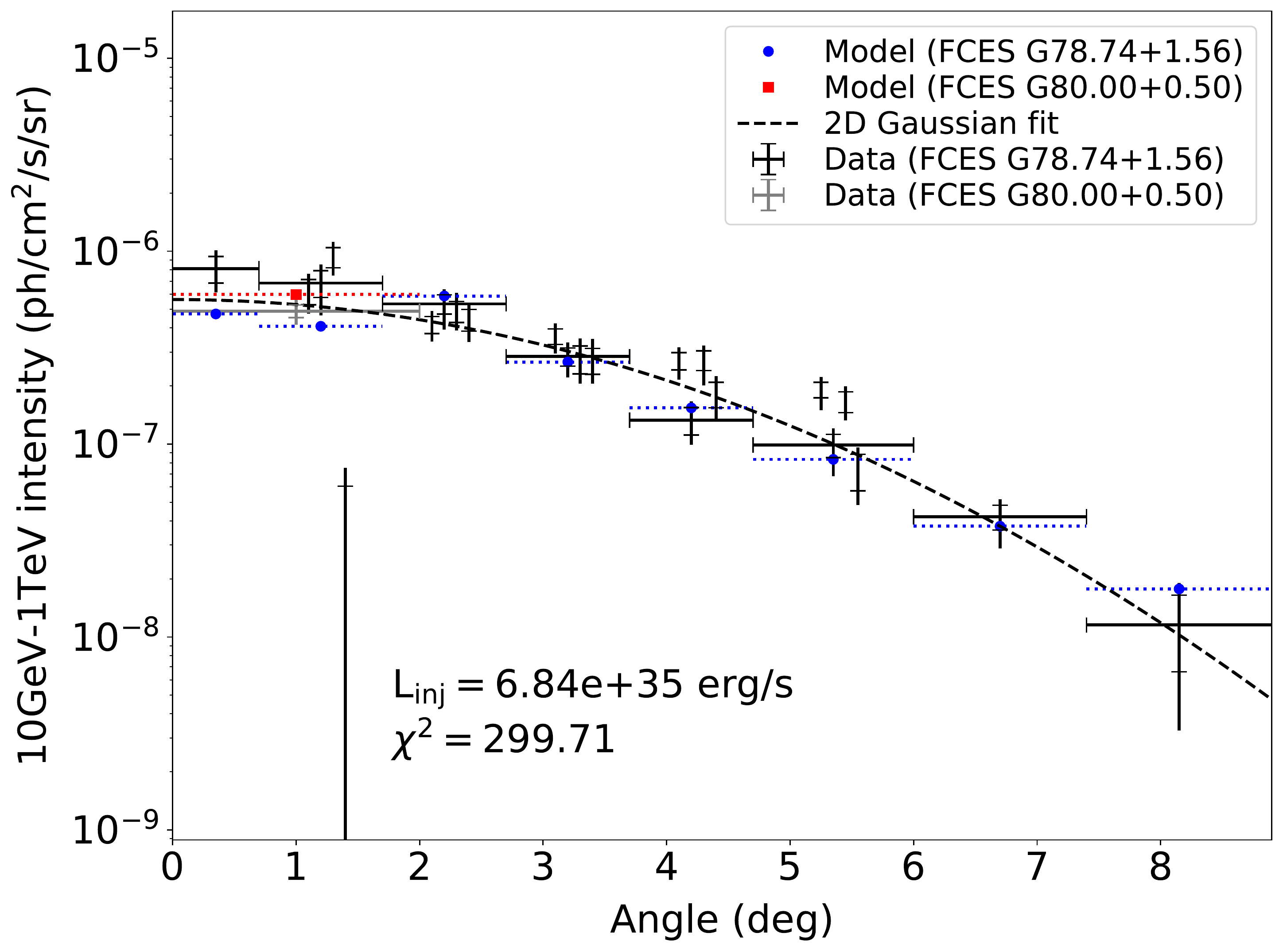}
\caption{Intensity radial profiles in three different gamma-ray energy bands for the \halo and \ig emission components, compared to predictions for model setup L1. The intensity distribution corresponding to the best-fit two-dimensional Gaussian model is displayed for comparison as a dotted line. The $\chi^2$ value correspond to the deviation from the 2D Gaussian fit.}
\label{fig:diffmodel:lepto:int:L1}
\end{figure}

The corresponding predicted intensity profiles are compared to the measurements in Figure~\ref{fig:diffmodel:lepto:int:L1} for model L1. As for the spectra, the fit is slightly degraded compared to that obtained with hadronic models. The best scenario is L1, with diffusion suppressed by two orders of magnitude with respect to the interstellar average and a diffusion time of 3\,Myr. A lower level of diffusion suppression results in too flat intensity profiles, undershooting the data in the inner regions and exceeding them at large distances from the injection point. As mentioned above, a smaller diffusion time does not help because it yields a too hard spectrum for \NNhalo.

We also tested the hypothesis that the observed emission actually is a pulsar halo, following the discovery of very extended gamma-ray emission around some middle-aged pulsars \citep{Abeysekara:2017b}. In such a scenario, PSR J2032$+$4127 appears as an interesting candidate because of its location close to the peak of the emission and characteristic age of $\sim$200~kyr. We used the phenomenological two-zone diffusion-loss halo model implementation presented in \citet{Martin:2022}, using as baseline key parameters: the spin-down power, estimated distance, and characteristic age of PSR J2032$+$4127 from the ATNF data base\footnote{\url{https://www.atnf.csiro.au/research/pulsar/psrcat/}}; a broken power law injection spectrum with indices 1.8 and 2.2 below and above a break energy of 500\,GeV respectively; an injection starting time of 40\,kyr; diffusion suppression by a factor of 50 within 50\,pc of the pulsar, with a power law dependence in rigidity with index $1/3$; a surrounding magnetic field with strength $B=3$\,$\upmu$G and the interstellar radiation field model from the original Cygnus cocoon paper \citep{ackermann2011}. We neglected the effect of proper motion on the emission morphology.

The fit of the predicted emission properties to the observed spectra and intensity profiles is relatively good (see Appendix~\ref{app:scenarios}), although not at the level of those obtained in scenarios H1-H4 and L1-L2. Yet, the implied present-day injection luminosity is of the order of $10^{36}$\, erg~s$^{-1}$, an order of magnitude larger than the spin-down power of PSR J2032$+$4127, which dismisses this pulsar as the possible source of the halo. This result seems to be robust against variations of the main model parameters. One cannot exclude, however, that another currently unknown pulsar with the right properties exists in this active star-forming region.


\subsection{The origin of the cocoon}


To summarise, our extended set of observables for \NNig and \NNhalo, including a radial profile for the extended component over nearly 10\degr, together with intensity measurements and emissivity estimates in three energy bands from 0.5\,GeV to 1\,TeV, can be accounted for reasonably well from a simple diffusion-loss model with a small number of free parameters. Several pretty different model setups seem to provide viable explanations of the observations, which suggests that more developed modelling frameworks and, more likely, additional observational data need to be considered in future studies.

An important result is that the data can be explained from one single population of injected particles. This population spans the full region of extended component \NNhalo, and gives rise to the central component \NNig by interacting with ionised gas in the innermost regions. Both hadronic and leptonic scenarios are viable, although it should be confirmed that leptonic scenarios are still valid in a more realistic modelling framework including non-uniform inverse-Compton losses in the strongly varying radiation fields of the region. All solutions have in common to require a flat particle spectrum at the source, with a power law index 2.0, which points to very recent acceleration. 

The solutions are, however, very different in terms of energetics and time scales involved. Setups H1 and L1 feature continuous injection, strong diffusion suppression in the region (by a factor 100 with respect to the large-scale interstellar average, over a spatial extent of more than 200\,pc), transport proceeding over several Myr (in agreement with age estimates for Cyg~OB2 and NGC~6910), and low acceleration efficiencies (at the sub-percent level in the hadronic scenario if Cyg~OB2 and NGC~6910 are the mechanical power source). Setups H2 and L2 are very similar with continuous injection, moderate diffusion suppression (by a factor 10 with respect to the large-scale interstellar average), a more recent injection and transport process over the last $0.3-1$\,Myr, and five-to-ten times higher acceleration efficiencies with respect to H1 and L1.

Setups H3 and H4 describe an even more recent event, with injection lasting 3 or 30\,kyr, transport proceeding over 10 or 100\,kyr in a medium where diffusion is not or only moderately suppressed, from a much more powerful source with properties that eventually seem relevant to an SN. Support for these scenarios would imply finding evidence of a middle-aged remnant in the region. The $\gamma$~Cygni SNR (G78.2$+$2.1), with its estimated age $\sim10$\,kyr, would be an interesting candidate source for scenario H3. Another option could be the remnant that resulted from the explosion giving birth to PSR J2032$+$4127, $\sim100-200$\,kyr ago. Its age is comparable to the diffusion time involved in scenario H4 and is high enough that the remnant has most likely gone undetectable by now. Interestingly, the typical injection time of 30\,kyr and diffusion suppression by a factor 10 in scenario H4 are reminiscent of the results obtained in \citet{Nava:2019} for the non-linear diffusion of $0.1-1$\,TeV CRs escaping from a supernova remnant in a hot ionised medium.

For comparable model setups, the difference in injection efficiency between leptonic and hadronic scenarios is of an order of magnitude at most. This means that mixed lepto-hadronic scenarios either imply a relatively high electron-to-proton ratio at injection, or a predominance of the hadronic contribution to the emission if the electron-to-proton ratio at injection is expected to have more classical values of $10^{-2}$ at most.

All the potential sources discussed above, Cyg~OB2, NGC~6910, the $\gamma$~Cygni SNR, and the unobserved SNR associated with PSR J2032$+$4127 are displaced with respect to the centroid of the emission (Figure~\ref{fig:morph_MWL}). This could suggest an inhomogenous transport scenario if one of these sources is indeed responsible for the origin of the particle population.

Possible improvements to the modelling presented here include the possibility of multiple and extended sources, for instance Cyg~OB2 and NGC~6910 releasing accelerated particles at their respective super-wind termination shocks, and a more complete transport scheme, including inhomogeneous diffusion and (or) energy losses over such a large volume, or the effect of advection. The latter point is particularly relevant in the case of high levels of diffusion suppression, as it may dominate the transport of the lowest-energy particles and alter their diffusion. The contribution from pre-existing CRs, for instance their leptonic emission from inverse-Compton scattering in the dense photon fields of the main clusters \citep{Orlando:2007} or their hadronic emission after reacceleration in the turbulent interior of the region \citep{tolksdorf2019}, certainly deserves a more sophisticated treatment than done here. Meanwhile, we can qualitatively compare our results to more sophisticated models of particle acceleration and transport at cluster wind termination shocks and in SBs from the literature.


\citet{morlino2021} presents a model of particle acceleration at the winds of star clusters that predicts a spectrum similar to the cocoon central component \NNig, which spans a region with size comparable to that of the wind termination shocks from Cyg~OB2 and NGC~6910 (see Figure~\ref{fig:morph_MWL}). The spatial distribution of low-energy particles in their model can be rather flat up to a few times the termination shock radius, which is comparable to the extended component \NNhalo. However, their model predicts that higher energy particles are more tightly confined around the shock, which is in contrast with our results of a harder spectrum for \NNhalo than \NNig. In addition, as already discussed above, there is no clear identification of a super-wind termination shock in the region, nor is it clear that there is a super-wind emanating from Cyg~OB2.

Alternatively, \citet{vieu2022} have shown that their model for particle acceleration and transport in SBs can reproduce the overall spectrum of the Cygnus cocoon measured by the LAT and HAWC in the case of efficient confinement in the bubble shell. In these conditions, they show that particle densities are rather uniform inside the SB, which might be in good agreement with the flat radial emissivity profile of the \NNhalo component. 
However, it is not obvious that their model can reproduce the morphological properties of the observed emission, especially its centrally peaked nature, if particles are efficiently trapped in an outer shell (the location of which remains unclear in Cygnus).

Recently, \citet{fornieri2022} presented a model of gamma-ray emission from Cygnus featuring two CR sources (Cyg~OB2 and the $\gamma$~Cygni SNR) and a description of particle transport that takes into account the detection of multiple plasma modes in the region \citep{plasmamodes}. As a consequence CR diffusion is predicted to be inhomogenous, which results in confinement for a long time in the central cavities where plasma modes are predominantly magnetosonic, and a more rapid diffusion in the nearby Alfv\'enic-dominated regions. However, their model does not take into account ionised gas in the central cavities, which seems required to explain the emissions observed from \NNig. Furthermore, they calculate a diffusion coefficient for physical parameters, such as gas density and temperature, that are not necessarily relevant for the entire gamma-ray emitting region. Overall, it is not clear if their model can explain the large extension of \NNhalo.

\subsection{\ob}\label{sec:ob}

The introduction of \ob (\NNob), significantly improves the likelihood of the model (Section ~\ref{sec:morphology}). However, the best-fit Gaussian model is only partially contained in our analysis region, and therefore the results may be subject to large uncertainties. A proper characterisation of this excess is left for future studies, and in this section we only provide general considerations on possibilities concerning its origin. There is no obvious correlation between \NNob and structures in the gas maps (Figure~\ref{fig:gasmaps} and \ref{fig:DNM}).

No SNRs are found overlapping with \NNob in SNRCat\footnote{\url{http://www.physics.umanitoba.ca/snr/SNRcat}} \citep{ferrand2012}. On the other hand, the ATNF 1.67 pulsar catalogue\footnote{\url{http://www.atnf.csiro.au/research/pulsar/psrcat/}} \citep{atnf2005} lists three pulsars within the $r_{68}$ area of \NNob with a spin-down power $> 10^{34}$~erg~s$^{-1}$. Among those PSR~J2111$+$4606, the closest to the source centroid at an offset of 3.3\degr, has a spin-down power of $1.4 \times 10^{36}$~erg~s$^{-1}$, a characteristic age of 17.5\,kyr, and an uncertain distance. It may be reminiscent of some middle-aged pulsars powering large gamma-ray sources such as HESS~J1825$-$137 \citep{hess1825,principe2020} or the pulsar halos around PSR~J0633+1746 or PSR~B0656+14 \citep{Abeysekara:2017b}. In the latter case, the offset of the pulsar from the centre of the emission can be explained by a combination of proper motion and time-dependent injection \citep{Zhang:2021}.

\NNob is bordered on the east by X-ray emission in the Cygnus SB \citep{cash1980}. The corresponding X-ray structure, dubbed as S-ARC~3 in \citet{uyaniker01} has been associated to the stellar association Cyg~OB4. However, the very existence of Cyg~OB4 is questioned based on parallax distances \citep{dezeeuw1999}. \citet{cantat-gaudin2020} reports ten stellar clusters with more than 100 members at a probability $>70\%$ within the $r_{68}$ area of \NNob. Among those, seven are located at distances from the Earth $< 1.8$~kpc, which would correspond to a physical size $< 200$~pc. Most of them are quite far away from the gamma-ray emission centroid, with the closest at 2.8\degr\ being NGC~7082 at 1.339~kpc from Earth and with an estimated age of 61~Myr, larger than typical stellar clusters with established detections in gamma rays \citep[for instance][]{tibaldo2021}.

The spectrum of \NNob has a particularly strong break with a steep slope at low energies and a hard spectrum above a few GeV that is strikingly similar to the one of \NNhalo. Establishing a physical connection between the two emission components is not obvious. \NNob may be produced by particles escaping the volume encompassed by \NNhalo, after an energy-dependent transport process that depleted the particle spectrum at the low-energy end. This is reminiscent of the illumination of neighbouring gas clouds by CRs escaping from a nearby source \citep[for instance][]{Tang2019}. Alternatively, steep slopes at low energies has been suggested as a signature of particles reacceleration in SBs \citep[for instance][]{tolksdorf2019}, and there may be radial gradients in such a mechanism.

\section{Summary and conclusions}\label{sec:conclusions}

We presented an analysis of the gamma-ray emission from the Cygnus region based on $\sim$13~years of \textit{Fermi}-LAT data. The extraction of the emission from the so-called Cygnus cocoon was performed from a dedicated modelling of interstellar emission from the region.
Compared to the analysis presented in \citet{ackermann2011}, we used almost seven times more data, produced with an improved reconstruction scheme corresponding to enhanced instrument performance. The data analysis is based on a much larger catalogue of gamma-ray sources and dedicated results for major sources in the fields. We also used improved gas tracer data and an iterative procedure to derive the dark neutral gas map.

As a result, the emission from the cocoon is now separated into two main components: first, a central component, \ig (\NNig), traced by a model for the distribution of ionised gas within the borders of photo-dissociation regions, and having a power law spectrum with index $2.19 \pm 0.03^{+0.00}_{-0.01}$; second, an extended component, \halo (\NNhalo), that can be modelled with a 2D Gaussian intensity distribution of extension $r_{68} = 4.4\degr \pm 0.1\degr\,^{+0.1\degr}_{-0.1\degr}$ and a smooth broken power law spectrum with spectral indices $1.67 \pm 0.05^{+0.02}_{-0.01}$ and $2.12 \pm 0.02^{+0.00}_{-0.01}$ below and above $3.0 \pm 0.6^{+0.0}_{-0.2}$~GeV, respectively. Emission from this component is significantly detected out to nearly 10\degr\ from the approximate centre of the star-forming region. Its total spectrum is significantly different from that of \NNig, and it exhibits significant spectral variations in azimuth in the innermost $\lesssim 3\degr$.

Two additional extended emission components were significantly detected during the analysis. Source \sw (\NNsw) overlaps with a bright arc of 8~$\upmu$m emission on the border of the central cavities in Cygnus X, and has a spectrum statistically compatible with \NNig. Although the spectral similarity and spatial proximity suggests a common origin, \NNsw does not show any obvious correlation with known gas structures. Another source, \ob (\NNob), is offset by several degrees with respect to Cygnus X and its spectrum is significantly different from all the other extended components studied, so a common origin seems unlikely. The centroid of \NNob lies on the edge of our analysis region, and therefore its current characterisation may be inaccurate. A proper study of this component is left for follow-up work.

The extended set of observables resulting from our analysis for the two brightest sources making up the cocoon, \NNig and \NNhalo, can be accounted for reasonably well from a simple diffusion-loss framework with a small number of free parameters, under several model setups. In all viable scenarios, one single population of non-thermal particles with a flat injection spectrum at the central source is sufficient and both hadronic and leptonic options are viable. Particles span the full extent of source \NNhalo as a result of diffusion, and give rise to source \NNig by interacting with ionised gas in the innermost regions. Possible solutions are very different in terms of energetics, transport conditions, and time scales involved. Some scenarios involve continuous injection during $\sim0.3-3$\,Myr, transport in a medium with moderately to strongly suppressed diffusion with respect to the large-scale interstellar average, and injection luminosities in the $10^{36}-10^{37}\mathrm{~erg~s^{-1}}$ range. They could describe a process by which the observed gamma-ray emission is powered by particle acceleration in the prominent star clusters Cyg~OB2 and NGC~6910. Alternatively, a hadronic solution exists involving a more recent event, with injection lasting $3-30$\,kyr and transport proceeding over 10\,kyr in a medium where diffusion is not or only moderately suppressed, and a much more powerful source with injection luminosity $\sim 10^{39}\mathrm{~erg~s^{-1}}$. Such a scenario seems more relevant to a single supernova explosion.

Possible improvements beyond this simple interpretation framework include accounting for multiple and extended sources, a more advanced description of particle transport in an inhomogeneous medium and including advection, and, last but not least, the description of physically motivated acceleration mechanisms. The observables extracted from our analysis are made available in machine-readable format and can be used in the future to perform detailed comparisons with more sophisticated models.

From the observational perspective, significant advances in gamma rays can be expected from instruments with improved sensitivity and angular resolution. The upcoming Cherenkov Telescope Array \citep{CTA-science} above a few tens of GeV will provide a sensitivity an order of magnitude better than previous ground-based instruments and an angular resolution reaching a few arcmin. Proposed space missions dedicated to the MeV to GeV domain \citep{astrogam,amego} may also improve a few times the angular resolution and by one or two orders of magnitude the sensitivity compared to \F, and provide observations in an energy range, the sub-MeV and MeV domain, poorly observed until now. Complementary advances in the characterisation of interstellar gas \citep[for instance][]{emig2022}, and of multi-wavelength and multi-messenger emission from the cocoon \citep[for instance][]{mizuno2015,yoast-hull2017} are also key to improving our understanding of particle acceleration and transport in this region.

\begin{acknowledgements}
The \textit{Fermi} LAT Collaboration acknowledges generous ongoing support
from a number of agencies and institutes that have supported both the
development and the operation of the LAT as well as scientific data analysis.
These include the National Aeronautics and Space Administration and the
Department of Energy in the United States, the Commissariat \`a l'Energie Atomique
and the Centre National de la Recherche Scientifique / Institut National de Physique
Nucl\'eaire et de Physique des Particules in France, the Agenzia Spaziale Italiana
and the Istituto Nazionale di Fisica Nucleare in Italy, the Ministry of Education,
Culture, Sports, Science and Technology (MEXT), High Energy Accelerator Research
Organization (KEK) and Japan Aerospace Exploration Agency (JAXA) in Japan, and
the K.~A.~Wallenberg Foundation, the Swedish Research Council and the
Swedish National Space Board in Sweden.
 
Additional support for science analysis during the operations phase is gratefully
acknowledged from the Istituto Nazionale di Astrofisica in Italy and the Centre
National d'\'Etudes Spatiales in France. This work performed in part under DOE
Contract DE-AC02-76SF00515.\\

This work was supported by the "Agence Nationale de la Recherche" through grant ANR-19-CE31-0014 (GAMALO project, PI: P. Martin).\\

This work makes use of NumPy \citep{harris2020}, AstroPy \citep{Astropy2022}, Matplotlib \citep{Hunter2007}, SciPy \citep{Scipy2020}, and the colourmaps in the CMasher package \citep{VdVelden2020}. \\

The authors would like to thank E. Orlando and M. Pesce-Rollins for their helpful comments on the manuscript, as well as I. A. Grenier for insightful conversations about the project.
\end{acknowledgements}

\noindent \textit{Note added after acceptance.} Our calculation of stellar cluster properties in Sect. 4.1 considered stars reaching 60~$M_\odot$, which corresponds to type O3. It was pointed out to us by G. Morlino that the most massive star observed in NGC~6910 is of type O9.6V \citep{kaur2020}, thus with a mass $< 20 M_\odot$, and that, based on the relationship between cluster mass and stellar mass upper limit from \citep{wk2004}, the most massive stars in that cluster are expected to reach $\sim$25~$M_\odot$. Such an upper limit on the stellar mass would yield a reduction of more than an order of magnitude in the cluster mechanical power. In this case, NGC~6910 could not be the only source of accelerated particles in scenario H2, as it would require $\gtrsim 100\%$ efficiency in the conversion of mechanical energy into particle energy, but remains a viable candidate source in scenario H1, with a $\sim$10\% conversion efficiency. For the more massive OB2 association the impact of the stellar mass upper limit from  \citep{wk2004} on the estimated cluster luminosity is only of a factor of a few and does not alter significantly our conclusions.

\bibliographystyle{aa} 
\bibliography{biblio}

\begin{thebibliography}{96}
\expandafter\ifx\csname natexlab\endcsname\relax\def\natexlab#1{#1}\fi

\bibitem[{{Abeysekara} {et~al.}(2017){Abeysekara}, {Albert}, {Alfaro},
  {Alvarez}, {{\'A}lvarez}, {Arceo}, {Arteaga-Vel{\'a}zquez}, {Avila Rojas},
  {Ayala Solares}, {Barber}, {Bautista-Elivar}, {Becerril}, {Belmont-Moreno},
  {BenZvi}, {Berley}, {Bernal}, {Braun}, {Brisbois}, {Caballero-Mora},
  {Capistr{\'a}n}, {Carrami{\~n}ana}, {Casanova}, {Castillo}, {Cotti},
  {Cotzomi}, {Couti{\~n}o de Le{\'o}n}, {De Le{\'o}n}, {De la Fuente},
  {Dingus}, {DuVernois}, {D{\'\i}az-V{\'e}lez}, {Ellsworth}, {Engel},
  {Enr{\'\i}quez-Rivera}, {Fiorino}, {Fraija}, {Garc{\'\i}a-Gonz{\'a}lez},
  {Garfias}, {Gerhardt}, {Gonz{\'a}lez Mu{\~n}oz}, {Gonz{\'a}lez}, {Goodman},
  {Hampel-Arias}, {Harding}, {Hern{\'a}ndez}, {Hern{\'a}ndez-Almada}, {Hinton},
  {Hona}, {Hui}, {H{\"u}ntemeyer}, {Iriarte}, {Jardin-Blicq}, {Joshi},
  {Kaufmann}, {Kieda}, {Lara}, {Lauer}, {Lee}, {Lennarz}, {Vargas},
  {Linnemann}, {Longinotti}, {Luis Raya}, {Luna-Garc{\'\i}a}, {L{\'o}pez-Coto},
  {Malone}, {Marinelli}, {Martinez}, {Martinez-Castellanos},
  {Mart{\'\i}nez-Castro}, {Mart{\'\i}nez-Huerta}, {Matthews}, {Mirand
  a-Romagnoli}, {Moreno}, {Mostaf{\'a}}, {Nellen}, {Newbold}, {Nisa},
  {Noriega-Papaqui}, {Pelayo}, {Pretz}, {P{\'e}rez-P{\'e}rez}, {Ren}, {Rho},
  {Rivi{\`e}re}, {Rosa-Gonz{\'a}lez}, {Rosenberg}, {Ruiz-Velasco}, {Salazar},
  {Salesa Greus}, {Sand oval}, {Schneider}, {Schoorlemmer}, {Sinnis}, {Smith},
  {Springer}, {Surajbali}, {Taboada}, {Tibolla}, {Tollefson}, {Torres},
  {Ukwatta}, {Vianello}, {Weisgarber}, {Westerhoff}, {Wisher}, {Wood},
  {Yapici}, {Yodh}, {Younk}, {Zepeda}, {Zhou}, {Guo}, {Hahn}, {Li}, \&
  {Zhang}}]{Abeysekara:2017b}
{Abeysekara}, A.~U., {Albert}, A., {Alfaro}, R., {et~al.} 2017, Science, 358,
  911

\bibitem[{{Abeysekara} {et~al.}(2021){Abeysekara}, {Albert}, {Alfaro},
  {Alvarez}, {Camacho}, {Arteaga-Vel{\'a}zquez}, {Arunbabu}, {Rojas},
  {Solares}, {Baghmanyan}, {Belmont-Moreno}, {BenZvi}, {Blandford}, {Brisbois},
  {Caballero-Mora}, {Capistr{\'a}n}, {Carrami{\~n}ana}, {Casanova}, {Cotti},
  {Le{\'o}n}, {De la Fuente}, {Hernandez}, {Dingus}, {DuVernois}, {Durocher},
  {D{\'\i}az-V{\'e}lez}, {Ellsworth}, {Engel}, {Espinoza}, {Fan}, {Fang},
  {Fleischhack}, {Fraija}, {Galv{\'a}n-G{\'a}mez}, {Garcia},
  {Garc{\'\i}a-Gonz{\'a}lez}, {Garfias}, {Giacinti}, {Gonz{\'a}lez}, {Goodman},
  {Harding}, {Hernandez}, {Hinton}, {Hona}, {Huang}, {Hueyotl-Zahuantitla},
  {H{\"u}ntemeyer}, {Iriarte}, {Jardin-Blicq}, {Joshi}, {Kieda}, {Lara}, {Lee},
  {Vargas}, {Linnemann}, {Longinotti}, {Luis-Raya}, {Lundeen}, {Malone},
  {Martinez}, {Martinez-Castellanos}, {Mart{\'\i}nez-Castro}, {Matthews},
  {Miranda-Romagnoli}, {Morales-Soto}, {Moreno}, {Mostaf{\'a}}, {Nayerhoda},
  {Nellen}, {Newbold}, {Nisa}, {Noriega-Papaqui}, {Olivera-Nieto}, {Omodei},
  {Peisker}, {P{\'e}rez Araujo}, {P{\'e}rez-P{\'e}rez}, {Ren}, {Rho},
  {Rosa-Gonz{\'a}lez}, {Ruiz-Velasco}, {Salazar}, {Greus}, {Sandoval},
  {Schneider}, {Schoorlemmer}, {Serna}, {Smith}, {Springer}, {Surajbali},
  {Tollefson}, {Torres}, {Torres-Escobedo}, {Ure{\~n}a-Mena}, {Weisgarber},
  {Werner}, {Willox}, {Zepeda}, {Zhou}, {De Le{\'o}n}, \&
  {{\'A}lvarez}}]{cocoon_hawc}
{Abeysekara}, A.~U., {Albert}, A., {Alfaro}, R., {et~al.} 2021, Nature
  Astronomy, 5, 465

\bibitem[{{Abramowski} {et~al.}(2015){Abramowski}, {Aharonian}, {Ait Benkhali},
  {Akhperjanian}, {Ang{\"u}ner}, {Backes}, {Balenderan}, {Balzer}, {Barnacka},
  {Becherini}, {Becker-Tjus}, {Berge}, {Bernhard}, {Bernl{\"o}hr}, {Birsin},
  {Biteau}, {B{\"o}ttcher}, {Boisson}, {Bolmont}, {Bordas}, {Bregeon}, {Brun},
  {Brun}, {Bryan}, {Bulik}, {Carrigan}, {Casanova}, {Chadwick}, {Chakraborty},
  {Chalme-Calvet}, {Chaves}, {Chr{\'e}tien}, {Colafrancesco}, {Cologna},
  {Conrad}, {Couturier}, {Cui}, {Dalton}, {Davids}, {Degrange}, {Deil}, {de
  Wilt}, {Djannati-Ata{\"\i}}, {Domainko}, {Donath}, {Drury}, {Dubus},
  {Dutson}, {Dyks}, {Dyrda}, {Edwards}, {Egberts}, {Eger}, {Espigat},
  {Farnier}, {Fegan}, {Feinstein}, {Fernandes}, {Fernandez}, {Fiasson},
  {Fontaine}, {F{\"o}rster}, {F{\"u}{\ss}ling}, {Gabici}, {Gajdus}, {Gallant},
  {Garrigoux}, {Giavitto}, {Giebels}, {Glicenstein}, {Gottschall}, {Grondin},
  {Grudzi{\'n}ska}, {Hadasch}, {H{\"a}ffner}, {Hahn}, {Harris}, {Heinzelmann},
  {Henri}, {Hermann}, {Hervet}, {Hillert}, {Hinton}, {Hofmann}, {Hofverberg},
  {Holler}, {Horns}, {Ivascenko}, {Jacholkowska}, {Jahn}, {Jamrozy}, {Janiak},
  {Jankowsky}, {Jung}, {Kastendieck}, {Katarzy{\'n}ski}, {Katz}, {Kaufmann},
  {Kh{\'e}lifi}, {Kieffer}, {Klepser}, {Klochkov}, {Klu{\'z}niak}, {Kolitzus},
  {Komin}, {Kosack}, {Krakau}, {Krayzel}, {Kr{\"u}ger}, {Laffon}, {Lamanna},
  {Lefaucheur}, {Lefranc}, {Lemi{\`e}re}, {Lemoine-Goumard}, {Lenain}, {Lohse},
  {Lopatin}, {Lu}, {Marandon}, {Marcowith}, {Marx}, {Maurin}, {Maxted},
  {Mayer}, {McComb}, {M{\'e}hault}, {Meintjes}, {Menzler}, {Meyer}, {Mitchell},
  {Moderski}, {Mohamed}, {Mor{\r{a}}}, {Moulin}, {Murach}, {de Naurois},
  {Niemiec}, {Nolan}, {Oakes}, {Odaka}, {Ohm}, {Opitz}, {Ostrowski}, {Oya},
  {Panter}, {Parsons}, {Paz Arribas}, {Pekeur}, {Pelletier}, {Perez},
  {Petrucci}, {Peyaud}, {Pita}, {Poon}, {P{\"u}hlhofer}, {Punch},
  {Quirrenbach}, {Raab}, {Reichardt}, {Reimer}, {Reimer}, {Renaud}, {de los
  Reyes}, {Rieger}, {Rob}, {Romoli}, {Rosier-Lees}, {Rowell}, {Rudak},
  {Rulten}, {Sahakian}, {Salek}, {Sanchez}, {Santangelo}, {Schlickeiser},
  {Sch{\"u}ssler}, {Schulz}, {Schwanke}, {Schwarzburg}, {Schwemmer}, {Sol},
  {Spanier}, {Spengler}, {Spies}, {Stawarz}, {Steenkamp}, {Stegmann},
  {Stinzing}, {Stycz}, {Sushch}, {Tavernet}, {Tavernier}, {Taylor}, {Terrier},
  {Tluczykont}, {Trichard}, {Valerius}, {van Eldik}, {van Soelen},
  {Vasileiadis}, {Veh}, {Venter}, {Viana}, {Vincent}, {Vink}, {V{\"o}lk},
  {Volpe}, {Vorster}, {Vuillaume}, {Wagner}, {Wagner}, {Wagner}, {Ward},
  {Weidinger}, {Weitzel}, {White}, {Wierzcholska}, {Willmann}, {W{\"o}rnlein},
  {Wouters}, {Yang}, {Zabalza}, {Zaborov}, {Zacharias}, {Zdziarski}, {Zech}, \&
  {Zechlin}}]{Abramowski:2015}
{Abramowski}, A., {Aharonian}, F., {Ait Benkhali}, F., {et~al.} 2015, Science,
  347, 406

\bibitem[{{Acero} {et~al.}(2016){Acero}, {Ackermann}, {Ajello}, {Albert},
  {Baldini}, {Ballet}, {Barbiellini}, {Bastieri}, {Bellazzini}, {Bissaldi},
  {Bloom}, {Bonino}, {Bottacini}, {Brandt}, {Bregeon}, {Bruel}, {Buehler},
  {Buson}, {Caliandro}, {Cameron}, {Caragiulo}, {Caraveo}, {Casandjian},
  {Cavazzuti}, {Cecchi}, {Charles}, {Chekhtman}, {Chiang}, {Chiaro}, {Ciprini},
  {Claus}, {Cohen-Tanugi}, {Conrad}, {Cuoco}, {Cutini}, {D'Ammando}, {de
  Angelis}, {de Palma}, {Desiante}, {Digel}, {Di Venere}, {Drell}, {Favuzzi},
  {Fegan}, {Ferrara}, {Focke}, {Franckowiak}, {Funk}, {Fusco}, {Gargano},
  {Gasparrini}, {Giglietto}, {Giordano}, {Giroletti}, {Glanzman}, {Godfrey},
  {Grenier}, {Guiriec}, {Hadasch}, {Harding}, {Hayashi}, {Hays}, {Hewitt},
  {Hill}, {Horan}, {Hou}, {Jogler}, {J{\'o}hannesson}, {Kamae}, {Kuss},
  {Landriu}, {Larsson}, {Latronico}, {Li}, {Li}, {Longo}, {Loparco},
  {Lovellette}, {Lubrano}, {Maldera}, {Malyshev}, {Manfreda}, {Martin},
  {Mayer}, {Mazziotta}, {McEnery}, {Michelson}, {Mirabal}, {Mizuno}, {Monzani},
  {Morselli}, {Nuss}, {Ohsugi}, {Omodei}, {Orienti}, {Orlando}, {Ormes},
  {Paneque}, {Pesce-Rollins}, {Piron}, {Pivato}, {Rain{\`o}}, {Rando},
  {Razzano}, {Razzaque}, {Reimer}, {Reimer}, {Remy}, {Renault},
  {S{\'a}nchez-Conde}, {Schaal}, {Schulz}, {Sgr{\`o}}, {Siskind}, {Spada},
  {Spandre}, {Spinelli}, {Strong}, {Suson}, {Tajima}, {Takahashi}, {Thayer},
  {Thompson}, {Tibaldo}, {Tinivella}, {Torres}, {Tosti}, {Troja}, {Vianello},
  {Werner}, {Wood}, {Wood}, {Zaharijas}, \& {Zimmer}}]{acero2016}
{Acero}, F., {Ackermann}, M., {Ajello}, M., {et~al.} 2016, \apjs, 223, 26

\bibitem[{{Ackermann} {et~al.}(2012{\natexlab{a}}){Ackermann}, {Ajello},
  {Allafort}, {Baldini}, {Ballet}, {Barbiellini}, {Bastieri}, {Belfiore},
  {Bellazzini}, {Berenji}, {Blandford}, {Bloom}, {Bonamente}, {Borgland},
  {Bottacini}, {Bregeon}, {Brigida}, {Bruel}, {Buehler}, {Buson}, {Caliandro},
  {Cameron}, {Caraveo}, {Casandjian}, {Cecchi}, {Chekhtman}, {Ciprini},
  {Claus}, {Cohen-Tanugi}, {de Angelis}, {de Palma}, {Dermer}, {Silva},
  {Drell}, {Dumora}, {Favuzzi}, {Fegan}, {Focke}, {Fortin}, {Fukazawa},
  {Fusco}, {Gargano}, {Germani}, {Giglietto}, {Giordano}, {Giroletti},
  {Glanzman}, {Godfrey}, {Grenier}, {Guillemot}, {Guiriec}, {Hadasch},
  {Hanabata}, {Harding}, {Hayashida}, {Hayashi}, {Hays}, {J{\'o}hannesson},
  {Johnson}, {Kamae}, {Katagiri}, {Kataoka}, {Kerr}, {Kn{\"o}dlseder}, {Kuss},
  {Lande}, {Latronico}, {Lee}, {Longo}, {Loparco}, {Lott}, {Lovellette},
  {Lubrano}, {Martin}, {Mazziotta}, {McEnery}, {Mehault}, {Michelson},
  {Mitthumsiri}, {Mizuno}, {Monte}, {Monzani}, {Morselli}, {Moskalenko},
  {Murgia}, {Naumann-Godo}, {Nolan}, {Norris}, {Nuss}, {Ohsugi}, {Okumura},
  {Omodei}, {Orlando}, {Ormes}, {Ozaki}, {Paneque}, {Parent}, {Pesce-Rollins},
  {Pierbattista}, {Piron}, {Porter}, {Rain{\`o}}, {Rando}, {Razzano}, {Reimer},
  {Reposeur}, {Ritz}, {Saz Parkinson}, {Sgr{\`o}}, {Siskind}, {Smith},
  {Spinelli}, {Strong}, {Takahashi}, {Tanaka}, {Thayer}, {Thayer}, {Thompson},
  {Tibaldo}, {Torres}, {Tosti}, {Tramacere}, {Troja}, {Uchiyama},
  {Vandenbroucke}, {Vasileiou}, {Vianello}, {Vitale}, {Waite}, {Wang}, {Winer},
  {Wood}, {Yang}, {Zimmer}, \& {Bontemps}}]{Ackermann2012}
{Ackermann}, M., {Ajello}, M., {Allafort}, A., {et~al.} 2012{\natexlab{a}},
  Astronomy and Astrophysics, 538, A71

\bibitem[{{Ackermann} {et~al.}(2011){Ackermann}, {Ajello}, {Allafort},
  {Baldini}, {Ballet}, {Barbiellini}, {Bastieri}, {Belfiore}, {Bellazzini},
  {Berenji}, {Blandford}, {Bloom}, {Bonamente}, {Borgland}, {Bottacini},
  {Brigida}, {Bruel}, {Buehler}, {Buson}, {Caliandro}, {Cameron}, {Caraveo},
  {Casandjian}, {Cecchi}, {Chekhtman}, {Cheung}, {Chiang}, {Ciprini}, {Claus},
  {Cohen-Tanugi}, {de Angelis}, {de Palma}, {Dermer}, {do Couto e Silva},
  {Drell}, {Dumora}, {Favuzzi}, {Fegan}, {Focke}, {Fortin}, {Fukazawa},
  {Fusco}, {Gargano}, {Germani}, {Giglietto}, {Giordano}, {Giroletti},
  {Glanzman}, {Godfrey}, {Grenier}, {Guillemot}, {Guiriec}, {Hadasch},
  {Hanabata}, {Harding}, {Hayashida}, {Hayashi}, {Hays}, {J{\'o}hannesson},
  {Johnson}, {Kamae}, {Katagiri}, {Kataoka}, {Kerr}, {Kn{\"o}dlseder}, {Kuss},
  {Lande}, {Latronico}, {Lee}, {Longo}, {Loparco}, {Lott}, {Lovellette},
  {Lubrano}, {Martin}, {Mazziotta}, {McEnery}, {Mehault}, {Michelson},
  {Mitthumsiri}, {Mizuno}, {Monte}, {Monzani}, {Morselli}, {Moskalenko},
  {Murgia}, {Naumann-Godo}, {Nolan}, {Norris}, {Nuss}, {Ohsugi}, {Okumura},
  {Orlando}, {Ormes}, {Ozaki}, {Paneque}, {Parent}, {Pesce-Rollins},
  {Pierbattista}, {Piron}, {Pohl}, {Prokhorov}, {Rain{\`o}}, {Rando},
  {Razzano}, {Reposeur}, {Ritz}, {Parkinson}, {Sgr{\`o}}, {Siskind}, {Smith},
  {Spinelli}, {Strong}, {Takahashi}, {Tanaka}, {Thayer}, {Thayer}, {Thompson},
  {Tibaldo}, {Torres}, {Tosti}, {Tramacere}, {Troja}, {Uchiyama},
  {Vandenbroucke}, {Vasileiou}, {Vianello}, {Vitale}, {Waite}, {Wang}, {Winer},
  {Wood}, {Yang}, {Zimmer}, \& {Bontemps}}]{ackermann2011}
{Ackermann}, M., {Ajello}, M., {Allafort}, A., {et~al.} 2011, Science, 334,
  1103

\bibitem[{{Ackermann} {et~al.}(2012{\natexlab{b}}){Ackermann}, {Ajello},
  {Atwood}, {Baldini}, {Ballet}, {Barbiellini}, {Bastieri}, {Bechtol},
  {Bellazzini}, {Berenji}, {Blandford}, {Bloom}, {Bonamente}, {Borgland},
  {Brandt}, {Bregeon}, {Brigida}, {Bruel}, {Buehler}, {Buson}, {Caliandro},
  {Cameron}, {Caraveo}, {Cavazzuti}, {Cecchi}, {Charles}, {Chekhtman},
  {Chiang}, {Ciprini}, {Claus}, {Cohen-Tanugi}, {Conrad}, {Cutini}, {de
  Angelis}, {de Palma}, {Dermer}, {Digel}, {Silva}, {Drell}, {Drlica-Wagner},
  {Falletti}, {Favuzzi}, {Fegan}, {Ferrara}, {Focke}, {Fortin}, {Fukazawa},
  {Funk}, {Fusco}, {Gaggero}, {Gargano}, {Germani}, {Giglietto}, {Giordano},
  {Giroletti}, {Glanzman}, {Godfrey}, {Grove}, {Guiriec}, {Gustafsson},
  {Hadasch}, {Hanabata}, {Harding}, {Hayashida}, {Hays}, {Horan}, {Hou},
  {Hughes}, {J{\'o}hannesson}, {Johnson}, {Johnson}, {Kamae}, {Katagiri},
  {Kataoka}, {Kn{\"o}dlseder}, {Kuss}, {Lande}, {Latronico}, {Lee},
  {Lemoine-Goumard}, {Longo}, {Loparco}, {Lott}, {Lovellette}, {Lubrano},
  {Mazziotta}, {McEnery}, {Michelson}, {Mitthumsiri}, {Mizuno}, {Monte},
  {Monzani}, {Morselli}, {Moskalenko}, {Murgia}, {Naumann-Godo}, {Norris},
  {Nuss}, {Ohsugi}, {Okumura}, {Omodei}, {Orlando}, {Ormes}, {Paneque},
  {Panetta}, {Parent}, {Pesce-Rollins}, {Pierbattista}, {Piron}, {Pivato},
  {Porter}, {Rain{\`o}}, {Rando}, {Razzano}, {Razzaque}, {Reimer}, {Reimer},
  {Sadrozinski}, {Sgr{\`o}}, {Siskind}, {Spandre}, {Spinelli}, {Strong},
  {Suson}, {Takahashi}, {Tanaka}, {Thayer}, {Thayer}, {Thompson}, {Tibaldo},
  {Tinivella}, {Torres}, {Tosti}, {Troja}, {Usher}, {Vandenbroucke},
  {Vasileiou}, {Vianello}, {Vitale}, {Waite}, {Wang}, {Winer}, {Wood}, {Wood},
  {Yang}, {Ziegler}, \& {Zimmer}}]{IC_model}
{Ackermann}, M., {Ajello}, M., {Atwood}, W.~B., {et~al.} 2012{\natexlab{b}},
  Astrophysical Journal, 750, 3

\bibitem[{{Aharonian} {et~al.}(2019){Aharonian}, {Yang}, \& {de O{\~n}a
  Wilhelmi}}]{aharonian2019}
{Aharonian}, F., {Yang}, R., \& {de O{\~n}a Wilhelmi}, E. 2019, Nature
  Astronomy, 3, 561

\bibitem[{{Astropy Collaboration} {et~al.}(2022){Astropy Collaboration},
  {Price-Whelan}, {Lim}, {Earl}, {Starkman}, {Bradley}, {Shupe}, {Patil},
  {Corrales}, {Brasseur}, {N{\"o}the}, {Donath}, {Tollerud}, {Morris},
  {Ginsburg}, {Vaher}, {Weaver}, {Tocknell}, {Jamieson}, {van Kerkwijk},
  {Robitaille}, {Merry}, {Bachetti}, {G{\"u}nther}, {Aldcroft},
  {Alvarado-Montes}, {Archibald}, {B{\'o}di}, {Bapat}, {Barentsen},
  {Baz{\'a}n}, {Biswas}, {Boquien}, {Burke}, {Cara}, {Cara}, {Conroy},
  {Conseil}, {Craig}, {Cross}, {Cruz}, {D'Eugenio}, {Dencheva}, {Devillepoix},
  {Dietrich}, {Eigenbrot}, {Erben}, {Ferreira}, {Foreman-Mackey}, {Fox},
  {Freij}, {Garg}, {Geda}, {Glattly}, {Gondhalekar}, {Gordon}, {Grant},
  {Greenfield}, {Groener}, {Guest}, {Gurovich}, {Handberg}, {Hart},
  {Hatfield-Dodds}, {Homeier}, {Hosseinzadeh}, {Jenness}, {Jones}, {Joseph},
  {Kalmbach}, {Karamehmetoglu}, {Ka{\l}uszy{\'n}ski}, {Kelley}, {Kern},
  {Kerzendorf}, {Koch}, {Kulumani}, {Lee}, {Ly}, {Ma}, {MacBride}, {Maljaars},
  {Muna}, {Murphy}, {Norman}, {O'Steen}, {Oman}, {Pacifici}, {Pascual},
  {Pascual-Granado}, {Patil}, {Perren}, {Pickering}, {Rastogi}, {Roulston},
  {Ryan}, {Rykoff}, {Sabater}, {Sakurikar}, {Salgado}, {Sanghi}, {Saunders},
  {Savchenko}, {Schwardt}, {Seifert-Eckert}, {Shih}, {Jain}, {Shukla}, {Sick},
  {Simpson}, {Singanamalla}, {Singer}, {Singhal}, {Sinha}, {Sip{\H{o}}cz},
  {Spitler}, {Stansby}, {Streicher}, {{\v{S}}umak}, {Swinbank}, {Taranu},
  {Tewary}, {Tremblay}, {Val-Borro}, {Van Kooten}, {Vasovi{\'c}}, {Verma}, {de
  Miranda Cardoso}, {Williams}, {Wilson}, {Winkel}, {Wood-Vasey}, {Xue},
  {Yoachim}, {Zhang}, {Zonca}, \& {Astropy Project Contributors}}]{Astropy2022}
{Astropy Collaboration}, {Price-Whelan}, A.~M., {Lim}, P.~L., {et~al.} 2022,
  \apj, 935, 167

\bibitem[{{Atoyan} {et~al.}(1995){Atoyan}, {Aharonian}, \&
  {V{\"o}lk}}]{Atoyan:1995}
{Atoyan}, A.~M., {Aharonian}, F.~A., \& {V{\"o}lk}, H.~J. 1995, \prd, 52, 3265

\bibitem[{{Atwood} {et~al.}(2013){Atwood}, {Albert}, {Baldini}, {Tinivella},
  {Bregeon}, {Pesce-Rollins}, {Sgr{\`o}}, {Bruel}, {Charles}, {Drlica-Wagner},
  {Franckowiak}, {Jogler}, {Rochester}, {Usher}, {Wood}, {Cohen-Tanugi}, \&
  {Zimmer}}]{Atwood2013}
{Atwood}, W., {Albert}, A., {Baldini}, L., {et~al.} 2013, in Fourth Fermi
  Symposium Proceedings, ed. T.~J. {Brandt}, N.~{Omodei}, \& C.~{Wilson-Hodge}
  (eConf C121028), {8--13}

\bibitem[{{Atwood} {et~al.}(2009){Atwood}, {Abdo}, {Ackermann}, {Althouse},
  {Anderson}, {Axelsson}, {Baldini}, {Ballet}, {Band}, {Barbiellini},
  {Bartelt}, {Bastieri}, {Baughman}, {Bechtol}, {B{\'e}d{\'e}r{\`e}de},
  {Bellardi}, {Bellazzini}, {Berenji}, {Bignami}, {Bisello}, {Bissaldi},
  {Blandford}, {Bloom}, {Bogart}, {Bonamente}, {Bonnell}, {Borgland},
  {Bouvier}, {Bregeon}, {Brez}, {Brigida}, {Bruel}, {Burnett}, {Busetto},
  {Caliandro}, {Cameron}, {Caraveo}, {Carius}, {Carlson}, {Casandjian},
  {Cavazzuti}, {Ceccanti}, {Cecchi}, {Charles}, {Chekhtman}, {Cheung},
  {Chiang}, {Chipaux}, {Cillis}, {Ciprini}, {Claus}, {Cohen-Tanugi},
  {Condamoor}, {Conrad}, {Corbet}, {Corucci}, {Costamante}, {Cutini}, {Davis},
  {Decotigny}, {DeKlotz}, {Dermer}, {de Angelis}, {Digel}, {do Couto e Silva},
  {Drell}, {Dubois}, {Dumora}, {Edmonds}, {Fabiani}, {Farnier}, {Favuzzi},
  {Flath}, {Fleury}, {Focke}, {Funk}, {Fusco}, {Gargano}, {Gasparrini},
  {Gehrels}, {Gentit}, {Germani}, {Giebels}, {Giglietto}, {Giommi}, {Giordano},
  {Glanzman}, {Godfrey}, {Grenier}, {Grondin}, {Grove}, {Guillemot}, {Guiriec},
  {Haller}, {Harding}, {Hart}, {Hays}, {Healey}, {Hirayama}, {Hjalmarsdotter},
  {Horn}, {Hughes}, {J{\'o}hannesson}, {Johansson}, {Johnson}, {Johnson},
  {Johnson}, {Johnson}, {Kamae}, {Katagiri}, {Kataoka}, {Kavelaars}, {Kawai},
  {Kelly}, {Kerr}, {Klamra}, {Kn{\"o}dlseder}, {Kocian}, {Komin}, {Kuehn},
  {Kuss}, {Landriu}, {Latronico}, {Lee}, {Lee}, {Lemoine-Goumard}, {Lionetto},
  {Longo}, {Loparco}, {Lott}, {Lovellette}, {Lubrano}, {Madejski}, {Makeev},
  {Marangelli}, {Massai}, {Mazziotta}, {McEnery}, {Menon}, {Meurer},
  {Michelson}, {Minuti}, {Mirizzi}, {Mitthumsiri}, {Mizuno}, {Moiseev},
  {Monte}, {Monzani}, {Moretti}, {Morselli}, {Moskalenko}, {Murgia},
  {Nakamori}, {Nishino}, {Nolan}, {Norris}, {Nuss}, {Ohno}, {Ohsugi}, {Omodei},
  {Orlando}, {Ormes}, {Paccagnella}, {Paneque}, {Panetta}, {Parent}, {Pearce},
  {Pepe}, {Perazzo}, {Pesce-Rollins}, {Picozza}, {Pieri}, {Pinchera}, {Piron},
  {Porter}, {Poupard}, {Rain{\`o}}, {Rando}, {Rapposelli}, {Razzano}, {Reimer},
  {Reimer}, {Reposeur}, {Reyes}, {Ritz}, {Rochester}, {Rodriguez}, {Romani},
  {Roth}, {Russell}, {Ryde}, {Sabatini}, {Sadrozinski}, {Sanchez}, {Sander},
  {Sapozhnikov}, {Parkinson}, {Scargle}, {Schalk}, {Scolieri}, {Sgr{\`o}},
  {Share}, {Shaw}, {Shimokawabe}, {Shrader}, {Sierpowska-Bartosik}, {Siskind},
  {Smith}, {Smith}, {Spandre}, {Spinelli}, {Starck}, {Stephens}, {Strickman},
  {Strong}, {Suson}, {Tajima}, {Takahashi}, {Takahashi}, {Tanaka}, {Tenze},
  {Tether}, {Thayer}, {Thayer}, {Thompson}, {Tibaldo}, {Tibolla}, {Torres},
  {Tosti}, {Tramacere}, {Turri}, {Usher}, {Vilchez}, {Vitale}, {Wang},
  {Watters}, {Winer}, {Wood}, {Ylinen}, \& {Ziegler}}]{LATpaper}
{Atwood}, W.~B., {Abdo}, A.~A., {Ackermann}, M., {et~al.} 2009, \apj, 697, 1071

\bibitem[{{Bartoli} {et~al.}(2014){Bartoli}, {Bernardini}, {Bi}, {Branchini},
  {Budano}, {Camarri}, {Cao}, {Cardarelli}, {Catalanotti}, {Chen}, {Chen},
  {Creti}, {Cui}, {Dai}, {D'Amone}, {Danzengluobu}, {De Mitri}, {D'Ettorre
  Piazzoli}, {Di Girolamo}, {Di Sciascio}, {Feng}, {Feng}, {Feng}, {Gou},
  {Guo}, {He}, {Hu}, {Hu}, {Iacovacci}, {Iuppa}, {Jia}, {Labaciren}, {Li},
  {Liguori}, {Liu}, {Liu}, {Liu}, {Lu}, {Ma}, {Ma}, {Mancarella}, {Mari},
  {Marsella}, {Martello}, {Mastroianni}, {Montini}, {Ning}, {Panareo},
  {Perrone}, {Pistilli}, {Ruggieri}, {Salvini}, {Santonico}, {Shen}, {Sheng},
  {Shi}, {Surdo}, {Tan}, {Vallania}, {Vernetto}, {Vigorito}, {Wang}, {Wu},
  {Wu}, {Xue}, {Yang}, {Yang}, {Yao}, {Yuan}, {Zha}, {Zhang}, {Zhang}, {Zhang},
  {Zhang}, {Zhao}, {Zhaxiciren}, {Zhaxisangzhu}, {Zhou}, {Zhu}, {Zhu}, {Zizzi},
  \& {ARGO-YBJ Collaboration}}]{cocoon_argo}
{Bartoli}, B., {Bernardini}, P., {Bi}, X.~J., {et~al.} 2014, \apj, 790, 152

\bibitem[{{Berlanas} {et~al.}(2020){Berlanas}, {Herrero}, {Comer{\'o}n},
  {Sim{\'o}n-D{\'\i}az}, {Lennon}, {Pasquali}, {Ma{\'\i}z Apell{\'a}niz},
  {Sota}, \& {Peller{\'\i}n}}]{berlanas2020}
{Berlanas}, S.~R., {Herrero}, A., {Comer{\'o}n}, F., {et~al.} 2020, \aap, 642,
  A168

\bibitem[{{Berlanas} {et~al.}(2019){Berlanas}, {Wright}, {Herrero}, {Drew}, \&
  {Lennon}}]{berlanas2019}
{Berlanas}, S.~R., {Wright}, N.~J., {Herrero}, A., {Drew}, J.~E., \& {Lennon},
  D.~J. 2019, \mnras, 484, 1838

\bibitem[{{Binns} {et~al.}(2008){Binns}, {Wiedenbeck}, {Arnould}, {Cummings},
  {de Nolfo}, {Goriely}, {Israel}, {Leske}, {Mewaldt}, {Stone}, \& {von
  Rosenvinge}}]{binns2008}
{Binns}, W.~R., {Wiedenbeck}, M.~E., {Arnould}, M., {et~al.} 2008, \nar, 52,
  427

\bibitem[{{Bluem} {et~al.}(2020){Bluem}, {Kaaret}, {Fuelberth}, {Zajczyk},
  {LaRocca}, {Ringuette}, {Jahoda}, \& {Kuntz}}]{bluem2020}
{Bluem}, J., {Kaaret}, P., {Fuelberth}, W., {et~al.} 2020, \apj, 905, 91

\bibitem[{{Bruel}(2021)}]{Bruel21}
{Bruel}, P. 2021, \aap, 656, A81

\bibitem[{{Bruel} {et~al.}(2018){Bruel}, {Burnett}, {Digel}, {Johannesson},
  {Omodei}, \& {Wood}}]{Bruel2018}
{Bruel}, P., {Burnett}, T.~H., {Digel}, S.~W., {et~al.} 2018, arXiv e-prints,
  arXiv:1810.11394

\bibitem[{Burnham \& Anderson(2002)}]{burnham2002}
Burnham, K. \& Anderson, D. 2002, Model selection and multimodel inference: a
  practical information-theoretic approach (Springer Verlag)

\bibitem[{{Bykov}(2001)}]{bykov2001}
{Bykov}, A.~M. 2001, \ssr, 99, 317

\bibitem[{{Bykov} {et~al.}(2020){Bykov}, {Marcowith}, {Amato}, {Kalyashova},
  {Kruijssen}, \& {Waxman}}]{bykov2020}
{Bykov}, A.~M., {Marcowith}, A., {Amato}, E., {et~al.} 2020, \ssr, 216, 42

\bibitem[{{Cantat-Gaudin} {et~al.}(2020){Cantat-Gaudin}, {Anders},
  {Castro-Ginard}, {Jordi}, {Romero-G{\'o}mez}, {Soubiran}, {Casamiquela},
  {Tarricq}, {Moitinho}, {Vallenari}, {Bragaglia}, {Krone-Martins}, \&
  {Kounkel}}]{cantat-gaudin2020}
{Cantat-Gaudin}, T., {Anders}, F., {Castro-Ginard}, A., {et~al.} 2020, \aap,
  640, A1

\bibitem[{{Cao} {et~al.}(2021){Cao}, {Aharonian}, {An}, {Axikegu}, {Bai},
  {Bao}, {Bastieri}, {Bi}, {Bi}, {Cai}, {Cai}, {Cao}, {Chang}, {Chang},
  {Chang}, {Chen}, {Chen}, {Chen}, {Chen}, {Chen}, {Chen}, {Chen}, {Chen},
  {Chen}, {Chen}, {Chen}, {Chen}, {Chen}, {Cheng}, {Cheng}, {Cui}, {Cui},
  {Cui}, {Dai}, {Dai}, {Dai}, {Danzengluobu}, {della Volpe}, {D'Ettorre
  Piazzoli}, {Dong}, {Fan}, {Fan}, {Fan}, {Fang}, {Fang}, {Feng}, {Feng},
  {Feng}, {Feng}, {Gao}, {Gao}, {Gao}, {Gao}, {Ge}, {Geng}, {Gong}, {Gou},
  {Gu}, {Guo}, {Guo}, {Guo}, {Guo}, {Han}, {He}, {He}, {He}, {He}, {He}, {He},
  {Heller}, {Hor}, {Hou}, {Hou}, {Hu}, {Hu}, {Hu}, {Hu}, {Huang}, {Huang},
  {Huang}, {Huang}, {Huang}, {Ji}, {Ji}, {Jia}, {Jiang}, {Jiang}, {Jin},
  {Kuleshov}, {Levochkin}, {Li}, {Li}, {Li}, {Li}, {Li}, {Li}, {Li}, {Li},
  {Li}, {Li}, {Li}, {Li}, {Li}, {Li}, {Li}, {Li}, {Li}, {Liang}, {Liang},
  {Lin}, {Liu}, {Liu}, {Liu}, {Liu}, {Liu}, {Liu}, {Liu}, {Liu}, {Liu}, {Liu},
  {Liu}, {Liu}, {Liu}, {Liu}, {Liu}, {Long}, {Lu}, {Lv}, {Ma}, {Ma}, {Ma},
  {Mao}, {Masood}, {Mitthumsiri}, {Montaruli}, {Nan}, {Pang},
  {Pattarakijwanich}, {Pei}, {Qi}, {Ruffolo}, {Rulev}, {S{\'a}iz}, {Shao},
  {Shchegolev}, {Sheng}, {Shi}, {Song}, {Stenkin}, {Stepanov}, {Sun}, {Sun},
  {Sun}, {Tam}, {Tang}, {Tian}, {Wang}, {Wang}, {Wang}, {Wang}, {Wang}, {Wang},
  {Wang}, {Wang}, {Wang}, {Wang}, {Wang}, {Wang}, {Wang}, {Wang}, {Wang},
  {Wang}, {Wang}, {Wang}, {Wang}, {Wang}, {Wang}, {Wei}, {Wei}, {Wei}, {Wen},
  {Wu}, {Wu}, {Wu}, {Wu}, {Wu}, {Xi}, {Xia}, {Xia}, {Xiang}, {Xiao}, {Xiao},
  {Xin}, {Xin}, {Xing}, {Xu}, {Xu}, {Xue}, {Yan}, {Yang}, {Yang}, {Yang},
  {Yang}, {Yang}, {Yang}, {Yang}, {Yao}, {Yao}, {Ye}, {Yin}, {Yin}, {You},
  {You}, {Yu}, {Yuan}, {Zeng}, {Zeng}, {Zeng}, {Zeng}, {Zha}, {Zhai}, {Zhang},
  {Zhang}, {Zhang}, {Zhang}, {Zhang}, {Zhang}, {Zhang}, {Zhang}, {Zhang},
  {Zhang}, {Zhang}, {Zhang}, {Zhang}, {Zhang}, {Zhang}, {Zhang}, {Zhang},
  {Zhang}, {Zhang}, {Zhao}, {Zhao}, {Zhao}, {Zhao}, {Zhao}, {Zheng}, {Zheng},
  {Zhou}, {Zhou}, {Zhou}, {Zhou}, {Zhou}, {Zhou}, {Zhu}, {Zhu}, {Zhu}, {Zhu},
  \& {Zuo}}]{cao2021}
{Cao}, Z., {Aharonian}, F.~A., {An}, Q., {et~al.} 2021, \nat, 594, 33

\bibitem[{{Casandjian}(2015)}]{casandjian2015}
{Casandjian}, J.-M. 2015, \apj, 806, 240

\bibitem[{{Cash} {et~al.}(1980){Cash}, {Charles}, {Bowyer}, {Walter},
  {Garmire}, \& {Riegler}}]{cash1980}
{Cash}, W., {Charles}, P., {Bowyer}, S., {et~al.} 1980, \apjl, 238, L71

\bibitem[{{Cherenkov Telescope Array Consortium} {et~al.}(2019){Cherenkov
  Telescope Array Consortium}, {Acharya}, {Agudo}, {Al Samarai}, {Alfaro},
  {Alfaro}, {Alispach}, {Alves Batista}, {Amans}, {Amato}, {Ambrosi},
  {Antolini}, {Antonelli}, {Aramo}, {Araya}, {Armstrong}, {Arqueros},
  {Arrabito}, {Asano}, {Ashley}, {Backes}, {Balazs}, {Balbo}, {Ballester},
  {Ballet}, {Bamba}, {Barkov}, {Barres de Almeida}, {Barrio}, {Bastieri},
  {Becherini}, {Belfiore}, {Benbow}, {Berge}, {Bernardini}, {Bernardini},
  {Bernardos}, {Bernl{\"o}hr}, {Bertucci}, {Biasuzzi}, {Bigongiari}, {Biland},
  {Bissaldi}, {Biteau}, {Blanch}, {Blazek}, {Boisson}, {Bolmont}, {Bonanno},
  {Bonardi}, {Bonavolont{\`a}}, {Bonnoli}, {Bosnjak}, {B{\"o}ttcher},
  {Braiding}, {Bregeon}, {Brill}, {Brown}, {Brun}, {Brunetti}, {Buanes},
  {Buckley}, {Bugaev}, {B{\"u}hler}, {Bulgarelli}, {Bulik}, {Burton},
  {Burtovoi}, {Busetto}, {Canestrari}, {Capalbi}, {Capitanio}, {Caproni},
  {Caraveo}, {C{\'a}rdenas}, {Carlile}, {Carosi}, {Carqu{\'\i}n}, {Carr},
  {Casanova}, {Cascone}, {Catalani}, {Catalano}, {Cauz}, {Cerruti}, {Chadwick},
  {Chaty}, {Chaves}, {Chen}, {Chen}, {Chernyakova}, {Chikawa}, {Christov},
  {Chudoba}, {Cie{\'s}lar}, {Coco}, {Colafrancesco}, {Colin}, {Conforti},
  {Connaughton}, {Conrad}, {Contreras}, {Cortina}, {Costa}, {Costantini},
  {Cotter}, {Covino}, {Crocker}, {Cuadra}, {Cuevas}, {Cumani}, {D'A{\`\i}},
  {D'Ammando}, {D'Avanzo}, {D'Urso}, {Daniel}, {Davids}, {Dawson}, {Dazzi}, {De
  Angelis}, {de C{\'a}ssia dos Anjos}, {De Cesare}, {De Franco}, {de Gouveia
  Dal Pino}, {de la Calle}, {de los Reyes Lopez}, {De Lotto}, {De Luca}, {De
  Lucia}, {de Naurois}, {de O{\~n}a Wilhelmi}, {De Palma}, {De Persio}, {de
  Souza}, {Deil}, {Del Santo}, {Delgado}, {della Volpe}, {Di Girolamo}, {Di
  Pierro}, {Di Venere}, {D{\'\i}az}, {Dib}, {Diebold}, {Djannati-Ata{\"\i}},
  {Dom{\'\i}nguez}, {Dominis Prester}, {Dorner}, {Doro}, {Drass}, {Dravins},
  {Dubus}, {Dwarkadas}, {Ebr}, {Eckner}, {Egberts}, {Einecke}, {Ekoume},
  {Els{\"a}sser}, {Ernenwein}, {Espinoza}, {Evoli}, {Fairbairn},
  {Falceta-Goncalves}, {Falcone}, {Farnier}, {Fasola}, {Fedorova}, {Fegan},
  {Fernandez-Alonso}, {Fern{\'a}ndez-Barral}, {Ferrand}, {Fesquet},
  {Filipovic}, {Fioretti}, {Fontaine}, {Fornasa}, {Fortson}, {Freixas
  Coromina}, {Fruck}, {Fujita}, {Fukazawa}, {Funk}, {F{\"u}{\ss}ling},
  {Gabici}, {Gadola}, {Gallant}, {Garcia}, {Garcia L{\'o}pez}, {Garczarczyk},
  {Gaskins}, {Gasparetto}, {Gaug}, {Gerard}, {Giavitto}, {Giglietto}, {Giommi},
  {Giordano}, {Giro}, {Giroletti}, {Giuliani}, {Glicenstein}, {Gnatyk},
  {Godinovic}, {Goldoni}, {G{\'o}mez-Vargas}, {Gonz{\'a}lez}, {Gonz{\'a}lez},
  {G{\"o}tz}, {Graham}, {Grandi}, {Granot}, {Green}, {Greenshaw}, {Griffiths},
  {Gunji}, {Hadasch}, {Hara}, {Hardcastle}, {Hassan}, {Hayashi}, {Hayashida},
  {Heller}, {Helo}, {Hermann}, {Hinton}, {Hnatyk}, {Hofmann}, {Holder},
  {Horan}, {H{\"o}randel}, {Horns}, {Horvath}, {Hovatta}, {Hrabovsky},
  {Hrupec}, {Humensky}, {H{\"u}tten}, {Iarlori}, {Inada}, {Inome}, {Inoue},
  {Inoue}, {Inoue}, {Iocco}, {Ioka}, {Iori}, {Ishio}, {Iwamura}, {Jamrozy},
  {Janecek}, {Jankowsky}, {Jean}, {Jung-Richardt}, {Jurysek}, {Kaaret},
  {Karkar}, {Katagiri}, {Katz}, {Kawanaka}, {Kazanas}, {Kh{\'e}lifi}, {Kieda},
  {Kimeswenger}, {Kimura}, {Kisaka}, {Knapp}, {Kn{\"o}dlseder}, {Koch},
  {Kohri}, {Komin}, {Kosack}, {Kraus}, {Krause}, {Krau{\ss}}, {Kubo}, {Kukec
  Mezek}, {Kuroda}, {Kushida}, {La Palombara}, {Lamanna}, {Lang}, {Lapington},
  {Le Blanc}, {Leach}, {Lees}, {Lefaucheur}, {Leigui de Oliveira}, {Lenain},
  {Lico}, {Limon}, {Lindfors}, {Lohse}, {Lombardi}, {Longo}, {L{\'o}pez},
  {L{\'o}pez-Coto}, {Lu}, {Lucarelli}, {Luque-Escamilla}, {Lyard}, {Maccarone},
  {Maier}, {Majumdar}, {Malaguti}, {Mandat}, {Maneva}, {Manganaro}, {Mangano},
  {Marcowith}, {Mar{\'\i}n}, {Markoff}, {Mart{\'\i}}, {Martin},
  {Mart{\'\i}nez}, {Mart{\'\i}nez}, {Masetti}, {Masuda}, {Maurin}, {Maxted},
  {Mazin}, {Medina}, {Melandri}, {Mereghetti}, {Meyer}, {Minaya}, {Mirabal},
  {Mirzoyan}, {Mitchell}, {Mizuno}, {Moderski}, {Mohammed}, {Mohrmann},
  {Montaruli}, {Moralejo}, {Morcuende-Parrilla}, {Mori}, {Morlino}, {Morris},
  {Morselli}, {Moulin}, {Mukherjee}, {Mundell}, {Murach}, {Muraishi}, {Murase},
  {Nagai}, {Nagataki}, {Nagayoshi}, {Naito}, {Nakamori}, {Nakamura}, {Niemiec},
  {Nieto}, {Niko{\l}ajuk}, {Nishijima}, {Noda}, {Nosek}, {Novosyadlyj},
  {Nozaki}, {O'Brien}, {Oakes}, {Ohira}, {Ohishi}, {Ohm}, {Okazaki}, {Okumura},
  {Ong}, {Orienti}, {Orito}, {Osborne}, {Ostrowski}, {Otte}, {Oya}, {Padovani},
  {Paizis}, {Palatiello}, {Palatka}, {Paoletti}, {Paredes}, {Pareschi},
  {Parsons}, {Pe'er}, {Pech}, {Pedaletti}, {Perri}, {Persic}, {Petrashyk},
  {Petrucci}, {Petruk}, {Peyaud}, {Pfeifer}, {Piano}, {Pisarski}, {Pita},
  {Pohl}, {Polo}, {Pozo}, {Prandini}, {Prast}, {Principe}, {Prokhorov},
  {Prokoph}, {Prouza}, {P{\"u}hlhofer}, {Punch}, {P{\"u}rckhauer}, {Queiroz},
  {Quirrenbach}, {Rain{\`o}}, {Razzaque}, {Reimer}, {Reimer}, {Reisenegger},
  {Renaud}, {Rezaeian}, {Rhode}, {Ribeiro}, {Rib{\'o}}, {Richtler}, {Rico},
  {Rieger}, {Riquelme}, {Rivoire}, {Rizi}, {Rodriguez}, {Rodriguez Fernandez},
  {Rodr{\'\i}guez V{\'a}zquez}, {Rojas}, {Romano}, {Romeo}, {Rosado}, {Rovero},
  {Rowell}, {Rudak}, {Rugliancich}, {Rulten}, {Sadeh}, {Safi-Harb}, {Saito},
  {Sakaki}, {Sakurai}, {Salina}, {S{\'a}nchez-Conde}, {Sandaker}, {Sandoval},
  {Sangiorgi}, {Sanguillon}, {Sano}, {Santander}, {Sarkar}, {Satalecka},
  {Saturni}, {Schioppa}, {Schlenstedt}, {Schneider}, {Schoorlemmer},
  {Schovanek}, {Schulz}, {Schussler}, {Schwanke}, {Sciacca}, {Scuderi},
  {Seitenzahl}, {Semikoz}, {Sergijenko}, {Servillat}, {Shalchi}, {Shellard},
  {Sidoli}, {Siejkowski}, {Sillanp{\"a}{\"a}}, {Sironi}, {Sitarek}, {Sliusar},
  {Slowikowska}, {Sol}, {Stamerra}, {Stani{\v{c}}}, {Starling}, {Stawarz},
  {Stefanik}, {Stephan}, {Stolarczyk}, {Stratta}, {Straumann}, {Suomijarvi},
  {Supanitsky}, {Tagliaferri}, {Tajima}, {Tavani}, {Tavecchio}, {Tavernet},
  {Tayabaly}, {Tejedor}, {Temnikov}, {Terada}, {Terrier}, {Terzic}, {Teshima},
  {Testa}, {Thoudam}, {Tian}, {Tibaldo}, {Tluczykont}, {Todero Peixoto},
  {Tokanai}, {Tomastik}, {Tonev}, {Tornikoski}, {Torres}, {Torresi}, {Tosti},
  {Tothill}, {Tovmassian}, {Travnicek}, {Trichard}, {Trifoglio}, {Troyano
  Pujadas}, {Tsujimoto}, {Umana}, {Vagelli}, {Vagnetti}, {Valentino},
  {Vallania}, {Valore}, {van Eldik}, {Vandenbroucke}, {Varner}, {Vasileiadis},
  {Vassiliev}, {V{\'a}zquez Acosta}, {Vecchi}, {Vega}, {Vercellone}, {Veres},
  {Vergani}, {Verzi}, {Vettolani}, {Viana}, {Vigorito}, {Villanueva}, {Voelk},
  {Vollhardt}, {Vorobiov}, {Vrastil}, {Vuillaume}, {Wagner}, {Wagner},
  {Walter}, {Ward}, {Warren}, {Watson}, {Werner}, {White}, {White},
  {Wierzcholska}, {Wilcox}, {Will}, {Williams}, {Wischnewski}, {Wood},
  {Yamamoto}, {Yamazaki}, {Yanagita}, {Yang}, {Yoshida}, {Yoshiike},
  {Yoshikoshi}, {Zacharias}, {Zaharijas}, {Zampieri}, {Zandanel}, {Zanin},
  {Zavrtanik}, {Zavrtanik}, {Zdziarski}, {Zech}, {Zechlin}, {Zhdanov},
  {Ziegler}, \& {Zorn}}]{CTA-science}
{Cherenkov Telescope Array Consortium}, {Acharya}, B.~S., {Agudo}, I., {et~al.}
  2019, {Science with the Cherenkov Telescope Array}

\bibitem[{Cox(2000)}]{cox2000}
Cox, A. 2000, {Allen's astrophysical quantities; 4th ed.} (New York, NY: AIP)

\bibitem[{{Dame}(2011)}]{dame2011}
{Dame}, T.~M. 2011, arXiv e-prints, arXiv:1101.1499

\bibitem[{{Dame} {et~al.}(2001){Dame}, {Hartmann}, \& {Thaddeus}}]{CfA_CO}
{Dame}, T.~M., {Hartmann}, D., \& {Thaddeus}, P. 2001, The Astrophysical
  Journal, 547, 792

\bibitem[{{de Angelis} {et~al.}(2018){de Angelis}, {Tatischeff}, {Grenier},
  {McEnery}, {Mallamaci}, {Tavani}, {Oberlack}, {Hanlon}, {Walter}, {Argan},
  {von Ballmoos}, {Bulgarelli}, {Bykov}, {Hernanz}, {Kanbach}, {Kuvvetli},
  {Pearce}, {Zdziarski}, {Conrad}, {Ghisellini}, {Harding}, {Isern}, {Leising},
  {Longo}, {Madejski}, {Martinez}, {Mazziotta}, {Paredes}, {Pohl}, {Rando},
  {Razzano}, {Aboudan}, {Ackermann}, {Addazi}, {Ajello}, {Albertus},
  {{\'A}lvarez}, {Ambrosi}, {Ant{\'o}n}, {Antonelli}, {Babic}, {Baibussinov},
  {Balbo}, {Baldini}, {Balman}, {Bambi}, {Barres de Almeida}, {Barrio},
  {Bartels}, {Bastieri}, {Bednarek}, {Bernard}, {Bernardini}, {Bernasconi},
  {Bertucci}, {Biland}, {Bissaldi}, {Boettcher}, {Bonvicini}, {Bosch-Ramon},
  {Bottacini}, {Bozhilov}, {Bretz}, {Branchesi}, {Brdar}, {Bringmann},
  {Brogna}, {Budtz J{\o}rgensen}, {Busetto}, {Buson}, {Busso}, {Caccianiga},
  {Camera}, {Campana}, {Caraveo}, {Cardillo}, {Carlson}, {Celestin},
  {Cerme{\~n}o}, {Chen}, {Cheung}, {Churazov}, {Ciprini}, {Coc},
  {Colafrancesco}, {Coleiro}, {Collmar}, {Coppi}, {Curado da Silva}, {Cutini},
  {D'Ammando}, {de Lotto}, {de Martino}, {De Rosa}, {Del Santo}, {Delgado},
  {Diehl}, {Dietrich}, {Dolgov}, {Dom{\'\i}nguez}, {Dominis Prester},
  {Donnarumma}, {Dorner}, {Doro}, {Dutra}, {Elsaesser}, {Fabrizio},
  {Fern{\'a}ndez-Barral}, {Fioretti}, {Foffano}, {Formato}, {Fornengo},
  {Foschini}, {Franceschini}, {Franckowiak}, {Funk}, {Fuschino}, {Gaggero},
  {Galanti}, {Gargano}, {Gasparrini}, {Gehrz}, {Giammaria}, {Giglietto},
  {Giommi}, {Giordano}, {Giroletti}, {Ghirlanda}, {Godinovic}, {Gouiff{\'e}s},
  {Grove}, {Hamadache}, {Hartmann}, {Hayashida}, {Hryczuk}, {Jean}, {Johnson},
  {Jos{\'e}}, {Kaufmann}, {Khelifi}, {Kiener}, {Kn{\"o}dlseder}, {Kole},
  {Kopp}, {Kozhuharov}, {Labanti}, {Lalkovski}, {Laurent}, {Limousin},
  {Linares}, {Lindfors}, {Lindner}, {Liu}, {Lombardi}, {Loparco},
  {L{\'o}pez-Coto}, {L{\'o}pez Moya}, {Lott}, {Lubrano}, {Malyshev},
  {Mankuzhiyil}, {Mannheim}, {March{\~a}}, {Marcian{\`o}}, {Marcote},
  {Mariotti}, {Marisaldi}, {McBreen}, {Mereghetti}, {Merle}, {Mignani},
  {Minervini}, {Moiseev}, {Morselli}, {Moura}, {Nakazawa}, {Nava}, {Nieto},
  {Orienti}, {Orio}, {Orlando}, {Orleanski}, {Paiano}, {Paoletti}, {Papitto},
  {Pasquato}, {Patricelli}, {P{\'e}rez-Garc{\'\i}a}, {Persic}, {Piano},
  {Pichel}, {Pimenta}, {Pittori}, {Porter}, {Poutanen}, {Prandini}, {Prantzos},
  {Produit}, {Profumo}, {Queiroz}, {Rain{\'o}}, {Raklev}, {Regis}, {Reichardt},
  {Rephaeli}, {Rico}, {Rodejohann}, {Rodriguez Fernandez}, {Roncadelli},
  {Roso}, {Rovero}, {Ruffini}, {Sala}, {S{\'a}nchez-Conde}, {Santangelo}, {Saz
  Parkinson}, {Sbarrato}, {Shearer}, {Shellard}, {Short}, {Siegert},
  {Siqueira}, {Spinelli}, {Stamerra}, {Starrfield}, {Strong}, {Str{\"u}mke},
  {Tavecchio}, {Taverna}, {Terzi{\'c}}, {Thompson}, {Tibolla}, {Torres},
  {Turolla}, {Ulyanov}, {Ursi}, {Vacchi}, {van den Abeele},
  {Vankova-Kirilovai}, {Venter}, {Verrecchia}, {Vincent}, {Wang}, {Weniger},
  {Wu}, {Zaharija{\v{s}}}, {Zampieri}, {Zane}, {Zimmer}, {Zoglauer}, \&
  {E-Astrogam Collaboration}}]{astrogam}
{de Angelis}, A., {Tatischeff}, V., {Grenier}, I.~A., {et~al.} 2018, Journal of
  High Energy Astrophysics, 19, 1

\bibitem[{{de Zeeuw} {et~al.}(1999){de Zeeuw}, {Hoogerwerf}, {de Bruijne},
  {Brown}, \& {Blaauw}}]{dezeeuw1999}
{de Zeeuw}, P.~T., {Hoogerwerf}, R., {de Bruijne}, J.~H.~J., {Brown}, A.~G.~A.,
  \& {Blaauw}, A. 1999, \aj, 117, 354

\bibitem[{{Delgado} \& {Alfaro}(2000)}]{delgado2000}
{Delgado}, A.~J. \& {Alfaro}, E.~J. 2000, \aj, 119, 1848

\bibitem[{{Dickey} {et~al.}(2009){Dickey}, {Strasser}, {Gaensler}, {Haverkorn},
  {Kavars}, {McClure-Griffiths}, {Stil}, \& {Taylor}}]{dickey2009}
{Dickey}, J.~M., {Strasser}, S., {Gaensler}, B.~M., {et~al.} 2009, \apj, 693,
  1250

\bibitem[{{Emig} {et~al.}(2022){Emig}, {White}, {Salas}, {Karim}, {van Weeren},
  {Teuben}, {Zavagno}, {Chiu}, {Haverkorn}, {Oonk}, {Orr{\'u}}, {Polderman},
  {Reich}, {R{\"o}ttgering}, \& {Tielens}}]{emig2022}
{Emig}, K.~L., {White}, G.~J., {Salas}, P., {et~al.} 2022, \aap, 664, A88

\bibitem[{{Evoli} {et~al.}(2019){Evoli}, {Aloisio}, \& {Blasi}}]{Evoli:2019}
{Evoli}, C., {Aloisio}, R., \& {Blasi}, P. 2019, \prd, 99, 103023

\bibitem[{{Fermi-LAT collaboration} {et~al.}(2022){Fermi-LAT collaboration},
  {:}, {Abdollahi}, {Acero}, {Baldini}, {Ballet}, {Bastieri}, {Bellazzini},
  {Berenji}, {Berretta}, {Bissaldi}, {Blandford}, {Bloom}, {Bonino}, {Brill},
  {Britto}, {Bruel}, {Burnett}, {Buson}, {Cameron}, {Caputo}, {Caraveo},
  {Castro}, {Chaty}, {Cheung}, {Chiaro}, {Cibrario}, {Ciprini},
  {Coronado-Blazquez}, {Crnogorcevic}, {Cutini}, {D'Ammando}, {De Gaetano},
  {Digel}, {Di Lalla}, {Dirirsa}, {Di Venere}, {Dominguez}, {Fallah Ramazani},
  {Fegan}, {Ferrara}, {Fiori}, {Fleischhack}, {Franckowiak}, {Fukazawa},
  {Funk}, {Fusco}, {Galanti}, {Gammaldi}, {Gargano}, {Garrappa}, {Gasparrini},
  {Giacchino}, {Giglietto}, {Giordano}, {Giroletti}, {Glanzman}, {Green},
  {Grenier}, {Grondin}, {Guillemot}, {Guiriec}, {Gustafsson}, {Harding},
  {Hays}, {Hewitt}, {Horan}, {Hou}, {Johannesson}, {Karwin}, {Kayanoki},
  {Kerr}, {Kuss}, {Landriu}, {Larsson}, {Latronico}, {Lemoine-Goumard}, {Li},
  {Liodakis}, {Longo}, {Loparco}, {Lott}, {Lubrano}, {Maldera}, {Malyshev},
  {Manfreda}, {Marti-Devesa}, {Mazziotta}, {Mereu}, {Meyer}, {Michelson},
  {Mirabal}, {Mitthumsiri}, {Mizuno}, {Moiseev}, {Monzani}, {Morselli},
  {Moskalenko}, {Negro}, {Nuss}, {Omodei}, {Orienti}, {Orlando}, {Paneque},
  {Pei}, {Perkins}, {Persic}, {Pesce-Rollins}, {Petrosian}, {Pillera}, {Poon},
  {Porter}, {Principe}, {Raino}, {Rando}, {Rani}, {Razzano}, {Razzaque},
  {Reimer}, {Reimer}, {Reposeur}, {Sanchez-Conde}, {Saz Parkinson}, {Scotton},
  {Serini}, {Sgro}, {Siskind}, {Smith}, {Spandre}, {Spinelli}, {Sueoka},
  {Suson}, {Tajima}, {Tak}, {Thayer}, {Thompson}, {Torres}, {Troja},
  {Valverde}, {Wood}, \& {Zaharijas}}]{4FGL-DR3}
{Fermi-LAT collaboration}, {:}, {Abdollahi}, S., {et~al.} 2022, arXiv e-prints,
  arXiv:2201.11184

\bibitem[{{Ferrand} \& {Marcowith}(2010)}]{ferrand2010}
{Ferrand}, G. \& {Marcowith}, A. 2010, \aap, 510, A101

\bibitem[{{Ferrand} \& {Safi-Harb}(2012)}]{ferrand2012}
{Ferrand}, G. \& {Safi-Harb}, S. 2012, Advances in Space Research, 49, 1313

\bibitem[{{Fornieri} \& {Zhang}(2022)}]{fornieri2022}
{Fornieri}, O. \& {Zhang}, H. 2022, \prd, 106, 103015

\bibitem[{{Gabici} {et~al.}(2019){Gabici}, {Evoli}, {Gaggero}, {Lipari},
  {Mertsch}, {Orlando}, {Strong}, \& {Vittino}}]{gabici2019}
{Gabici}, S., {Evoli}, C., {Gaggero}, D., {et~al.} 2019, International Journal
  of Modern Physics D, 28, 1930022

\bibitem[{{G{\'e}nolini} {et~al.}(2019){G{\'e}nolini}, {Boudaud}, {Batista},
  {Caroff}, {Derome}, {Lavalle}, {Marcowith}, {Maurin}, {Poireau}, {Poulin},
  {Rosier}, {Salati}, {Serpico}, \& {Vecchi}}]{Genolini:2019}
{G{\'e}nolini}, Y., {Boudaud}, M., {Batista}, P.~I., {et~al.} 2019, \prd, 99,
  123028

\bibitem[{{Grenier} {et~al.}(2005){Grenier}, {Casandjian}, \&
  {Terrier}}]{grenier2005}
{Grenier}, I.~A., {Casandjian}, J.-M., \& {Terrier}, R. 2005, Science, 307,
  1292

\bibitem[{{Gupta} {et~al.}(2018){Gupta}, {Nath}, \& {Sharma}}]{gupta2018}
{Gupta}, S., {Nath}, B.~B., \& {Sharma}, P. 2018, \mnras, 479, 5220

\bibitem[{{H.~E.~S.~S. Collaboration} {et~al.}(2019){H.~E.~S.~S.
  Collaboration}, {Abdalla}, {Aharonian}, {Ait Benkhali}, {Ang{\"u}ner},
  {Arakawa}, {Arcaro}, {Armand}, {Arrieta}, {Backes}, {Barnard}, {Becherini},
  {Becker Tjus}, {Berge}, {Bernl{\"o}hr}, {Blackwell}, {B{\"o}ttcher},
  {Boisson}, {Bolmont}, {Bonnefoy}, {Bordas}, {Bregeon}, {Brun}, {Brun},
  {Bryan}, {B{\"u}chele}, {Bulik}, {Bylund}, {Capasso}, {Caroff}, {Carosi},
  {Casanova}, {Cerruti}, {Chakraborty}, {Chand}, {Chandra}, {Chaves}, {Chen},
  {Colafrancesco}, {Condon}, {Davids}, {Deil}, {Devin}, {deWilt}, {Dirson},
  {Djannati-Ata{\"\i}}, {Dmytriiev}, {Donath}, {Doroshenko}, {Drury}, {Dyks},
  {Egberts}, {Emery}, {Ernenwein}, {Eschbach}, {Fegan}, {Fiasson}, {Fontaine},
  {Funk}, {F{\"u}{\ss}ling}, {Gabici}, {Gallant}, {Gat{\'e}}, {Giavitto},
  {Glawion}, {Glicenstein}, {Gottschall}, {Grondin}, {Hahn}, {Haupt},
  {Heinzelmann}, {Henri}, {Hermann}, {Hinton}, {Hofmann}, {Hoischen}, {Holch},
  {Holler}, {Horns}, {Huber}, {Iwasaki}, {Jacholkowska}, {Jamrozy},
  {Jankowsky}, {Jankowsky}, {Jouvin}, {Jung-Richardt}, {Kastendieck},
  {Katarzy{\'n}ski}, {Katsuragawa}, {Katz}, {Kerszberg}, {Khangulyan},
  {Kh{\'e}lifi}, {King}, {Klepser}, {Klu{\'z}niak}, {Komin}, {Kosack}, {Kraus},
  {Lamanna}, {Lau}, {Lefaucheur}, {Lemi{\`e}re}, {Lemoine-Goumard}, {Lenain},
  {Leser}, {Lohse}, {L{\'o}pez-Coto}, {Lypova}, {Malyshev}, {Marandon},
  {Marcowith}, {Mariaud}, {Mart{\'\i}-Devesa}, {Marx}, {Maurin}, {Meintjes},
  {Mitchell}, {Moderski}, {Mohamed}, {Mohrmann}, {Moore}, {Moulin}, {Murach},
  {Nakashima}, {de Naurois}, {Ndiyavala}, {Niederwanger}, {Niemiec}, {Oakes},
  {O'Brien}, {Odaka}, {Ohm}, {Ostrowski}, {Oya}, {Panter}, {Parsons},
  {Perennes}, {Petrucci}, {Peyaud}, {Piel}, {Pita}, {Poireau}, {Priyana Noel},
  {Prokhorov}, {Prokoph}, {P{\"u}hlhofer}, {Punch}, {Quirrenbach}, {Raab},
  {Rauth}, {Reimer}, {Reimer}, {Renaud}, {Rieger}, {Rinchiuso}, {Romoli},
  {Rowell}, {Rudak}, {Ruiz-Velasco}, {Sahakian}, {Saito}, {Sanchez},
  {Santangelo}, {Sasaki}, {Schlickeiser}, {Sch{\"u}ssler}, {Schulz}, {Schutte},
  {Schwanke}, {Schwemmer}, {Seglar-Arroyo}, {Senniappan}, {Seyffert}, {Shafi},
  {Shilon}, {Shiningayamwe}, {Simoni}, {Sinha}, {Sol}, {Specovius},
  {Spir-Jacob}, {Stawarz}, {Steenkamp}, {Stegmann}, {Steppa}, {Takahashi},
  {Tavernet}, {Tavernier}, {Taylor}, {Terrier}, {Tibaldo}, {Tiziani},
  {Tluczykont}, {Trichard}, {Tsirou}, {Tsuji}, {Tuffs}, {Uchiyama}, {van der
  Walt}, {van Eldik}, {van Rensburg}, {van Soelen}, {Vasileiadis}, {Veh},
  {Venter}, {Vincent}, {Vink}, {Voisin}, {V{\"o}lk}, {Vuillaume}, {Wadiasingh},
  {Wagner}, {Wagner}, {White}, {Wierzcholska}, {Yang}, {Yoneda}, {Zaborov},
  {Zacharias}, {Zanin}, {Zdziarski}, {Zech}, {Zefi}, {Ziegler}, {Zorn}, \&
  {{\.Z}ywucka}}]{hess1825}
{H.~E.~S.~S. Collaboration}, {Abdalla}, H., {Aharonian}, F., {et~al.} 2019,
  \aap, 621, A116

\bibitem[{Harris {et~al.}(2020)Harris, Millman, van~der Walt, Gommers,
  Virtanen, Cournapeau, Wieser, Taylor, Berg, Smith, Kern, Picus, Hoyer, van
  Kerkwijk, Brett, Haldane, del R{\'{i}}o, Wiebe, Peterson,
  G{\'{e}}rard-Marchant, Sheppard, Reddy, Weckesser, Abbasi, Gohlke, \&
  Oliphant}]{harris2020}
Harris, C.~R., Millman, K.~J., van~der Walt, S.~J., {et~al.} 2020, Nature, 585,
  357

\bibitem[{{HI4PI Collaboration} {et~al.}(2016){HI4PI Collaboration}, {Ben
  Bekhti}, {Fl{\"o}er}, {Keller}, {Kerp}, {Lenz}, {Winkel}, {Bailin},
  {Calabretta}, {Dedes}, {Ford}, {Gibson}, {Haud}, {Janowiecki}, {Kalberla},
  {Lockman}, {McClure-Griffiths}, {Murphy}, {Nakanishi}, {Pisano}, \&
  {Staveley-Smith}}]{HI4PI}
{HI4PI Collaboration}, {Ben Bekhti}, N., {Fl{\"o}er}, L., {et~al.} 2016,
  Astronomy and Astrophysics, 594, A116

\bibitem[{{Ho} {et~al.}(2017){Ho}, {Ng}, {Lyne}, {Stappers}, {Coe}, {Halpern},
  {Johnson}, \& {Steele}}]{ho2017}
{Ho}, W. C.~G., {Ng}, C.~Y., {Lyne}, A.~G., {et~al.} 2017, \mnras, 464, 1211

\bibitem[{Hunter(2007)}]{Hunter2007}
Hunter, J.~D. 2007, Computing in Science \& Engineering, 9, 90

\bibitem[{{Kaur} {et~al.}(2020){Kaur}, {Sharma}, {Dewangan}, {Ojha},
  {Durgapal}, \& {Panwar}}]{kaur2020}
{Kaur}, H., {Sharma}, S., {Dewangan}, L.~K., {et~al.} 2020, \apj, 896, 29

\bibitem[{{Krause} \& {Diehl}(2014)}]{Krause:2014}
{Krause}, M. G.~H. \& {Diehl}, R. 2014, \apjl, 794, L21

\bibitem[{{Lancaster} {et~al.}(2021){Lancaster}, {Ostriker}, {Kim}, \&
  {Kim}}]{lancaster2021}
{Lancaster}, L., {Ostriker}, E.~C., {Kim}, J.-G., \& {Kim}, C.-G. 2021, \apj,
  914, 89

\bibitem[{{Leahy} {et~al.}(2013){Leahy}, {Green}, \& {Ranasinghe}}]{leahy2013}
{Leahy}, D.~A., {Green}, K., \& {Ranasinghe}, S. 2013, \mnras, 436, 968

\bibitem[{{Leahy} {et~al.}(2020){Leahy}, {Ranasinghe}, \&
  {Gelowitz}}]{leahy2020}
{Leahy}, D.~A., {Ranasinghe}, S., \& {Gelowitz}, M. 2020, \apjs, 248, 16

\bibitem[{{Li}(2022)}]{cocoon_lhaaso}
{Li}, C. 2022, in 37th International Cosmic Ray Conference. 12-23 July 2021.
  Berlin, 843

\bibitem[{{Lyne} {et~al.}(2015){Lyne}, {Stappers}, {Keith}, {Ray}, {Kerr},
  {Camilo}, \& {Johnson}}]{lyne2015}
{Lyne}, A.~G., {Stappers}, B.~W., {Keith}, M.~J., {et~al.} 2015, \mnras, 451,
  581

\bibitem[{{MAGIC Collaboration} {et~al.}(2020){MAGIC Collaboration}, {Acciari},
  {Ansoldi}, {Antonelli}, {Arbet Engels}, {Baack}, {Babi{\'c}}, {Banerjee},
  {Barres de Almeida}, {Barrio}, {Becerra Gonz{\'a}lez}, {Bednarek},
  {Bellizzi}, {Bernardini}, {Berti}, {Besenrieder}, {Bhattacharyya},
  {Bigongiari}, {Biland}, {Blanch}, {Bonnoli}, {Bo{\v{s}}njak}, {Busetto},
  {Carosi}, {Ceribella}, {Cerruti}, {Chai}, {Chilingarian}, {Cikota}, {Colak},
  {Colin}, {Colombo}, {Contreras}, {Cortina}, {Covino}, {D'Elia}, {Da Vela},
  {Dazzi}, {De Angelis}, {De Lotto}, {Delfino}, {Delgado}, {Depaoli}, {Di
  Pierro}, {Di Venere}, {Do Souto Espi{\~n}eira}, {Dominis Prester}, {Donini},
  {Dorner}, {Doro}, {Elsaesser}, {Fallah Ramazani}, {Fattorini}, {Ferrara},
  {Foffano}, {Fonseca}, {Font}, {Fruck}, {Fukami}, {Garc{\'\i}a L{\'o}pez},
  {Garczarczyk}, {Gasparyan}, {Gaug}, {Giglietto}, {Giordano}, {Gliwny},
  {Godinovi{\'c}}, {Green}, {Hadasch}, {Hahn}, {Herrera}, {Hoang}, {Hrupec},
  {H{\"u}tten}, {Inada}, {Inoue}, {Ishio}, {Iwamura}, {Jouvin}, {Kajiwara},
  {Karjalainen}, {Kerszberg}, {Kobayashi}, {Kubo}, {Kushida}, {Lamastra},
  {Lelas}, {Leone}, {Lindfors}, {Lombardi}, {Longo}, {L{\'o}pez},
  {L{\'o}pez-Coto}, {L{\'o}pez-Oramas}, {Loporchio}, {Machado de Oliveira
  Fraga}, {Masuda}, {Maggio}, {Majumdar}, {Makariev}, {Mallamaci}, {Maneva},
  {Manganaro}, {Mannheim}, {Maraschi}, {Mariotti}, {Mart{\'\i}nez}, {Mazin},
  {Mender}, {Mi{\'c}anovi{\'c}}, {Miceli}, {Miener}, {Minev}, {Miranda},
  {Mirzoyan}, {Molina}, {Moralejo}, {Morcuende}, {Moreno}, {Moretti},
  {Munar-Adrover}, {Neustroev}, {Nigro}, {Nilsson}, {Ninci}, {Nishijima},
  {Noda}, {Nogu{\'e}s}, {Nozaki}, {Ohtani}, {Oka}, {Otero-Santos},
  {Palatiello}, {Paneque}, {Paoletti}, {Paredes}, {Pavleti{\'c}}, {Pe{\~n}il},
  {Peresano}, {Persic}, {Prada Moroni}, {Prandini}, {Puljak}, {Rhode},
  {Rib{\'o}}, {Rico}, {Righi}, {Rugliancich}, {Saha}, {Sahakyan}, {Saito},
  {Sakurai}, {Satalecka}, {Schleicher}, {Schmidt}, {Schweizer}, {Sitarek},
  {{\v{S}}nidari{\'c}}, {Sobczynska}, {Spolon}, {Stamerra}, {Strom}, {Strzys},
  {Suda}, {Suri{\'c}}, {Takahashi}, {Tavecchio}, {Temnikov}, {Terzi{\'c}},
  {Teshima}, {Torres-Alb{\`a}}, {Tosti}, {van Scherpenberg}, {Vanzo}, {Vazquez
  Acosta}, {Ventura}, {Verguilov}, {Vigorito}, {Vitale}, {Vovk}, {Will},
  {Zari{\'c}}, {authors}, {:}, {Celli}, \& {Morlino}}]{MAGIC20}
{MAGIC Collaboration}, {Acciari}, V.~A., {Ansoldi}, S., {et~al.} 2020, arXiv
  e-prints, arXiv:2010.15854

\bibitem[{{Manchester} {et~al.}(2005){Manchester}, {Hobbs}, {Teoh}, \&
  {Hobbs}}]{atnf2005}
{Manchester}, R.~N., {Hobbs}, G.~B., {Teoh}, A., \& {Hobbs}, M. 2005, \aj, 129,
  1993

\bibitem[{{Martin} {et~al.}(2022){Martin}, {Marcowith}, \&
  {Tibaldo}}]{Martin:2022}
{Martin}, P., {Marcowith}, A., \& {Tibaldo}, L. 2022, \aap, 665, A132

\bibitem[{{Martins} {et~al.}(2005){Martins}, {Schaerer}, \&
  {Hillier}}]{martins2005}
{Martins}, F., {Schaerer}, D., \& {Hillier}, D.~J. 2005, \aap, 436, 1049

\bibitem[{{McEnery} \& {Amego Team}(2020)}]{amego}
{McEnery}, J. \& {Amego Team}. 2020, in American Astronomical Society Meeting
  Abstracts, Vol. 235, American Astronomical Society Meeting Abstracts \#235,
  372.15

\bibitem[{{Mizuno} {et~al.}(2015){Mizuno}, {Tanabe}, {Takahashi}, {Hayashi},
  {Yamazaki}, {Grenier}, \& {Tibaldo}}]{mizuno2015}
{Mizuno}, T., {Tanabe}, T., {Takahashi}, H., {et~al.} 2015, \apj, 803, 74

\bibitem[{{Mori}(2009)}]{Mori:2009}
{Mori}, M. 2009, Astroparticle Physics, 31, 341

\bibitem[{{Morlino} {et~al.}(2021){Morlino}, {Blasi}, {Peretti}, \&
  {Cristofari}}]{morlino2021}
{Morlino}, G., {Blasi}, P., {Peretti}, E., \& {Cristofari}, P. 2021, \mnras,
  504, 6096

\bibitem[{{Nava} {et~al.}(2019){Nava}, {Recchia}, {Gabici}, {Marcowith},
  {Brahimi}, \& {Ptuskin}}]{Nava:2019}
{Nava}, L., {Recchia}, S., {Gabici}, S., {et~al.} 2019, \mnras, 484, 2684

\bibitem[{{Orlando}(2018)}]{Orlando:2018}
{Orlando}, E. 2018, \mnras, 475, 2724

\bibitem[{{Orlando} \& {Strong}(2007)}]{Orlando:2007}
{Orlando}, E. \& {Strong}, A.~W. 2007, \apss, 309, 359

\bibitem[{{Planck Collaboration} {et~al.}(2016{\natexlab{a}}){Planck
  Collaboration}, {Adam}, {Ade}, {Aghanim}, {Alves}, {Arnaud}, {Ashdown},
  {Aumont}, {Baccigalupi}, {Banday}, {Barreiro}, {Bartlett}, {Bartolo},
  {Battaner}, {Benabed}, {Beno{\^\i}t}, {Benoit-L{\'e}vy}, {Bernard},
  {Bersanelli}, {Bielewicz}, {Bock}, {Bonaldi}, {Bonavera}, {Bond}, {Borrill},
  {Bouchet}, {Boulanger}, {Bucher}, {Burigana}, {Butler}, {Calabrese},
  {Cardoso}, {Catalano}, {Challinor}, {Chamballu}, {Chary}, {Chiang},
  {Christensen}, {Clements}, {Colombi}, {Colombo}, {Combet}, {Couchot},
  {Coulais}, {Crill}, {Curto}, {Cuttaia}, {Danese}, {Davies}, {Davis}, {de
  Bernardis}, {de Rosa}, {de Zotti}, {Delabrouille}, {D{\'e}sert}, {Dickinson},
  {Diego}, {Dole}, {Donzelli}, {Dor{\'e}}, {Douspis}, {Ducout}, {Dupac},
  {Efstathiou}, {Elsner}, {En{\ss}lin}, {Eriksen}, {Falgarone}, {Fergusson},
  {Finelli}, {Forni}, {Frailis}, {Fraisse}, {Franceschi}, {Frejsel},
  {Galeotta}, {Galli}, {Ganga}, {Ghosh}, {Giard}, {Giraud-H{\'e}raud},
  {Gjerl{\o}w}, {Gonz{\'a}lez-Nuevo}, {G{\'o}rski}, {Gratton}, {Gregorio},
  {Gruppuso}, {Gudmundsson}, {Hansen}, {Hanson}, {Harrison}, {Helou},
  {Henrot-Versill{\'e}}, {Hern{\'a}ndez-Monteagudo}, {Herranz}, {Hildebrandt},
  {Hivon}, {Hobson}, {Holmes}, {Hornstrup}, {Hovest}, {Huffenberger}, {Hurier},
  {Jaffe}, {Jaffe}, {Jones}, {Juvela}, {Keih{\"a}nen}, {Keskitalo}, {Kisner},
  {Kneissl}, {Knoche}, {Kunz}, {Kurki-Suonio}, {Lagache},
  {L{\"a}hteenm{\"a}ki}, {Lamarre}, {Lasenby}, {Lattanzi}, {Lawrence}, {Le
  Jeune}, {Leahy}, {Leonardi}, {Lesgourgues}, {Levrier}, {Liguori}, {Lilje},
  {Linden-V{\o}rnle}, {L{\'o}pez-Caniego}, {Lubin}, {Mac{\'\i}as-P{\'e}rez},
  {Maggio}, {Maino}, {Mandolesi}, {Mangilli}, {Maris}, {Marshall}, {Martin},
  {Mart{\'\i}nez-Gonz{\'a}lez}, {Masi}, {Matarrese}, {McGehee}, {Meinhold},
  {Melchiorri}, {Mendes}, {Mennella}, {Migliaccio}, {Mitra},
  {Miville-Desch{\^e}nes}, {Moneti}, {Montier}, {Morgante}, {Mortlock}, {Moss},
  {Munshi}, {Murphy}, {Naselsky}, {Nati}, {Natoli}, {Netterfield},
  {N{\o}rgaard-Nielsen}, {Noviello}, {Novikov}, {Novikov}, {Orlando},
  {Oxborrow}, {Paci}, {Pagano}, {Pajot}, {Paladini}, {Paoletti}, {Partridge},
  {Pasian}, {Patanchon}, {Pearson}, {Perdereau}, {Perotto}, {Perrotta},
  {Pettorino}, {Piacentini}, {Piat}, {Pierpaoli}, {Pietrobon}, {Plaszczynski},
  {Pointecouteau}, {Polenta}, {Pratt}, {Pr{\'e}zeau}, {Prunet}, {Puget},
  {Rachen}, {Reach}, {Rebolo}, {Reinecke}, {Remazeilles}, {Renault}, {Renzi},
  {Ristorcelli}, {Rocha}, {Rosset}, {Rossetti}, {Roudier},
  {Rubi{\~n}o-Mart{\'\i}n}, {Rusholme}, {Sandri}, {Santos}, {Savelainen},
  {Savini}, {Scott}, {Seiffert}, {Shellard}, {Spencer}, {Stolyarov}, {Stompor},
  {Strong}, {Sudiwala}, {Sunyaev}, {Sutton}, {Suur-Uski}, {Sygnet}, {Tauber},
  {Terenzi}, {Toffolatti}, {Tomasi}, {Tristram}, {Tucci}, {Tuovinen}, {Umana},
  {Valenziano}, {Valiviita}, {Van Tent}, {Vielva}, {Villa}, {Wade}, {Wandelt},
  {Wehus}, {Wilkinson}, {Yvon}, {Zacchei}, \& {Zonca}}]{freefree}
{Planck Collaboration}, {Adam}, R., {Ade}, P.~A.~R., {et~al.}
  2016{\natexlab{a}}, Astronomy and Astrophysics, 594, A10

\bibitem[{{Planck Collaboration} {et~al.}(2016{\natexlab{b}}){Planck
  Collaboration}, {Aghanim, N.}, {Ashdown, M.}, {Aumont, J.}, {Baccigalupi,
  C.}, {Ballardini, M.}, {Banday, A. J.}, {Barreiro, R. B.}, {Bartolo, N.},
  {Basak, S.}, {Benabed, K.}, {Bernard, J.-P.}, {Bersanelli, M.}, {Bielewicz,
  P.}, {Bonavera, L.}, {Bond, J. R.}, {Borrill, J.}, {Bouchet, F. R.},
  {Boulanger, F.}, {Burigana, C.}, {Calabrese, E.}, {Cardoso, J.-F.}, {Carron,
  J.}, {Chiang, H. C.}, {Colombo, L. P. L.}, {Comis, B.}, {Couchot, F.},
  {Coulais, A.}, {Crill, B. P.}, {Curto, A.}, {Cuttaia, F.}, {de Bernardis,
  P.}, {de Zotti, G.}, {Delabrouille, J.}, {Di Valentino, E.}, {Dickinson, C.},
  {Diego, J. M.}, {Dor\'e, O.}, {Douspis, M.}, {Ducout, A.}, {Dupac, X.},
  {Dusini, S.}, {Elsner, F.}, {En\ss{}lin, T. A.}, {Eriksen, H. K.},
  {Falgarone, E.}, {Fantaye, Y.}, {Finelli, F.}, {Forastieri, F.}, {Frailis,
  M.}, {Fraisse, A. A.}, {Franceschi, E.}, {Frolov, A.}, {Galeotta, S.},
  {Galli, S.}, {Ganga, K.}, {G\'enova-Santos, R. T.}, {Gerbino, M.}, {Ghosh,
  T.}, {Giraud-H\'eraud, Y.}, {Gonz\'alez-Nuevo, J.}, {G\'orski, K. M.},
  {Gruppuso, A.}, {Gudmundsson, J. E.}, {Hansen, F. K.}, {Helou, G.},
  {Henrot-Versill\'e, S.}, {Herranz, D.}, {Hivon, E.}, {Huang, Z.}, {Jaffe, A.
  H.}, {Jones, W. C.}, {Keih\"anen, E.}, {Keskitalo, R.}, {Kiiveri, K.},
  {Kisner, T. S.}, {Krachmalnicoff, N.}, {Kunz, M.}, {Kurki-Suonio, H.},
  {Lamarre, J.-M.}, {Langer, M.}, {Lasenby, A.}, {Lattanzi, M.}, {Lawrence, C.
  R.}, {Le Jeune, M.}, {Levrier, F.}, {Lilje, P. B.}, {Lilley, M.}, {Lindholm,
  V.}, {L\'opez-Caniego, M.}, {Ma, Y.-Z.}, {Mac\'{\i}as-P\'erez, J. F.},
  {Maggio, G.}, {Maino, D.}, {Mandolesi, N.}, {Mangilli, A.}, {Maris, M.},
  {Martin, P. G.}, {Mart\'{\i}nez-Gonz\'alez, E.}, {Matarrese, S.}, {Mauri,
  N.}, {McEwen, J. D.}, {Melchiorri, A.}, {Mennella, A.}, {Migliaccio, M.},
  {Miville-Desch\^enes, M.-A.}, {Molinari, D.}, {Moneti, A.}, {Montier, L.},
  {Morgante, G.}, {Moss, A.}, {Natoli, P.}, {Oxborrow, C. A.}, {Pagano, L.},
  {Paoletti, D.}, {Patanchon, G.}, {Perdereau, O.}, {Perotto, L.}, {Pettorino,
  V.}, {Piacentini, F.}, {Plaszczynski, S.}, {Polastri, L.}, {Polenta, G.},
  {Puget, J.-L.}, {Rachen, J. P.}, {Racine, B.}, {Reinecke, M.}, {Remazeilles,
  M.}, {Renzi, A.}, {Rocha, G.}, {Rosset, C.}, {Rossetti, M.}, {Roudier, G.},
  {Rubi\~no-Mart\'{\i}n, J. A.}, {Ruiz-Granados, B.}, {Salvati, L.}, {Sandri,
  M.}, {Savelainen, M.}, {Scott, D.}, {Sirignano, C.}, {Sirri, G.}, {Soler, J.
  D.}, {Spencer, L. D.}, {Suur-Uski, A.-S.}, {Tauber, J. A.}, {Tavagnacco, D.},
  {Tenti, M.}, {Toffolatti, L.}, {Tomasi, M.}, {Tristram, M.}, {Trombetti, T.},
  {Valiviita, J.}, {Van Tent, F.}, {Vielva, P.}, {Villa, F.}, {Vittorio, N.},
  {Wandelt, B. D.}, {Wehus, I. K.}, {Zacchei, A.}, \& {Zonca, A.}}]{GNILC}
{Planck Collaboration}, {Aghanim, N.}, {Ashdown, M.}, {et~al.}
  2016{\natexlab{b}}, Astronomy and Astrophysics, 596, A109

\bibitem[{{Popescu} {et~al.}(2017){Popescu}, {Yang}, {Tuffs}, {Natale},
  {Rushton}, \& {Aharonian}}]{Popescu:2017}
{Popescu}, C.~C., {Yang}, R., {Tuffs}, R.~J., {et~al.} 2017, \mnras, 470, 2539

\bibitem[{{Principe} {et~al.}(2020){Principe}, {Mitchell}, {Caroff}, {Hinton},
  {Parsons}, \& {Funk}}]{principe2020}
{Principe}, G., {Mitchell}, A.~M.~W., {Caroff}, S., {et~al.} 2020, \aap, 640,
  A76

\bibitem[{{Protassov} {et~al.}(2002){Protassov}, {van Dyk}, {Connors},
  {Kashyap}, \& {Siemiginowska}}]{protassov2002}
{Protassov}, R., {van Dyk}, D.~A., {Connors}, A., {Kashyap}, V.~L., \&
  {Siemiginowska}, A. 2002, \apj, 571, 545

\bibitem[{{Ray} {et~al.}(2011){Ray}, {Kerr}, {Parent}, {Abdo}, {Guillemot},
  {Ransom}, {Rea}, {Wolff}, {Makeev}, {Roberts}, {Camilo}, {Dormody}, {Freire},
  {Grove}, {Gwon}, {Harding}, {Johnston}, {Keith}, {Kramer}, {Michelson},
  {Romani}, {Saz Parkinson}, {Thompson}, {Weltevrede}, {Wood}, \&
  {Ziegler}}]{ray2011}
{Ray}, P.~S., {Kerr}, M., {Parent}, D., {et~al.} 2011, \apjs, 194, 17

\bibitem[{{Remy} {et~al.}(2017){Remy}, {Grenier}, {Marshall}, \&
  {Casandjian}}]{remy2017}
{Remy}, Q., {Grenier}, I.~A., {Marshall}, D.~J., \& {Casandjian}, J.~M. 2017,
  \aap, 601, A78

\bibitem[{{Saz Parkinson} {et~al.}(2010){Saz Parkinson}, {Dormody}, {Ziegler},
  {Ray}, {Abdo}, {Ballet}, {Baring}, {Belfiore}, {Burnett}, {Caliandro},
  {Camilo}, {Caraveo}, {de Luca}, {Ferrara}, {Freire}, {Grove}, {Gwon},
  {Harding}, {Johnson}, {Johnson}, {Johnston}, {Keith}, {Kerr},
  {Kn{\"o}dlseder}, {Makeev}, {Marelli}, {Michelson}, {Parent}, {Ransom},
  {Reimer}, {Romani}, {Smith}, {Thompson}, {Watters}, {Weltevrede}, {Wolff}, \&
  {Wood}}]{saz-parkinson2010}
{Saz Parkinson}, P.~M., {Dormody}, M., {Ziegler}, M., {et~al.} 2010, \apj, 725,
  571

\bibitem[{{Schlickeiser}(2002)}]{Schlickeiser:2002}
{Schlickeiser}, R. 2002, {Cosmic Ray Astrophysics}, ed. {Schlickeiser, R.}
  (Springer)

\bibitem[{{Tang}(2019)}]{Tang2019}
{Tang}, X. 2019, \mnras, 482, 3843

\bibitem[{{Tatischeff} {et~al.}(2021){Tatischeff}, {Raymond}, {Duprat},
  {Gabici}, \& {Recchia}}]{tatischeff2021}
{Tatischeff}, V., {Raymond}, J.~C., {Duprat}, J., {Gabici}, S., \& {Recchia},
  S. 2021, \mnras, 508, 1321

\bibitem[{{Taylor} {et~al.}(2003){Taylor}, {Gibson}, {Peracaula}, {Martin},
  {Landecker}, {Brunt}, {Dewdney}, {Dougherty}, {Gray}, {Higgs}, {Kerton},
  {Knee}, {Kothes}, {Purton}, {Uyaniker}, {Wallace}, {Willis}, \&
  {Durand}}]{CGPS}
{Taylor}, A.~R., {Gibson}, S.~J., {Peracaula}, M., {et~al.} 2003, Astronomical
  Journal, 125, 3145

\bibitem[{{Tibaldo} {et~al.}(2015){Tibaldo}, {Digel}, {Casandjian},
  {Franckowiak}, {Grenier}, {J{\'o}hannesson}, {Marshall}, {Moskalenko},
  {Negro}, {Orlando}, {Porter}, {Reimer}, \& {Strong}}]{Tibaldo2015}
{Tibaldo}, L., {Digel}, S.~W., {Casandjian}, J.~M., {et~al.} 2015,
  Astrophysical Journal, 807, 161

\bibitem[{{Tibaldo} {et~al.}(2021){Tibaldo}, {Gaggero}, \&
  {Martin}}]{tibaldo2021}
{Tibaldo}, L., {Gaggero}, D., \& {Martin}, P. 2021, Universe, 7, 141

\bibitem[{{Tolksdorf} {et~al.}(2019){Tolksdorf}, {Grenier}, {Joubaud}, \&
  {Schlickeiser}}]{tolksdorf2019}
{Tolksdorf}, T., {Grenier}, I.~A., {Joubaud}, T., \& {Schlickeiser}, R. 2019,
  \apj, 879, 66

\bibitem[{{Trotta} {et~al.}(2011){Trotta}, {J{\'o}hannesson}, {Moskalenko},
  {Porter}, {Ruiz de Austri}, \& {Strong}}]{Trotta:2011}
{Trotta}, R., {J{\'o}hannesson}, G., {Moskalenko}, I.~V., {et~al.} 2011, \apj,
  729, 106

\bibitem[{{Tutone} {et~al.}(2021){Tutone}, {Ballet}, {Acero}, {D'A{\`\i}}, \&
  {Cusumano}}]{Tutone21}
{Tutone}, A., {Ballet}, J., {Acero}, F., {D'A{\`\i}}, A., \& {Cusumano}, G.
  2021, \aap, 656, A139

\bibitem[{{Uyan{\i}ker} {et~al.}(2001){Uyan{\i}ker}, {F{\"u}rst}, {Reich},
  {Aschenbach}, \& {Wielebinski}}]{uyaniker01}
{Uyan{\i}ker}, B., {F{\"u}rst}, E., {Reich}, W., {Aschenbach}, B., \&
  {Wielebinski}, R. 2001, \aap, 371, 675

\bibitem[{{van der Velden}(2020)}]{VdVelden2020}
{van der Velden}, E. 2020, The Journal of Open Source Software, 5, 2004

\bibitem[{{Vieu} {et~al.}(2022){Vieu}, {Gabici}, {Tatischeff}, \&
  {Ravikularaman}}]{vieu2022}
{Vieu}, T., {Gabici}, S., {Tatischeff}, V., \& {Ravikularaman}, S. 2022,
  \mnras, 512, 1275

\bibitem[{{Vink} {et~al.}(2000){Vink}, {de Koter}, \& {Lamers}}]{vink2000}
{Vink}, J.~S., {de Koter}, A., \& {Lamers}, H.~J.~G.~L.~M. 2000, \aap, 362, 295

\bibitem[{Virtanen {et~al.}(2020)Virtanen, Gommers, Oliphant, Haberland, Reddy,
  Cournapeau, Burovski, Peterson, Weckesser, Bright, {van der Walt}, Brett,
  Wilson, Millman, Mayorov, Nelson, Jones, Kern, Larson, Carey, Polat, Feng,
  Moore, {VanderPlas}, Laxalde, Perktold, Cimrman, Henriksen, Quintero, Harris,
  Archibald, Ribeiro, Pedregosa, {van Mulbregt}, \& {SciPy 1.0
  Contributors}}]{Scipy2020}
Virtanen, P., Gommers, R., Oliphant, T.~E., {et~al.} 2020, Nature Methods, 17,
  261

\bibitem[{{Weaver} {et~al.}(1977){Weaver}, {McCray}, {Castor}, {Shapiro}, \&
  {Moore}}]{weaver1977}
{Weaver}, R., {McCray}, R., {Castor}, J., {Shapiro}, P., \& {Moore}, R. 1977,
  \apj, 218, 377

\bibitem[{Weidner \& Kroupa(2004)}]{wk2004}
Weidner, C. \& Kroupa, P. 2004, Monthly Notices of the Royal Astronomical
  Society, 348, 187

\bibitem[{{Wright} {et~al.}(2010){Wright}, {Drake}, {Drew}, \&
  {Vink}}]{wright2010}
{Wright}, N.~J., {Drake}, J.~J., {Drew}, J.~E., \& {Vink}, J.~S. 2010, \apj,
  713, 871

\bibitem[{{Yoast-Hull} {et~al.}(2017){Yoast-Hull}, {Gallagher}, {Halzen},
  {Kheirandish}, \& {Zweibel}}]{yoast-hull2017}
{Yoast-Hull}, T.~M., {Gallagher}, J.~S., {Halzen}, F., {Kheirandish}, A., \&
  {Zweibel}, E.~G. 2017, \prd, 96, 043011

\bibitem[{{Zabalza}(2015)}]{Zabalza:2015}
{Zabalza}, V. 2015, in International Cosmic Ray Conference, Vol.~34, 34th
  International Cosmic Ray Conference (ICRC2015), 922

\bibitem[{{Zhang} {et~al.}(2020){Zhang}, {Chepurnov}, {Yan}, {Makwana},
  {Santos-Lima}, \& {Appleby}}]{plasmamodes}
{Zhang}, H., {Chepurnov}, A., {Yan}, H., {et~al.} 2020, Nature Astronomy, 4,
  1001

\bibitem[{{Zhang} {et~al.}(2021){Zhang}, {Liu}, {Chen}, \& {Wang}}]{Zhang:2021}
{Zhang}, Y., {Liu}, R.-Y., {Chen}, S.~Z., \& {Wang}, X.-Y. 2021, \apj, 922, 130

\end{thebibliography}

\begin{appendix}

\section{Data analysis}
\label{app:analysis}

We provide in this appendix a series of technical details about the data analysis.

\subsection{Preliminary model optimisation}\label{app:prelopt}

The preliminary optimisation of the emission model went through the following steps:
\begin{enumerate}
\item  a simultaneous fit of the normalisation of bright sources with $TS > 10^4$ and predicted photons counts $> 500$;
\item an iterative fit of the normalisation and spectral shape of all the sources in the ROI by order of intensity and significance (method \texttt{optimize} of \texttt{Fermipy});
\item a further simultaneous fit of the normalisation of sources with $TS > 10^4$ and predicted photons counts $> 500$, and also of the spectral shape of sources with the same $TS$ condition and predicted photons counts $> 1000$.
\end{enumerate}

The thresholds on predicted photon counts and $TS$ were chosen to optimise all parameters for the brightest sources in the ROI (three pulsars, Cygnus Loop, $\gamma$~Cygni). We are more restrictive for the spectral shape to reduce the number of free parameters in the analysis, which otherwise can become unstable.

\subsection{Morphological fits}\label{app:morph}

In the morphological characterisation stages, the optimisation of the related components is performed in a two-step process.
\begin{enumerate}
\item Extension optimisation: we perform a first scan over a coarse grid of extension values, followed by a second finer scan around the first optimum and then the fit of a parabola to determine the final extension value and its uncertainty.
\item Position optimisation: we compute a map of log-likelihood values over a position grid of $2\degr\times 2\degr$ with a $0.2\degr$ binning and determine the best-fit position and its uncertainty from the fit of an ellipse.
\end{enumerate}

\subsection{Spectral models}\label{app:specmodels}

In the spectral characterisation of the various components of the cocoon, we consider the following models: a simple power law (PL) of expression
\begin{equation}
    \frac{dN}{dE} = N_0 \times \left( \frac{E}{E_0} \right)^{-\gamma},
\end{equation}
with $N_0$ flux at the reference energy $E_0$ and $\gamma$ spectral index; a log-parabola (LP) of expression
\begin{equation}
     \frac{dN}{dE} = N_0 \times \left( \frac{E}{E_0} \right)^{-\left[\alpha + \beta \ln \left( \frac{E}{E_0} \right) \right]},
\end{equation}
with $N_0$ flux at the reference energy $E_0$, $\alpha$ slope parameter, and $\beta$ curvature parameter; and a smooth broken power law (SBPL) of expression
\begin{equation}
     \frac{dN}{dE} = N_0 \times \left( \frac{E}{E_0} \right)^{-\gamma_1} \left[ 1 + \left( \frac{E}{E_b} \right)^{\frac{\gamma_2 - \gamma_1}{\kappa}} \right]^{-\kappa},
\end{equation}
with $N_0$ flux at the reference energy $E_0$, $E_b$ break energy, $\gamma_1$ spectral index at energies $\ll E_b$, $\gamma_2$ spectral index at energies $\gg E_b$, and $\kappa$ the smoothing parameter fixed to 0.2.

\section{Additional results on the spectro-morphological analysis }
\label{app:TSvar}

\begin{figure}[!h]
    \centering
    \includegraphics[width=0.5\textwidth]{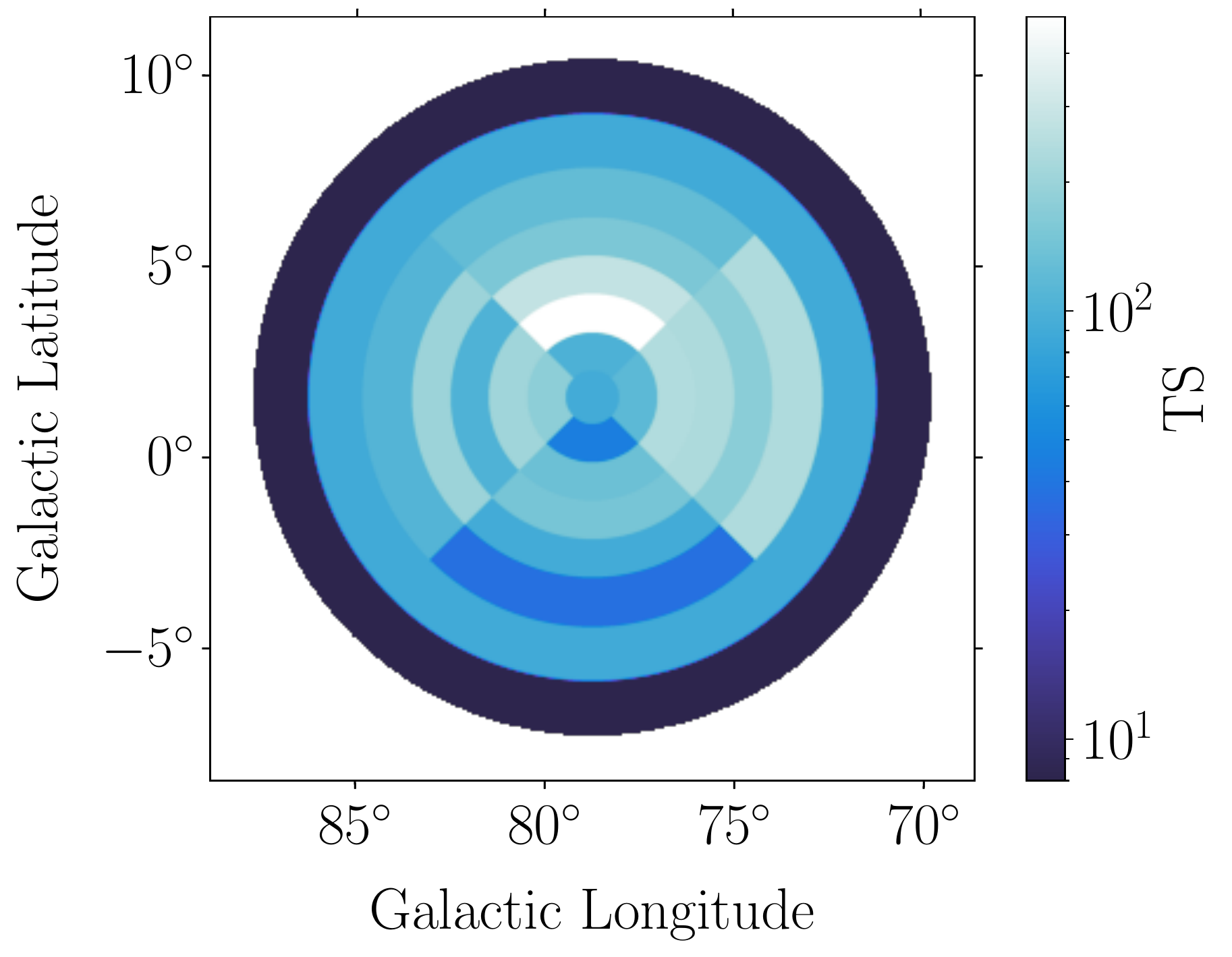}
    \caption{TS values in rings, segments, and disk for the best decomposition of \halo.}
    \label{fig:TSvar}
\end{figure}

Figure~\ref{fig:TSvar} shows the TS values for the best decomposition of \NNhalo, as in step D described in Section ~\ref{sec:rings}. The figure illustrates how the decomposition was designed to conserve a minimum TS of $25$ except for the largest ring where the requirement could not be met. 

\begin{figure}[!h]
    \centering
    \includegraphics[width=0.5\textwidth]{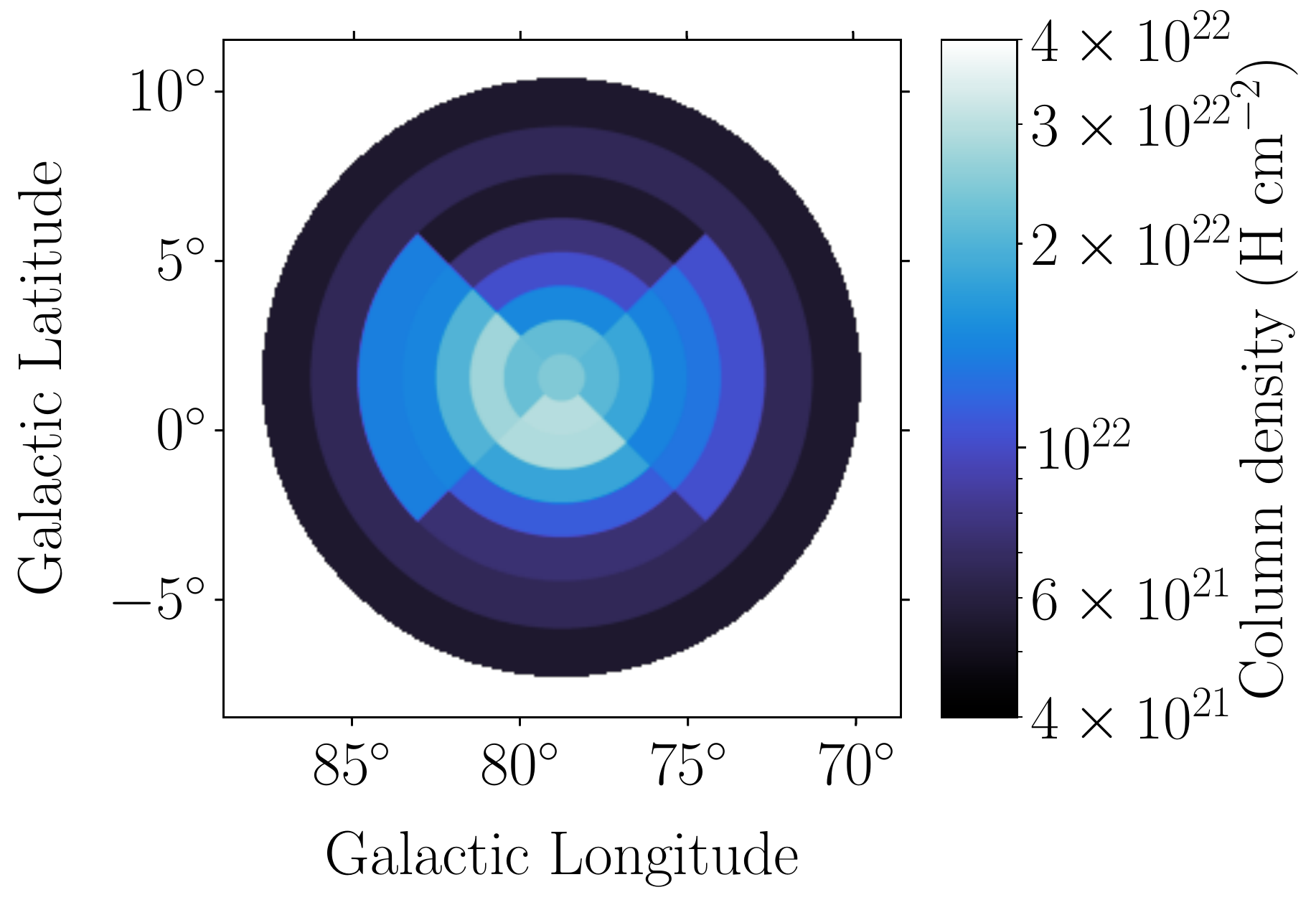}
    \caption{Total neutral gas column density in the local arm as in Figure~\ref{fig:totcoldens} reprojected onto the disk, rings, and segments used for the spectro-morphological analysis in section~\ref{sec:rings}.}
    \label{fig:ringsgas}
\end{figure}

We used the intensities derived in Section ~\ref{sec:rings} to derive emissivities. To do so, we divided the flux associated with each segment, ring, or disk by the total neutral gas column density in the local arm (atomic, molecular and DNM) integrated over solid angle for each segment, ring, or disk area (shown in Figure~\ref{fig:ringsgas}). 
See Section ~\ref{sec:discussion:landscape} for a discussion on uncertainties in the gas column densities relevant for the emissivity computation.

\begin{figure}[!htb]
    \centering
    \includegraphics[width=0.5\textwidth]{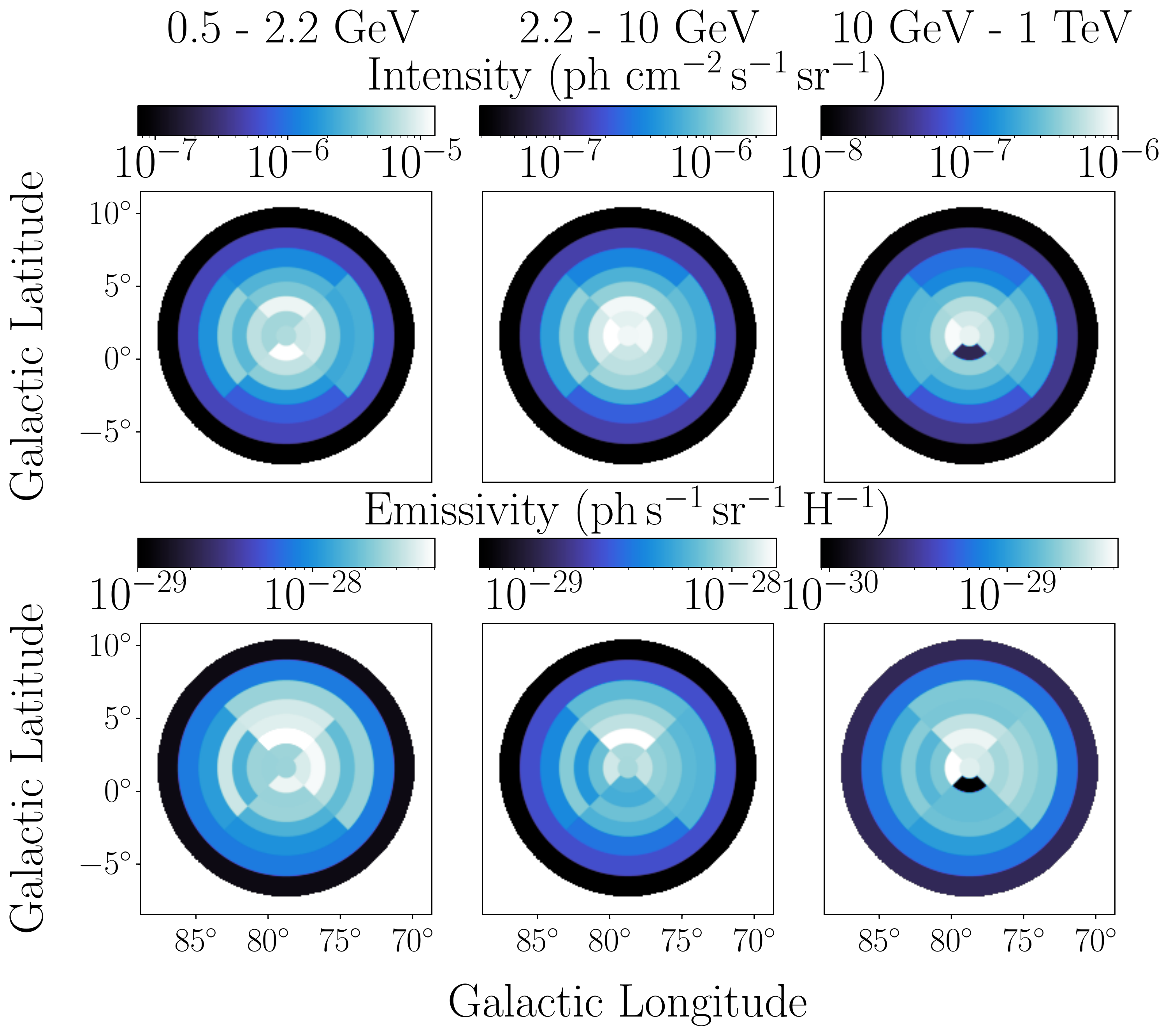}
    \caption{Intensity and emissivity maps in three different energy ranges for component \halo, according to the decomposition in rings and segments of step D (see Section ~\ref{sec:rings} for details).}
    \label{fig:ringsmaps}
\end{figure}

Figure~\ref{fig:ringsmaps} shows maps of intensities and emissivities for \NNhalo as decomposed in step D in section and in three different energy bands $0.5 - 2.236$ GeV, $2.236 - 10$ GeV and $10 - 1000$ GeV. Radial profiles of intensity and emissivity in the three energy bands are shown in Section  \ref{sec:discussion}, where they are used for quantitative interpretation of the results.

\FloatBarrier

\section{Diffusion-loss model framework}
\label{app:framework}

In Section  \ref{sec:discussion:model}, we introduce a simple diffusion-loss framework to account for the observed properties of the Cygnus cocoon. We provide here the full formalism of this framework.

The transport equation governing the evolution of the particle distribution $f$ in momentum $p$, position $r$, and time $t$ is:
\begin{equation}
\frac{\partial f}{\partial t} = \nabla \cdot (D \, . \, \nabla f ) - \frac{\partial}{\partial p} \left[ \dot{p} \, . \, f \right] + Q,
\end{equation}
where $D$ is a spatial diffusion coefficient, $\dot{p}$ a momentum loss term, and $Q$ a source term. The momentum loss term $\dot{p}$ includes losses that are uniform in space and arise from radiative processes, hadronic interactions for accelerated protons, and Bremsstrahlung, inverse-Compton scattering, and synchrotron radiation for accelerated electrons \citep{Schlickeiser:2002}. The source term $Q(p,t)$ is assumed to be point-like in space and to have a constant power-law with exponential cut-off spectral shape:
\begin{equation}
Q(p,t) = Q_0 \, . \, \left( \frac{p}{10\mathrm{\,GeV/c}} \right)^{-\alpha} \, . \, e^{-p / p_{\rm cut}} \, . \, e^{-t / t_{\rm inj}}.
\end{equation}
Particles are injected with a power law spectrum in momentum with index $\alpha$. The cut-off momentum is set to a high value of 1\,PeV/$c$ that our gamma-ray data are not sensitive to. To investigate the possibility that injection is not constant in time, and may have occurred over a finite duration some time ago followed by a longer diffusion time, we implemented (somewhat arbitrarily) an exponential decay of the source term. Integrating the source term $Q(p,t)$ over particle energies yields the time-dependent injection luminosity $L_{\rm inj}(t)$.
The diffusion coefficient $D(p)$ assumed to be constant in space and time is defined as:
\begin{equation}
D(p) = \beta \, . \, D_0 \, . \, \left( \frac{p}{10\mathrm{\,GeV/c}} \right)^\delta \mathrm{\quad with \quad} \delta = 1/3,
\end{equation}
where $\beta=v/c$ with $v$ the velocity of a particle and $c$ the velocity of light. The power-law dependence in momentum, with an index corresponding to a Kolmogorov scaling for the magnetic turbulence spectrum, is adapted from models of large-scale CR propagation in our Galaxy \citep{Trotta:2011,Orlando:2018}. We note that alternative expressions were considered recently in the light of more accurate direct CR measurements at Earth \citep{Evoli:2019,Genolini:2019}, and that the considered deviations from a pure power law may have consequences in the energy range we are interested in here.

The solution to the transport equation is obtained as \citep{Atoyan:1995}:
\begin{equation}
f(p,r,t) = \int_{{\rm max}[0,\,t_{\rm diff}-t_{\rm cool}(p_{\rm cut}, p)]}^{t_{\rm diff}} \frac{\dot{p}(p_0)}{\dot{p}(p)} \, . \, \frac{Q(p_0,t_0)}{\pi^{3/2} r_d^3} \, . \, e^{-r^2/r_d^2} \, . \, dt_0, \\
\label{eq:soleq}
\end{equation}
with $r_d$ diffusion distance. 
For a present-day momentum $p$, the integration runs either over the full injection and transport history, or over the recent period spanning a cooling time $t_{\rm cool}$ from the cut-off momentum down to $p$, with cooling time
\begin{equation}
t_{\rm cool}(p_0,p) = \int_{p}^{p_0} \frac{-dp}{\dot{p}}.
\label{eq:tcool}
\end{equation}

This three-dimensional spherically symmetric distribution of particles is integrated along the line of sight for any angular offset from the centre $\theta$ and given the distance $d$ to the source:
\begin{equation}
\Phi(p,\theta,t) = 2 \, . \, d^2  \, . \,  \int_{0}^{\infty}{f \left( p,\sqrt{\theta^2 d^2+\ell^2},t \right) d\ell}.
\end{equation}
The resulting angular distribution of particles is then used to compute non-thermal emissions at any point in the region of interest. To this aim, we used the naima package \citep{Zabalza:2015} in the approximation of isotropic radiation fields in the case of inverse-Compton scattering and using a nuclear enhancement factor of 1.845 in the case of pion decay \citep{Mori:2009}.

\section{Possible diffusion-loss scenarios}
\label{app:scenarios}

\begin{figure*}[!]
\centering
    \subfloat{\includegraphics[width=0.45\textwidth]{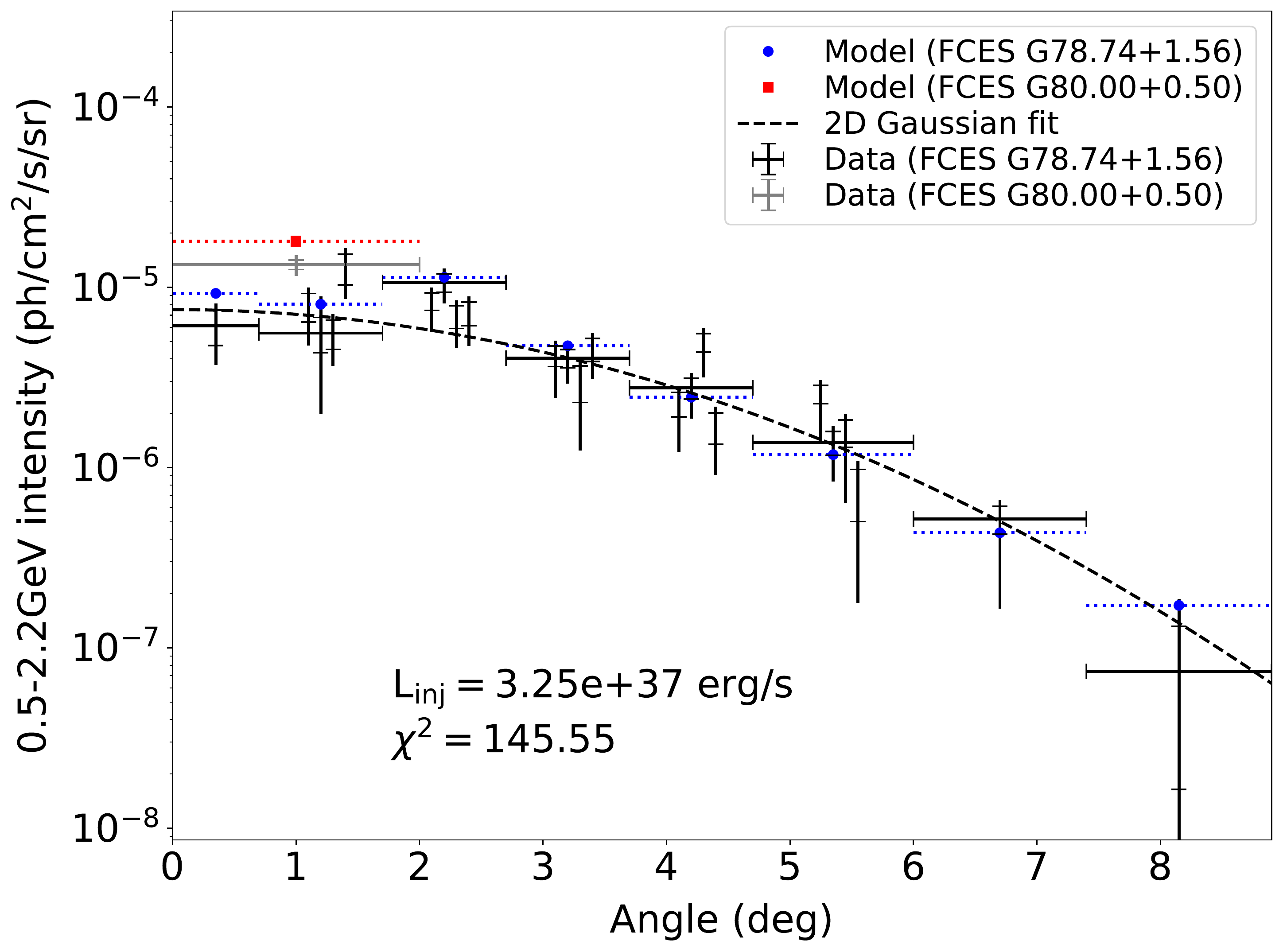}}
    \qquad
    \subfloat{\includegraphics[width=0.45\textwidth]{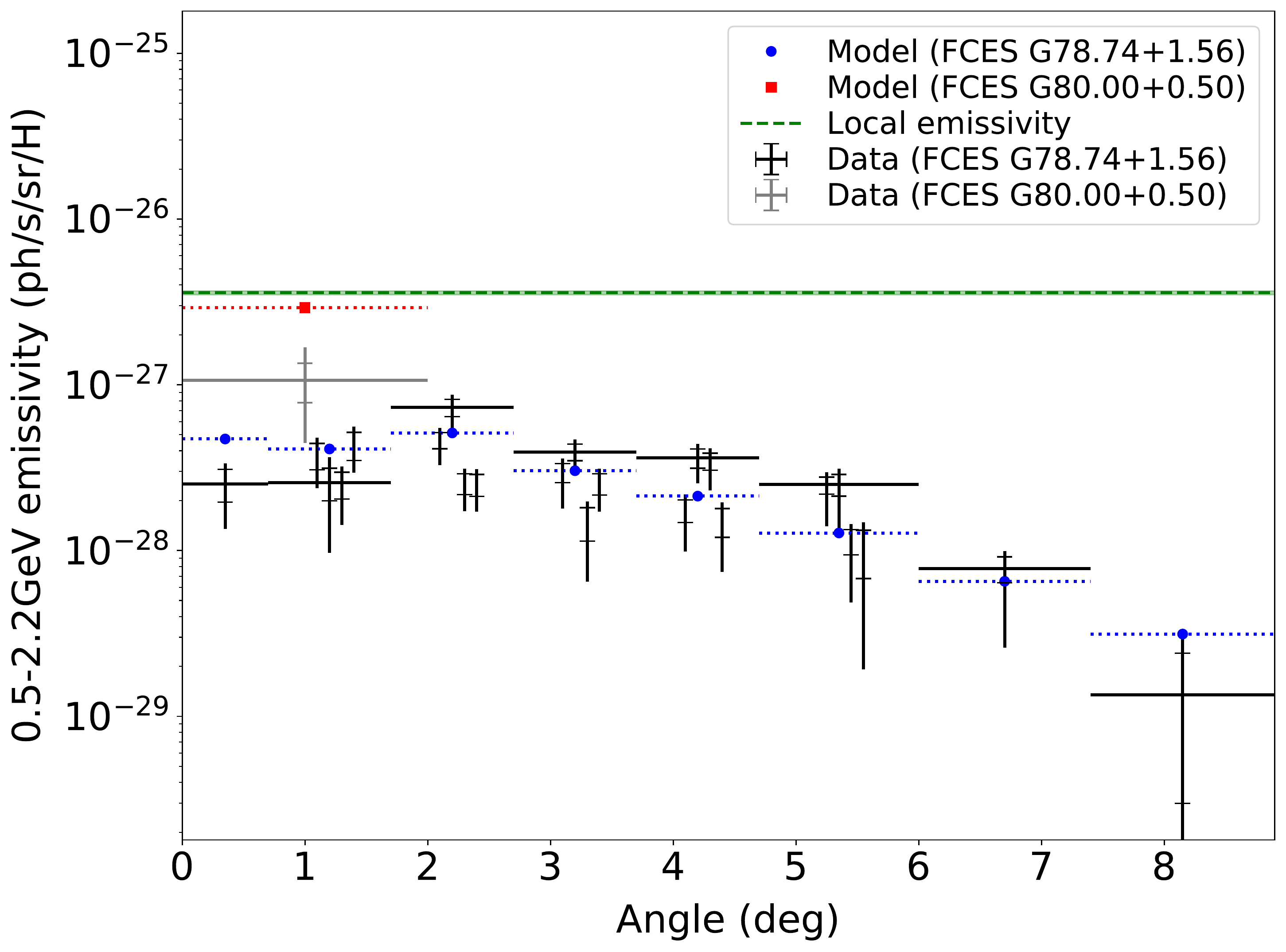}}
    \qquad
    \subfloat{\includegraphics[width=0.45\textwidth]{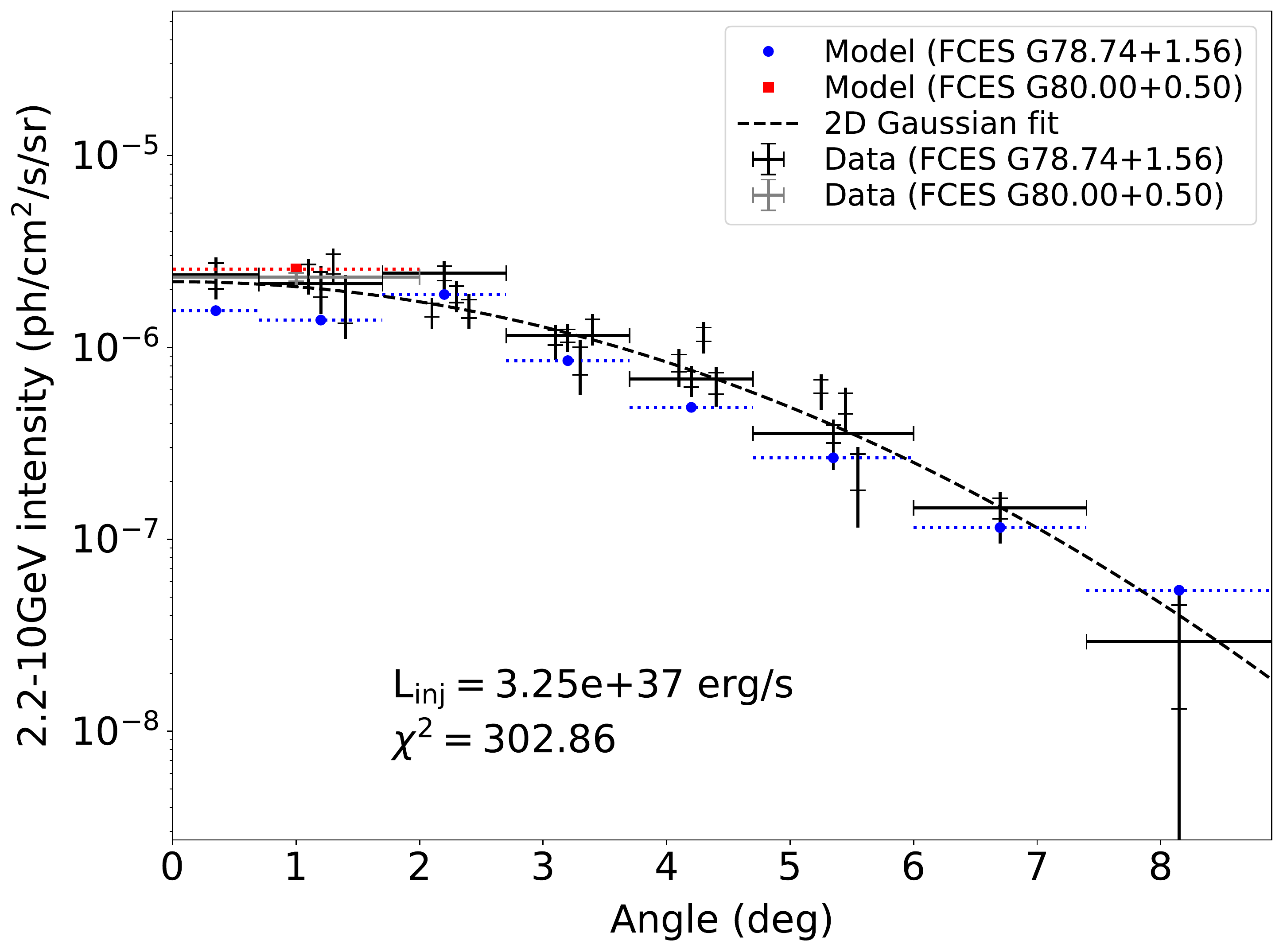}}
    \qquad
    \subfloat{\includegraphics[width=0.45\textwidth]{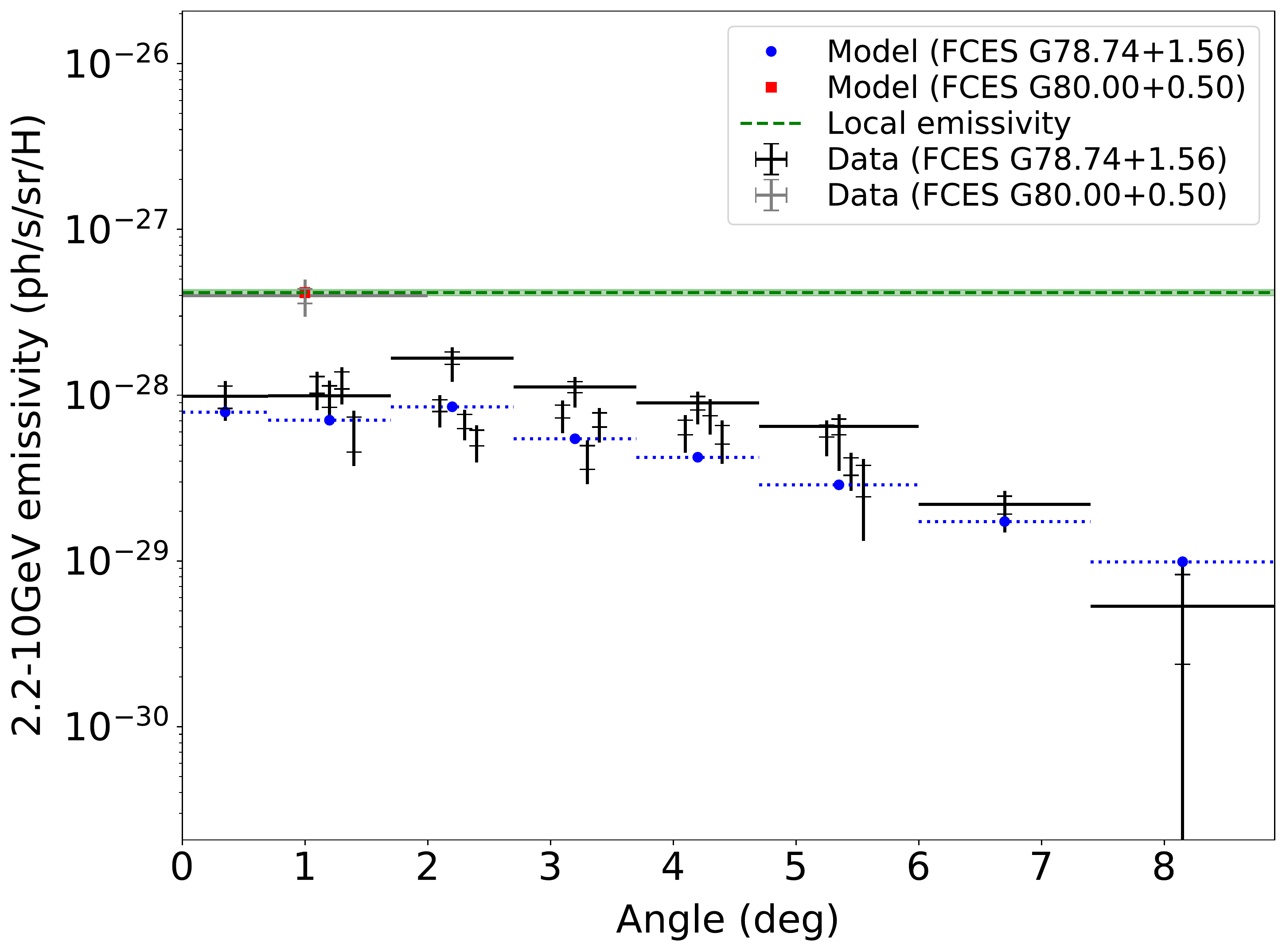}}
    \qquad
    \subfloat{\includegraphics[width=0.45\textwidth]{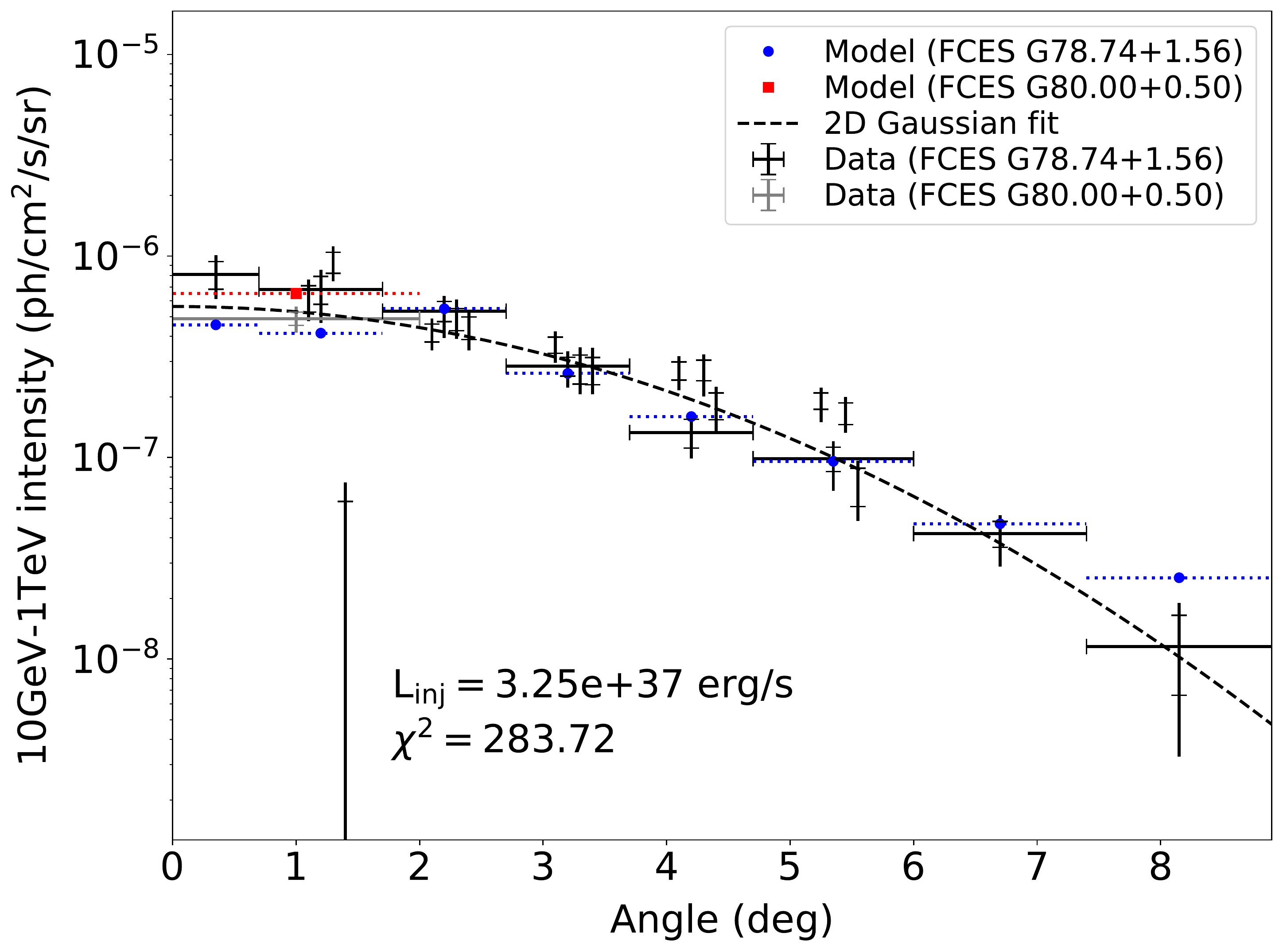}}
    \qquad
    \subfloat{\includegraphics[width=0.45\textwidth]{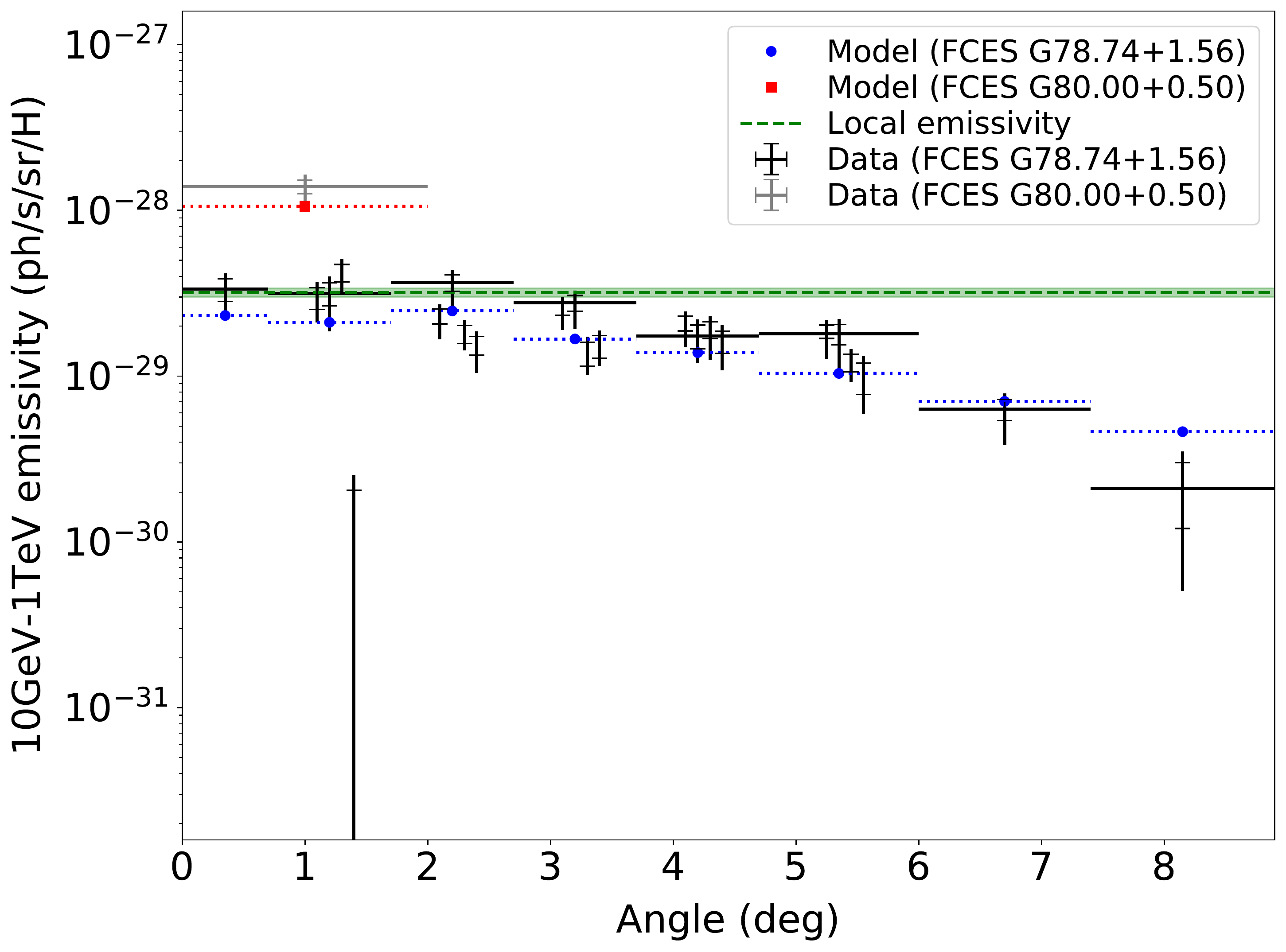}}
    \caption{Same as Figure \ref{fig:diffmodel:hadro:intemiss:H1}, for model setup H2.}
\label{fig:diffmodel:hadro:intemiss:H2}
\end{figure*}

\begin{figure*}[!]
\centering
    \subfloat{\includegraphics[width=0.45\textwidth]{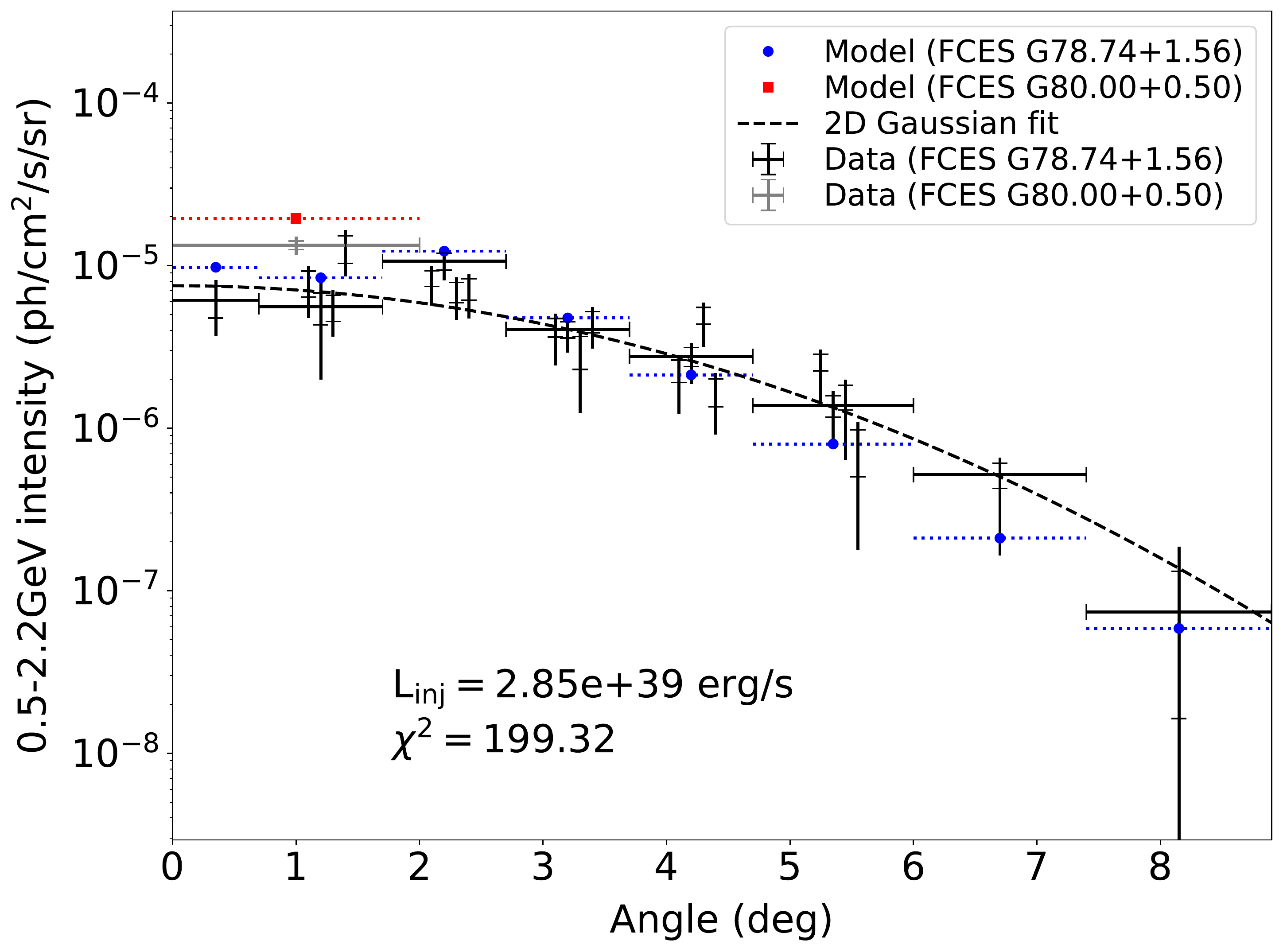}}
    \qquad
    \subfloat{\includegraphics[width=0.45\textwidth]{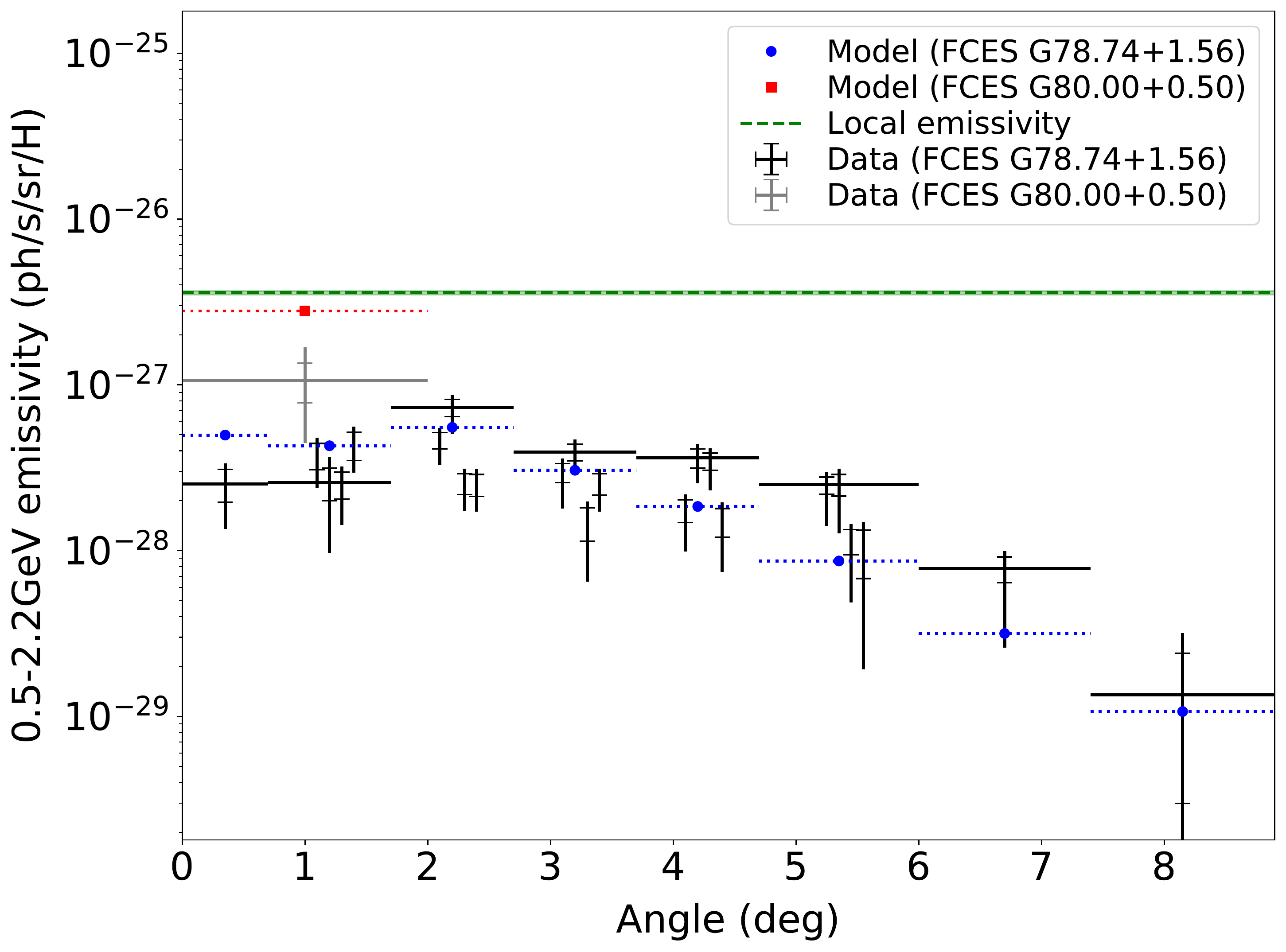}}
    \qquad
    \subfloat{\includegraphics[width=0.45\textwidth]{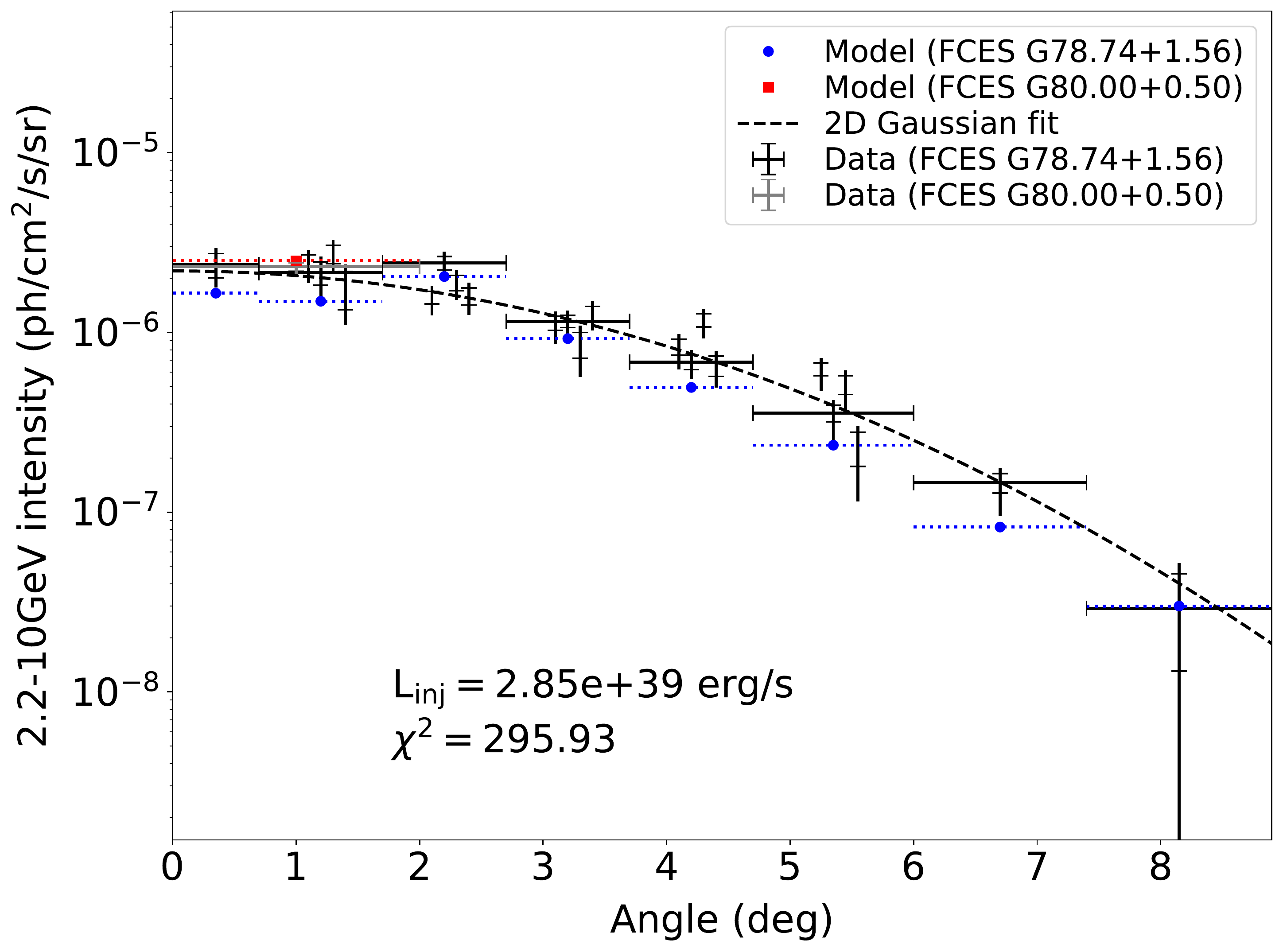}}
    \qquad
    \subfloat{\includegraphics[width=0.45\textwidth]{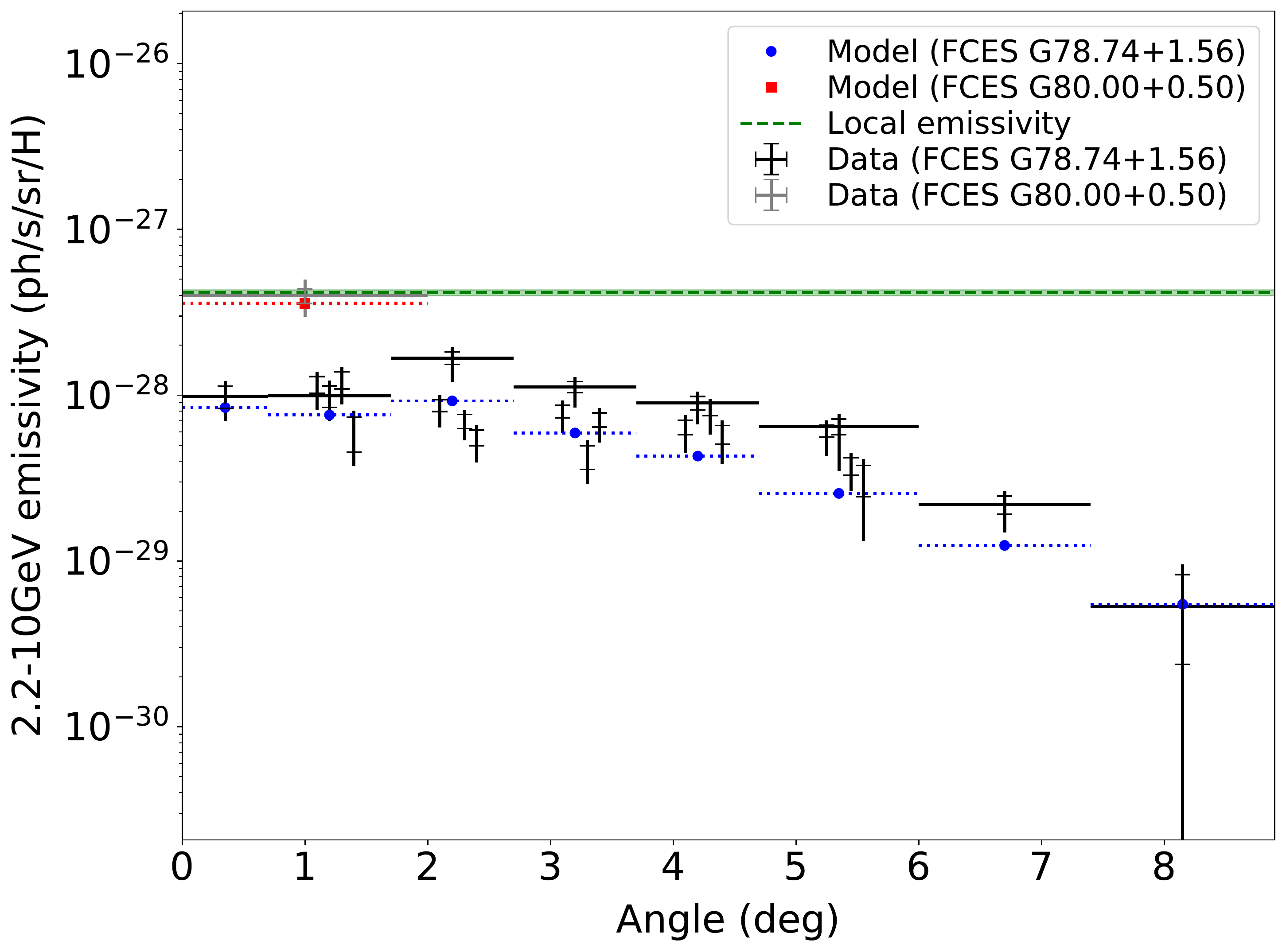}}
    \qquad
    \subfloat{\includegraphics[width=0.45\textwidth]{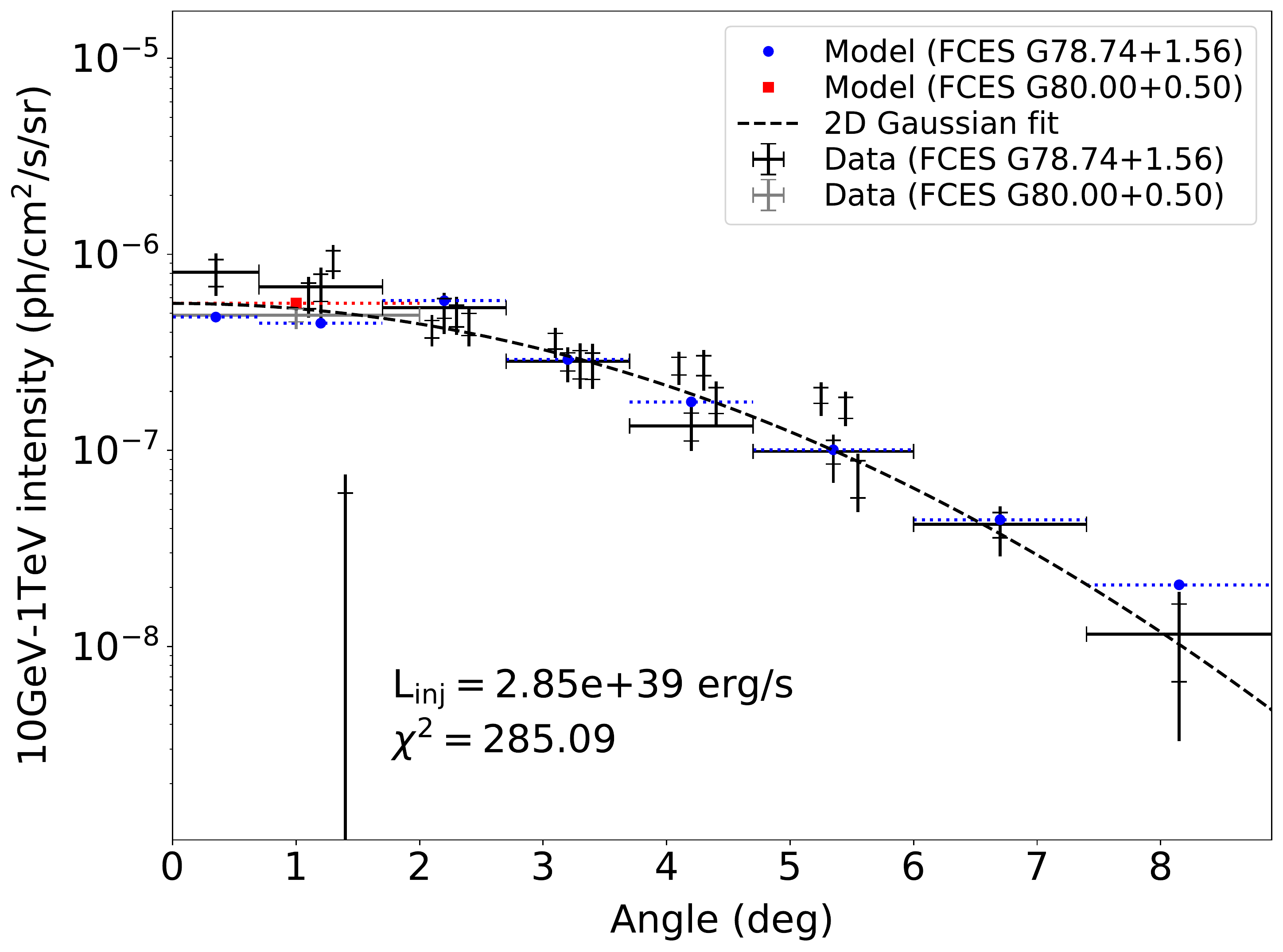}}
    \qquad
    \subfloat{\includegraphics[width=0.45\textwidth]{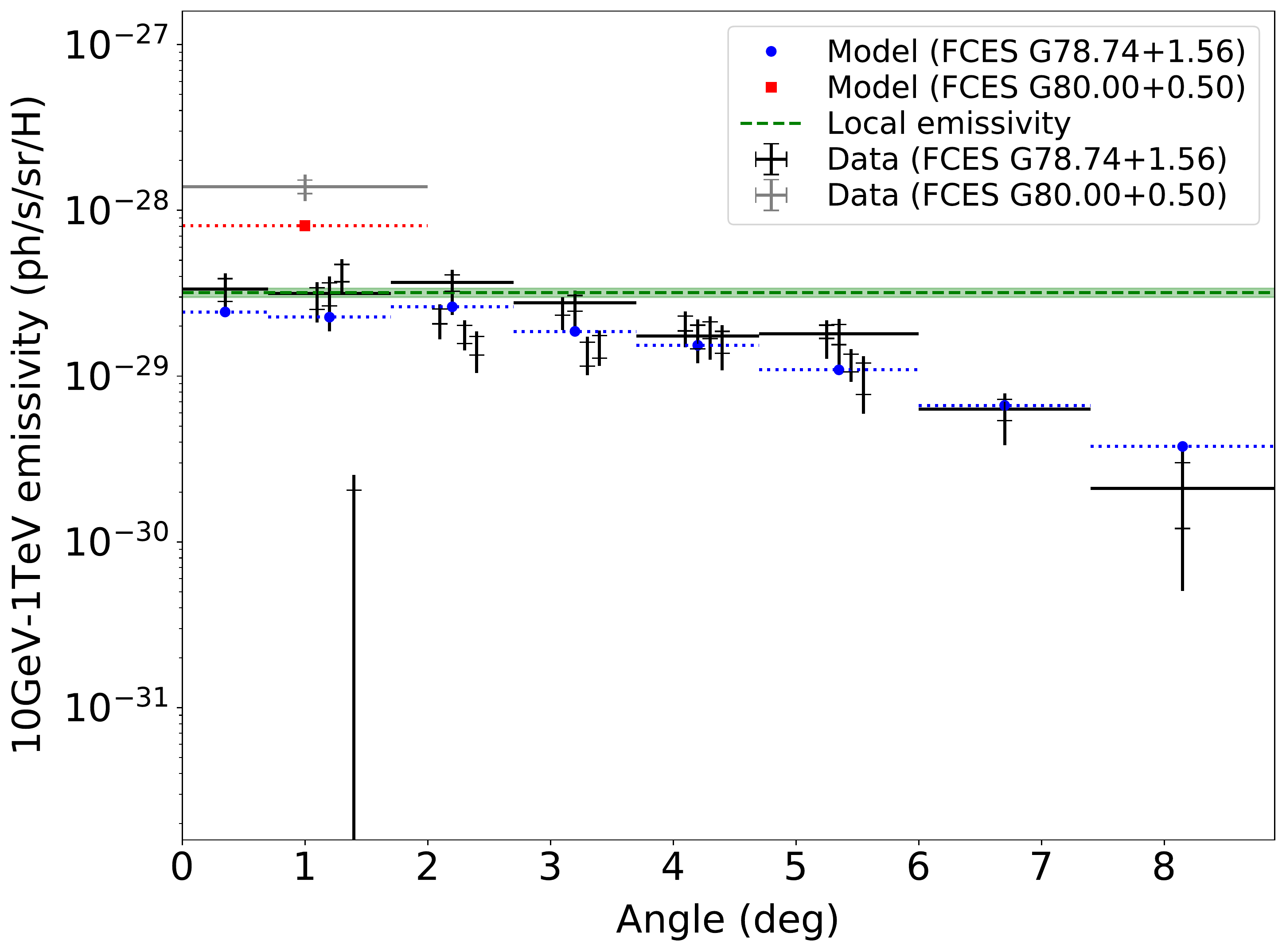}}
    \caption{Same as Figure \ref{fig:diffmodel:hadro:intemiss:H1}, for model setup H3 or H4.}
\label{fig:diffmodel:hadro:intemiss:H3}
\end{figure*}

\begin{figure}[!]
\centering
\includegraphics[width=0.9\columnwidth]{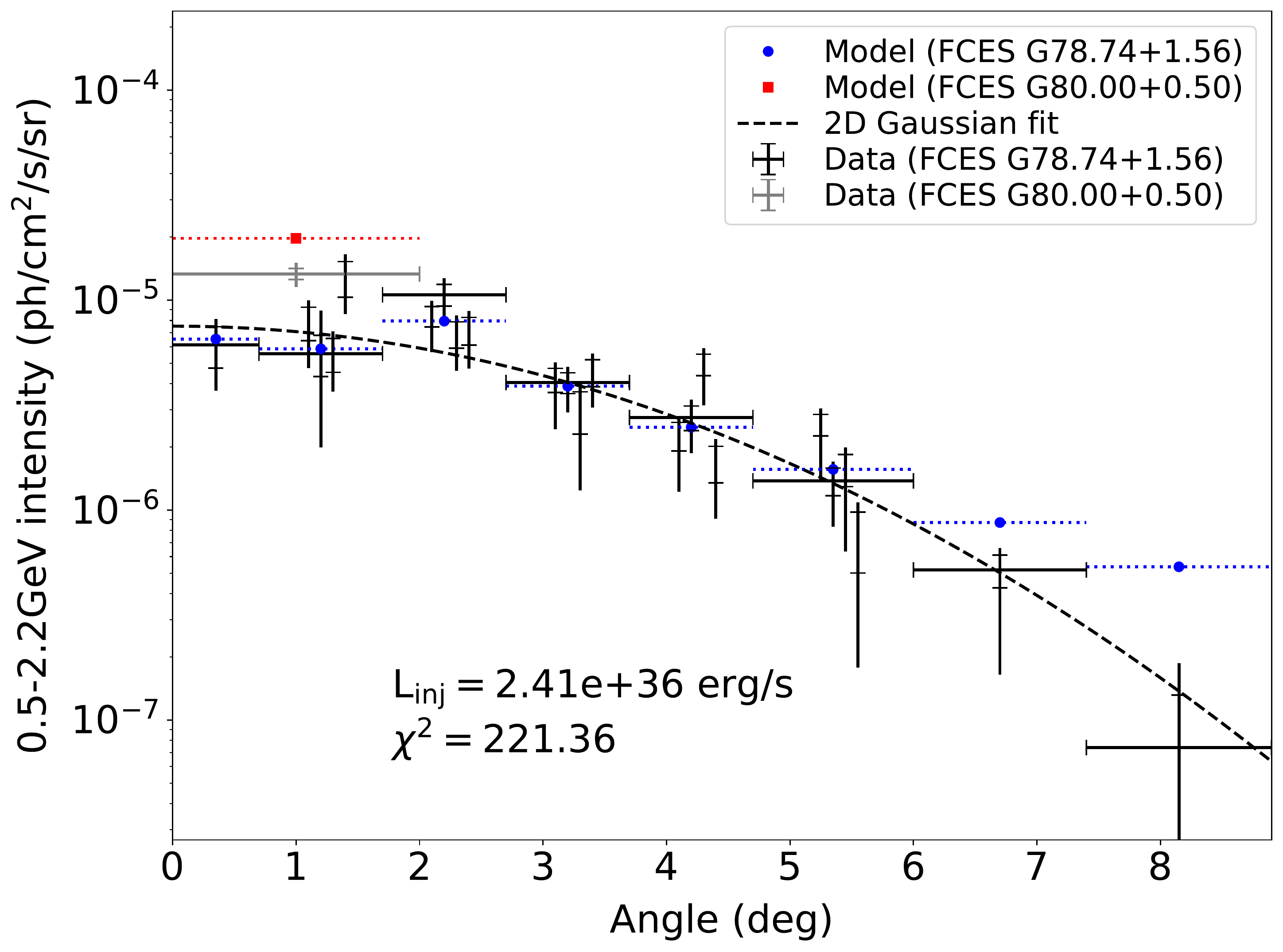} \\
\includegraphics[width=0.9\columnwidth]{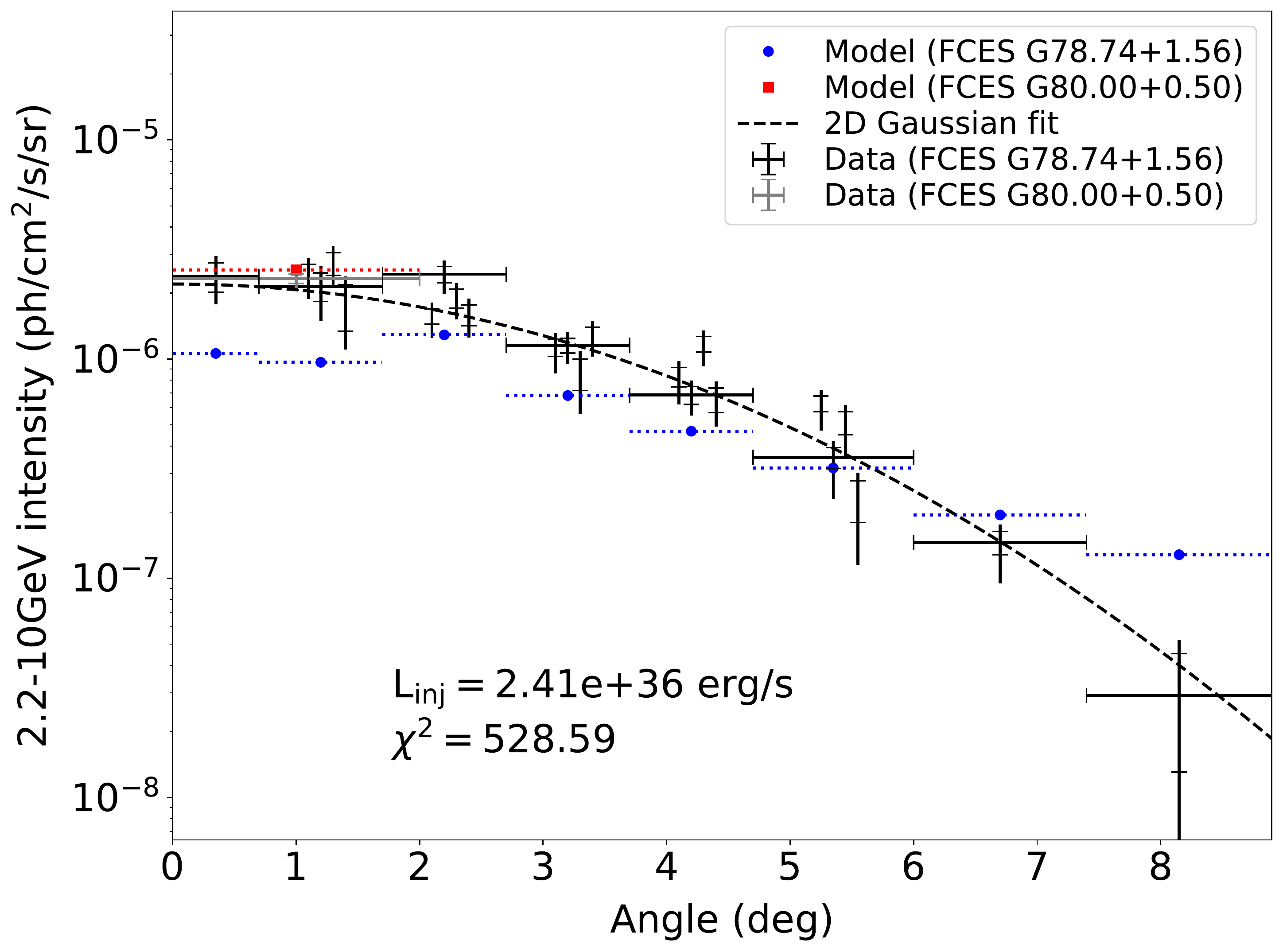} \\
\includegraphics[width=0.9\columnwidth]{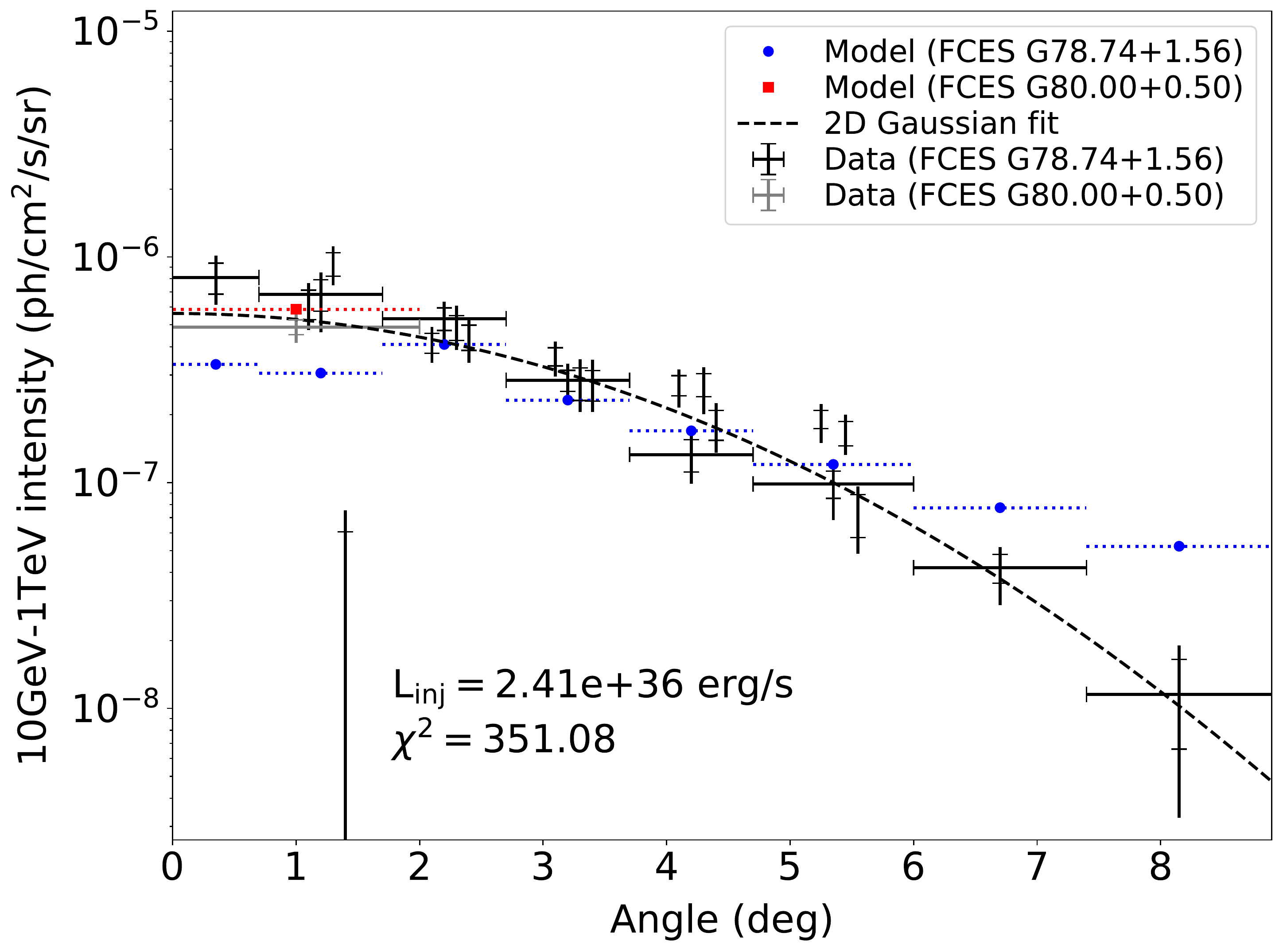}
\caption{Same as Figure \ref{fig:diffmodel:lepto:int:L1} for model setup L2.}
\label{fig:diffmodel:lepto:int:L2}
\end{figure}

\begin{figure}[!]
\centering
\includegraphics[width=0.9\columnwidth]{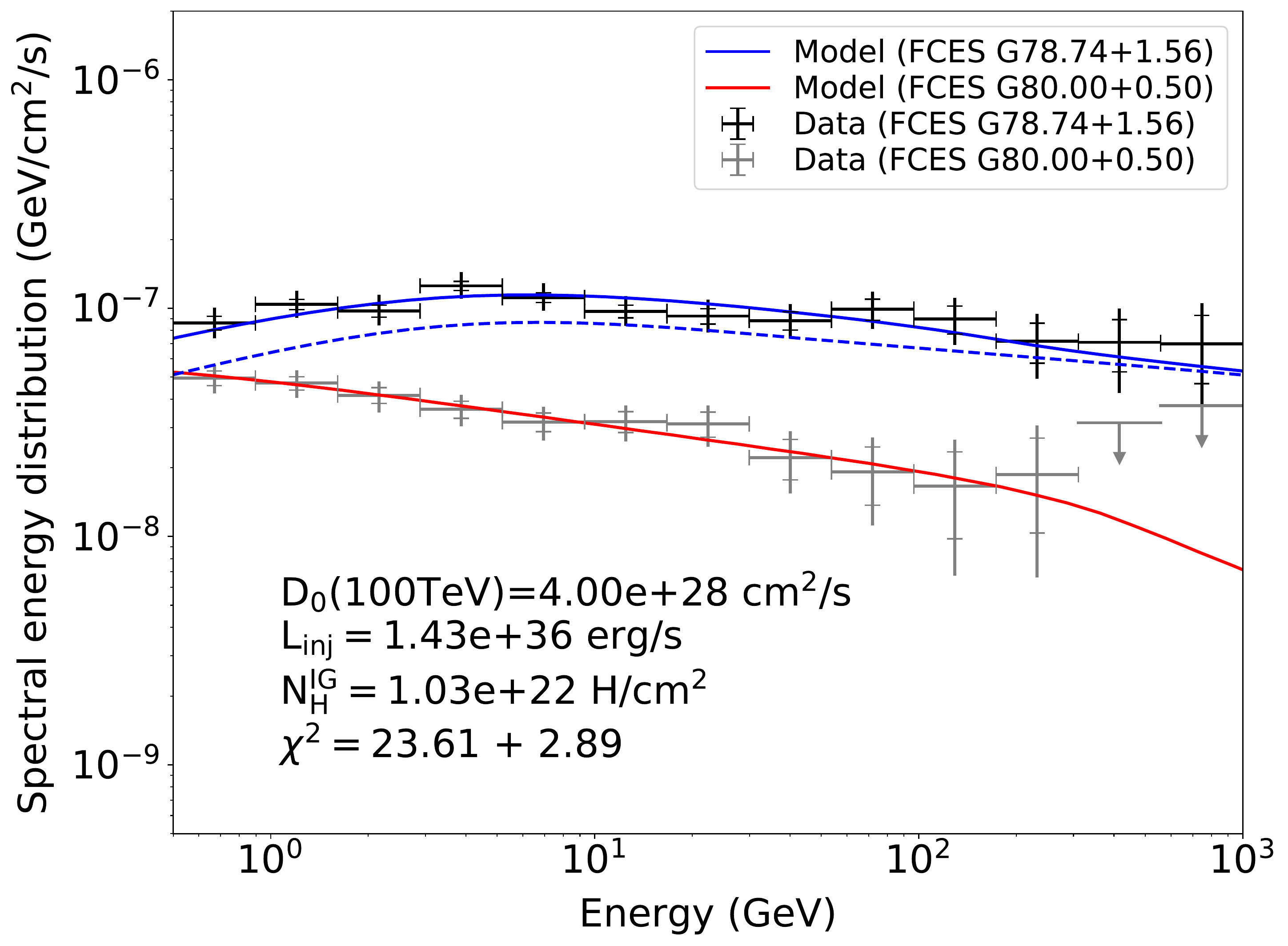}
\caption{Same as Figure \ref{fig:diffmodel:lepto:spec}  for the pulsar halo scenario.}
\label{fig:diffmodel:halo:spec}
\end{figure}

\begin{figure}[!]
\includegraphics[width=0.9\columnwidth]{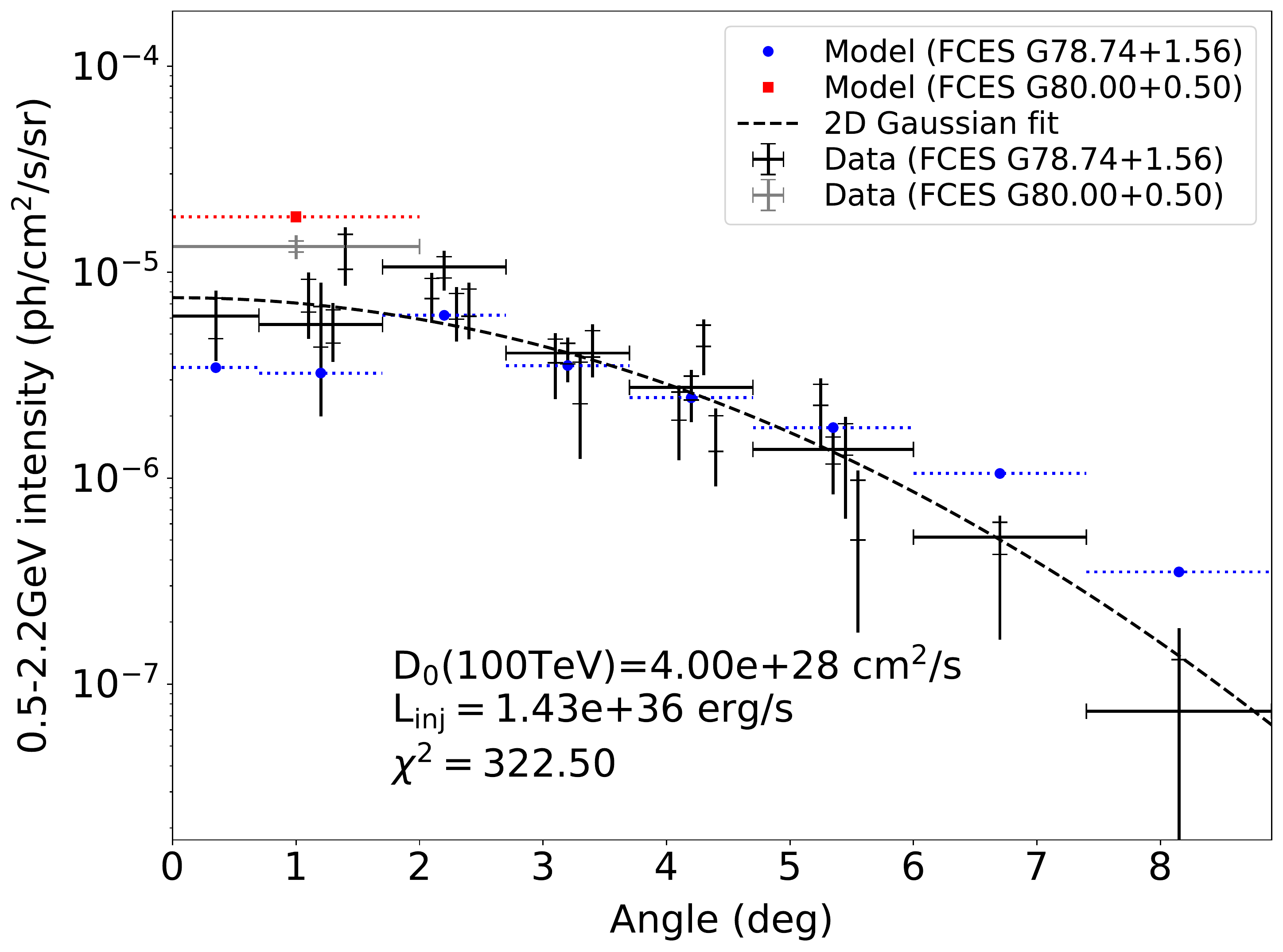} \\
\includegraphics[width=0.9\columnwidth]{plot_fit_intensity_profile_diffnorm4e+28_500MeV-2236MeV_nolocal-in-sed.pdf} \\
\includegraphics[width=0.9\columnwidth]{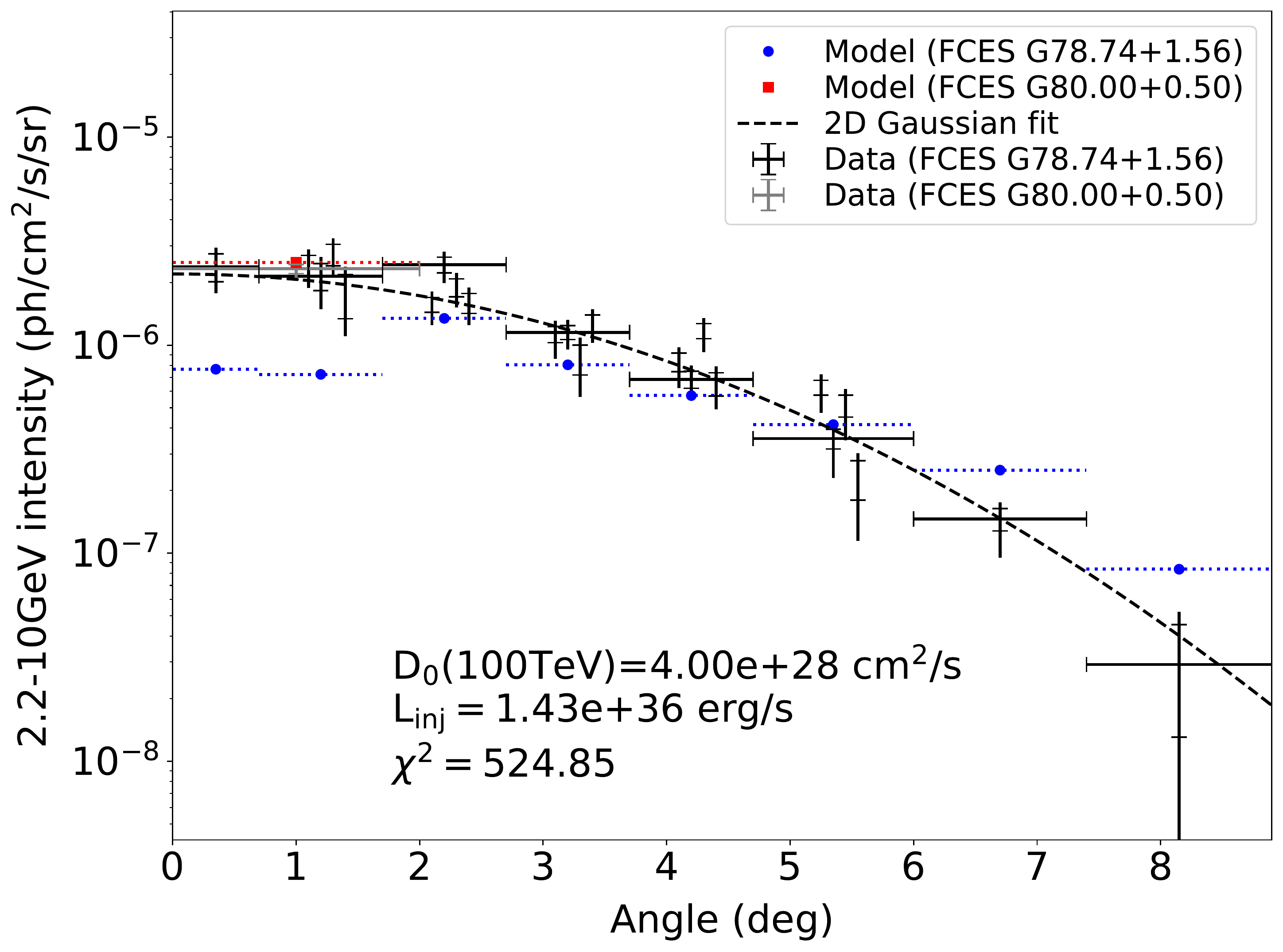}
\caption{Same as Figure \ref{fig:diffmodel:lepto:int:L1} for the pulsar halo scenario.}
\label{fig:diffmodel:psrhalo:int}
\end{figure}

We display here the results obtained for some of the diffusion scenarios considered in the interpretation of the results (see Section  \ref{sec:discussion:results}). The intensity and emissivity profiles for hadronic scenarios H2 and H3 or H4 are presented in Figs. \ref{fig:diffmodel:hadro:intemiss:H2} and \ref{fig:diffmodel:hadro:intemiss:H3}, and the intensity profiles for model leptonic scenario L2 are presented in Figure \ref{fig:diffmodel:lepto:int:L2}. The full set of results for the leptonic pulsar halo scenario is presented in Figs. \ref{fig:diffmodel:halo:spec} and \ref{fig:diffmodel:psrhalo:int}.

\end{appendix}

\end{document}